\pdfoutput=1

\documentclass{article}

\usepackage{arxiv}
\usepackage{natbib}
\usepackage[utf8]{inputenc} 
\usepackage[T1]{fontenc}    
\usepackage{url}            
\usepackage{booktabs}       
\usepackage{amsfonts}       
\usepackage{nicefrac}       
\usepackage{microtype}      
\usepackage{lipsum}
\usepackage{graphicx}
\usepackage{amsmath}
\usepackage{float}
\usepackage{mathtools}
\usepackage{amssymb}
\usepackage{algorithm}
\usepackage{algpseudocode}
\usepackage{amsfonts}
\usepackage{array}
\usepackage{booktabs}
\usepackage{arydshln}

\usepackage{multirow}

\usepackage{tabularx}

\usepackage{amsthm}
\usepackage{mathrsfs}
\usepackage{xcolor}
\usepackage{textcomp}
\usepackage{bm}
\usepackage{placeins}
\usepackage{threeparttable}
\usepackage[title]{appendix}

\newcommand{\bmhead}[1]{\medskip\noindent\textbf{#1}\par\nobreak\smallskip}
\providecommand{\backmatter}{}

\newcommand{\tridotT}{
  \overset{
    \smash{
      \lower0.6ex\hbox{$
        \mkern2mu
        \substack{
          \textstyle\cdot\phantom{\,\cdot}\\[-0.6ex]
          \textstyle\cdot\,\textstyle\cdot
        }
        \mkern16mu
      $}
    }
  }{T_{l}^{\mu\nu}}
}

\newcommand{\tridotTr}{
  \overset{
    \smash{
      \lower0.3ex\hbox{$
        \substack{
          \textstyle\cdot\phantom{\,\cdot}\\[-0.6ex]
          \textstyle\cdot\,\textstyle\cdot
        }
      $}
    }
  }{T_{r}}
}

\title{From Compact Orientation Distributions to Texture Evolution:\\
GSH-Initialized Orientation-Aware Deep Material Networks}

\author{%
  Ting-Ju Wei \\
  Special Melted Products Ltd \\
  President Way, Sheffield S4 7UR, United Kingdom \\
  \texttt{Ting-Ju.Wei@smp.ltd} \\
  \And
  Chuin-Shan Chen\thanks{Corresponding author.} \\
  Department of Civil Engineering \\
  Department of Materials Science and Engineering \\
  National Taiwan University \\
  No.~1, Sec.~4, Roosevelt Road, Taipei 10617, Taiwan \\
  \texttt{dchen@ntu.edu.tw} \\
}

\begin{document}
\maketitle
\begin{abstract}
The orientation-aware deep material network (ODMN) is a micromechanics-based surrogate for full-field crystal-plasticity simulations of representative volume elements (RVEs). The conventional ODMN learns the crystallographic orientations together with the interaction directions, so that a complex texture may require many material nodes and a high online cost. This study introduces an ODMN initialized by generalized spherical harmonics (GSH), termed the GSH-ODMN. The initial RVE texture is compacted into the \(2^N\) weighted orientations of a network of depth \(N\) by a volume-weighted GSH procedure. They are assigned to the material nodes and kept fixed, so that offline training identifies only the interaction directions. The best GSH truncation order increases with the network depth. The online Newton system is solved with a tree-structured lower-upper (LU) factorization that exploits the binary-tree sparsity of the network, so that the cost per iteration scales as \(O(2^N N^2)\) instead of \(O(8^N)\) for a dense solution. The framework is verified against full-field direct numerical simulations of two Inconel 625 (IN625) RVEs, one with a nearly random texture and one with a rolling texture, under isothermal and thermomechanical load-unload-reload paths. Compared with the conventional ODMN, the GSH-ODMN improves the predictions of the macroscopic stress and the texture evolution. At \(N=8\), the normalized texture-difference index between the predicted and reference textures remains below \(0.03\) in all cases. Finally, the GSH-ODMN is embedded at every integration point of a two-scale L-bracket model under cold and hot bending cycles, at an estimated cost three to four orders of magnitude below that of a concurrent full-field analysis.
\end{abstract}

\keywords{Deep Material Networks \and Crystal Plasticity \and Crystallographic Texture Evolution \and Orientation Distribution Functions \and Generalized Spherical Harmonics}

\section{Introduction}\label{sec1}

Multiscale simulation links microstructural physics to the macroscopic response of engineering materials. The link is especially important for polycrystalline materials, whose macroscopic behavior arises from the heterogeneous deformation of crystallographically anisotropic grains and depends on crystallographic orientation, grain morphology, and deformation history. Crystal plasticity (CP) describes the grain-scale constitutive response in terms of crystallographic slip and lattice rotation~\cite{PEIRCE19831951, LEBENSOHN19932611, ROTERS20101152, roters2011crystal}. It predicts the anisotropic macroscopic response together with grain-scale quantities such as local stress and strain, slip-system activity, and texture evolution~\cite{zhao2004texture, BECKER19952701}. Homogenizing the grain-scale fields over a representative volume element (RVE) yields the effective constitutive response required for structural-scale simulations~\cite{sachs1928ableitung, taylor1938plastic, terada2000simulation, roters2011crystal}.

Many homogenization schemes have been developed for polycrystalline materials. The classical Taylor model assumes a uniform deformation gradient in all grains and predicts macroscopic responses and deformation textures at low cost~\cite{taylor1938plastic}. The uniform-strain assumption, however, neglects grain-to-grain interactions and overconstrains the kinematics. Mean-field approaches, in particular self-consistent schemes, improve on the Taylor approximation by treating each grain as an inclusion embedded in an effective medium~\cite{tome2023material,LEBENSOHN19932611,berveiller1978extension}. They balance efficiency and accuracy, but they do not resolve intragranular field fluctuations or the interactions among neighboring grains. Grain-cluster formulations partly address this limitation by including local interactions among neighboring grains~\cite{tjahjanto2015multiscale}. Their predictions, however, can depend strongly on the prescribed cluster topology and on the penalty parameters that regularize grain-boundary incompatibility. These model-specific choices may require calibration, which limits transferability across microstructures and loading conditions.

Full-field CP methods overcome many of these limitations by discretizing the RVE in space and solving the corresponding boundary-value problem. The CP finite-element method (CPFEM) and the CP fast-Fourier-transform (CPFFT) method resolve the heterogeneous stress, strain, slip-system activity, and lattice rotation within grains and across grain boundaries~\cite{eisenlohr2013spectral,shanthraj2015numerically,roters2019damask,lebensohn2020spectral}. Full-field methods are accurate but computationally demanding. The nonlinear constitutive equations must be integrated at a large number of material points, and mechanical equilibrium must be solved iteratively over the entire RVE at every loading increment. The burden is most severe in concurrent multiscale simulations, in which an RVE boundary-value problem is solved at each integration point of the macroscopic model~\cite{terada2001class,watanabe2008characterization,gierden2022review,han2020microstructure}. The cost can become prohibitive for industrial process simulations with large deformation, complex loading paths, and many macroscopic degrees of freedom, which limits the use of full-field CP models in forming and machining processes such as deep drawing and turning. Micro--macro decoupling schemes reduce this cost by identifying the parameters of an assumed macroscopic constitutive model in advance from numerical material tests on the RVE~\cite{terada2013method}.

Data-driven surrogate models have also emerged as efficient alternatives to full-field CP homogenization. Feed-forward neural networks have been used to predict homogenized constitutive responses or spatially resolved deformation fields from microstructural descriptors and loading conditions~\cite{ibragimova2021new}. Convolutional neural networks have been used to approximate CPFEM responses of aluminum alloys~\cite{IBRAGIMOVA2022103374}. Recurrent neural networks, such as linearized minimal state cells, reproduce path-dependent material behavior and have replaced repeated CPFFT simulations in two-scale analyses, for example of dog-bone specimens under uniaxial loading~\cite{BONATTI2022104697,BONATTI2022103430}. More recently, hybrid architectures that combine graph and recurrent neural networks have been proposed to predict the homogenized response of polycrystalline materials under the Taylor assumption while accounting for grain-level interactions~\citep{hu2024temporal}. These surrogates greatly reduce the computational cost, but many of them learn a direct mapping from microstructure and loading history to the macroscopic stress. They therefore require a large number of nonlinear CP simulations to sample the relevant microstructures and loading paths. In addition, the material state is often represented by hidden variables without a micromechanical interpretation, which makes it difficult to evolve history-dependent internal variables and crystallographic texture in a physically consistent manner. Their accuracy may also deteriorate for microstructures, loading paths, or constitutive models outside the training domain.

In contrast to purely data-driven surrogates, deep material networks (DMNs) provide a physics-based framework for computational homogenization~\cite{liu2019deep,gajek2020micromechanics,noels2022micromechanics,WeiOverview}. A DMN is rooted in classical homogenization theory and has a hierarchical tree architecture, in which the leaf nodes represent the constituent phases and the internal building blocks perform analytical homogenization operations. The network parameters therefore have geometric and physical interpretations, such as constituent volume fractions and equilibrium directions. DMNs are trained offline on linear-elastic data only and then predict nonlinear material responses in online simulations~\cite{srinivas2026rapid,wan2024decoding}. Originally developed for short-fiber composites~\cite{liu2019deep,liu2019exploring,gajek2020micromechanics,noels2022micromechanics,wei2025foundation,jean2024graph}, DMNs have since been extended to other heterogeneous materials, including porous materials~\cite{noels2022interaction,haddad2026deep} and woven composites~\cite{shin2024deep2}, and to multiphysics homogenization, including thermomechanical coupling~\cite{shin2024deep,li2024micromechanics} and, more recently, electromechanical coupling in composite materials~\cite{wei2027deep}. For polycrystalline materials, the orientation-aware interaction-based DMN (ODMN) incorporates crystallographic orientations into the network and predicts texture evolution~\cite{wei2025orientation,wei2026parametric,wei2026efficientnonlinearmultiscaleprediction}.

The current ODMN has two limitations. First, an accurate prediction of texture evolution requires many material nodes to represent the initial distribution of crystallographic orientations. The larger network reduces the computational speed-up and limits the efficiency of the ODMN in large industrial simulations. Second, the existing formulation is restricted to mechanical homogenization and cannot account for thermomechanical loading. This matters in deformation processes in which the constitutive response depends strongly on temperature, as is common in thermally activated, dislocation-mediated plasticity.

To address the first limitation, this work adopts the texture-compaction procedure of Marki and Knezevic~\cite{marki2024nonlinear}, which is based on generalized spherical harmonics (GSH), to construct a compact set of weighted crystallographic orientations that represents the initial orientation distribution. The procedure optimizes the orientations and their weights in the space of GSH coefficients, so that the initial texture is represented with far fewer material nodes. The orientation-aware mechanism of the ODMN is reformulated in terms of Bunge Euler angles, the convention commonly used for texture description in materials science, and the matrix inverse required in the original formulation is eliminated. To address the second limitation, the ODMN is extended to thermomechanical loading, in which a prescribed macroscopic temperature history drives thermal expansion and temperature-dependent elasticity at the material nodes. In addition, the Newton system of the online stage is solved with a tree-structured lower-upper (LU) factorization that exploits the binary-tree sparsity of the network, so that the cost per iteration scales as $O(2^N N^2)$ instead of $O(8^N)$ for a dense solution, for a network of depth $N$. The resulting framework, referred to as the GSH-ODMN, is verified against CPFFT simulations of two RVEs, one with equiaxed grains and a nearly random texture and one with flattened grains and a rolling texture, along load-unload-reload (LUR) paths with and without a temperature cycle. Finally, the GSH-ODMN is applied to a two-scale thermomechanical simulation of an L-bracket in the commercial finite-element software Abaqus.

The remainder of this paper is organized as follows. Section~\ref{sec2} presents the GSH-based texture compaction, the revised orientation-aware formulation, and the thermomechanical online prediction. Section~\ref{sec3} verifies the framework against CPFFT simulations of two polycrystalline RVEs with different grain morphologies and initial textures. Section~\ref{sec4} applies it to a two-scale thermomechanical simulation of an L-bracket. Section~\ref{sec5} draws conclusions.

\section{Methodology}\label{sec2}

\begin{figure*}[tp]
\centering
\includegraphics[width=\textwidth]{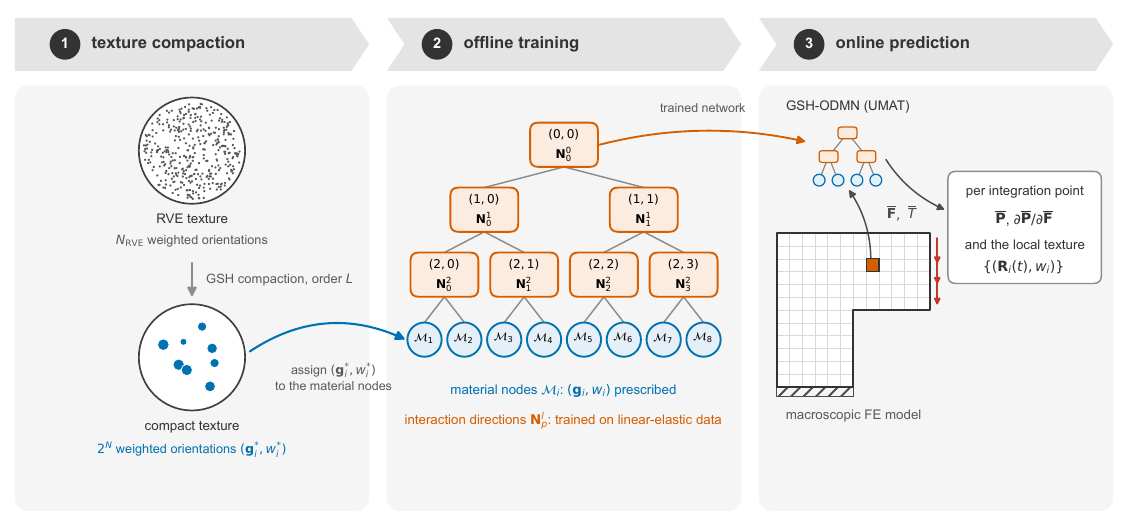}
\caption{GSH-ODMN framework in three stages. (1) GSH compaction of the RVE texture to $2^N$ weighted orientations $(\mathbf g_i^{*},w_i^{*})$. (2) Offline training of the binary-tree network (depth $N=3$ shown). The material nodes $\mathcal M_i$ are prescribed by the compact texture, and the interaction directions $\bm N_p^l$ are the only trained parameters. (3) Online evaluation of the trained network as a user material subroutine (UMAT) at every integration point of the macroscopic finite-element model.}
\label{fig:framework}
\end{figure*}

This section presents the GSH-ODMN. Section~\ref{sec21} introduces the ODMN architecture and its parameterization in terms of material nodes and homogenization building blocks. Section~\ref{sec22} describes the GSH-based texture compaction and its use for initializing the orientations and weights of the material nodes. Section~\ref{sec23} gives the revised orientation-aware mechanism. Section~\ref{sec24} covers the offline training, and Section~\ref{sec25} the nonlinear online prediction under thermomechanical loading. Fig.~\ref{fig:framework} summarizes the three stages of the framework and the notation used below.

\subsection{ODMN architecture and parameterization}
\label{sec21}

The ODMN architecture follows the hierarchical binary-tree formulation of our previous work~\cite{wei2025orientation}. The network consists of hierarchically connected homogenization building blocks, with material nodes attached to the terminal building blocks. Each material node $\mathcal{M}_i$ represents a subdomain $\Omega_i$ of the RVE domain $\Omega_{\mathrm{RVE}}$ and carries a volume fraction and a crystallographic orientation. For an ODMN of depth $N$, the $2^N$ subdomains form a partition of the RVE,
\begin{equation}
\Omega_{\mathrm{RVE}}
=
\bigcup_{i=1}^{2^N} \Omega_i,
\qquad
\Omega_i \cap \Omega_j = \varnothing
\quad \text{for } i \neq j.
\end{equation}
The volume fraction of material node $\mathcal{M}_i$ is
\begin{equation}
w_i =
\frac{|\Omega_i|}{|\Omega_{\mathrm{RVE}}|},
\qquad
\sum_{i=1}^{2^N} w_i = 1,
\end{equation}
where $|\cdot|$ denotes the volume of a domain. Each material node is also assigned a crystallographic orientation $\mathbf{g}_i$. The set of weighted orientations
\begin{equation}
\left\{
\left(\mathbf{g}_i,w_i\right)
\right\}_{i=1}^{2^N}
\end{equation}
is a discrete representation of the initial orientation distribution function (ODF).

The material-node subdomains are homogenized from the bottom of the tree upward until the root building block is reached, which gives the effective response of the RVE. Each building block combines the responses of its two child subdomains along an interaction direction $\bm N$. This direction is a unit vector parameterized by two dimensionless angular variables, $\theta$ and $\phi$,
\begin{equation}\label{eq:direction_vector}
\bm N =
\begin{bmatrix}
\cos(2\pi\phi)\sin(\pi\theta) \\
\sin(2\pi\phi)\sin(\pi\theta) \\
\cos(\pi\theta)
\end{bmatrix}.
\end{equation}

Each homogenization building block is identified by its level $l$ and position $p$ within the hierarchical binary tree. The root building block is located at level $l=0$. Level $l$ contains $2^l$ building blocks, indexed by $p=0,\ldots,2^l-1$. An ODMN of depth $N$ therefore contains $N$ levels of building blocks, $l=0,\ldots,N-1$, and $2^N-1$ building blocks in total. The two children of building block $(l,p)$ are $(l+1,2p)$ and $(l+1,2p+1)$. For each building block $(l,p)$, let $\mathcal{J}^{0}_{l,p}$ and $\mathcal{J}^{1}_{l,p}$ denote the sets of material nodes that belong to its first and second child subdomains. These sets follow recursively from the binary-tree topology,
\begin{equation} \label{eq:J0_defined} 
\mathcal{J}^{0}_{l,p} = \begin{cases} 
\mathcal{J}^{0}_{l+1,\,2p} \cup \mathcal{J}^{1}_{l+1,\,2p}, & l<N-1, \\[1ex] \left\{\mathcal{M}_{2p+1}\right\}, & l=N-1, \end{cases} \end{equation}
and 
\begin{equation} \label{eq:J1_defined} \mathcal{J}^{1}_{l,p} = \begin{cases} \mathcal{J}^{0}_{l+1,\,2p+1} \cup \mathcal{J}^{1}_{l+1,\,2p+1}, & l<N-1, \\[1ex] \left\{\mathcal{M}_{2p+2}\right\}, & l=N-1. \end{cases} \end{equation} At the terminal level $l=N-1$, each child subdomain is represented by a single material node. At higher levels, $\mathcal{J}^{0}_{l,p}$ and $\mathcal{J}^{1}_{l,p}$ contain all material nodes that descend from the corresponding child branches. The network topology and the indexing convention are described in detail in our previous ODMN work~\cite{wei2025orientation}.

The complete parameter set of an ODMN of depth $N$ is
\begin{equation}\label{eq:parameters}
\begin{aligned}
\mathcal{F} =\;
& \left\{
w_i, \mathbf{g}_i
\;\middle|\;
1\le i\le 2^N
\right\}
\\
& \cup
\left\{
\theta_p^l,\phi_p^l
\;\middle|\;
0\le l<N,\;
0\le p<2^l
\right\},
\end{aligned}
\end{equation}
where the first set contains the volume fractions and orientations of the material nodes, and the second set contains the angular variables $\theta_p^l$ and $\phi_p^l$ that define the interaction direction $\bm N_p^l$ of each building block $(l,p)$ through Eq.~\eqref{eq:direction_vector}.

\subsection{GSH-based texture compaction and ODMN initialization}
\label{sec22}

In the conventional ODMN, the orientations and weights of the material nodes are trainable parameters identified during offline training. Here, a compact set of weighted orientations is instead constructed from the texture of the RVE with the GSH compaction procedure of Marki and Knezevic~\cite{marki2024nonlinear}. The resulting orientations and weights are assigned to the material nodes as their crystallographic orientations and volume fractions and are held fixed during offline training.

Let the texture of the RVE be described by the ODF $f_{\mathrm{RVE}}(\mathbf{g})$, where $\mathbf{g}$ denotes a crystallographic orientation. The ODF is a normalized probability density over the orientation space $\mathcal{G}$,
\begin{equation}
f_{\mathrm{RVE}}(\mathbf{g})\,d\mathbf{g}
=
\frac{dV}{V_{\mathrm{RVE}}},
\qquad
\int_{\mathcal{G}}
f_{\mathrm{RVE}}(\mathbf{g})\,d\mathbf{g}
=
1,
\label{eq:odf_definition}
\end{equation}
where $V_{\mathrm{RVE}}=|\Omega_{\mathrm{RVE}}|$ is the RVE volume and $dV$ is the material volume whose orientations lie within the orientation-space element $d\mathbf{g}$. The orientation is parameterized by the Bunge Euler angles,
\begin{equation}
\mathbf{g}
=
(\varphi_1,\Phi,\varphi_2),
\end{equation}
with the normalized invariant volume element of the orientation space
\begin{equation}
d\mathbf{g}
=
\frac{1}{8\pi^2}
\sin\Phi\,
d\varphi_1\,d\Phi\,d\varphi_2,
\qquad
\int_{\mathcal{G}} d\mathbf{g}=1.
\end{equation}
With this normalization, a random texture corresponds to $f_{\mathrm{RVE}}=1$, and the ODF is expressed in multiples of a random distribution (m.r.d.)~\cite{bunge2013texture}.

The ODF is expanded in GSH basis functions. Before crystal and sample symmetries are taken into account, these functions are
\begin{equation}
T_l^{mn}(\mathbf{g})
=
e^{im\varphi_2}
P_l^{mn}(\cos\Phi)
e^{in\varphi_1},
\label{eq:gsh_basis}
\end{equation}
where $i$ in the exponents is the imaginary unit, $P_l^{mn}$ is the generalized Legendre function, $l$ is the harmonic degree, and $m$ and $n$ are the harmonic orders, with $-l\le m,n\le l$~\cite{bunge2013texture}. Throughout this subsection, $l$ denotes the harmonic degree, as in the texture literature. It is unrelated to the tree level of Section~\ref{sec21}.

Crystal and sample symmetries reduce the number of independent GSH components by combining symmetry-equivalent basis functions. Following the notation of
Marki and Knezevic~\cite{marki2024nonlinear}, the GSH basis that incorporates both symmetries is
\begin{equation}
\tridotT(\mathbf{g})
=
\sum_{m(\mu)}
\sum_{n(\nu)}
\ddot{A}_l^{m\mu}
\dot{A}_l^{n\nu}
T_l^{mn}(\mathbf{g}),
\label{eq:sym_gsh}
\end{equation}
where $\ddot{A}_l^{m\mu}$ and $\dot{A}_l^{n\nu}$ are the symmetry coefficients of the crystal and sample symmetries, respectively, and the sums run over the orders $m$ and $n$ that contribute to the symmetric functions with the independent indices $\mu$ and $\nu$. The dots above $T$ and $A$ belong to the symmetry notation of Bunge~\cite{bunge2013texture} and are not time derivatives.

With the symmetry-adapted basis, the ODF truncated at order $L$ is \begin{equation}
f(\mathbf{g}) = \sum_{l=0}^{L} \sum_{\mu=1}^{M(l)} \sum_{\nu=1}^{K(l)} C_l^{\mu\nu} \tridotT(\mathbf{g}), \label{eq:gsh_expansion} 
\end{equation} 
where $C_l^{\mu\nu}$ are the GSH expansion coefficients, and $M(l)$ and $K(l)$ are the numbers of independent symmetric functions of degree $l$ for the crystal and sample symmetries, respectively~\cite{bunge2013texture}. In this work, cubic crystal symmetry and triclinic sample symmetry are assumed, so that $K(l)=2l+1$. From the orthogonality of the GSH basis, the expansion coefficients are
\begin{equation} 
C_l^{\mu\nu} = (2l+1) \int_{\mathcal{G}} f(\mathbf{g}) \left[ \tridotT(\mathbf{g}) \right]^{*} \,d\mathbf{g}, \label{eq:gsh_coeff_integral} \end{equation} 
where $(\cdot)^*$ denotes the complex conjugate.

For a numerically generated RVE, the crystallographic texture is available as a finite set of weighted orientations,
\begin{equation}
\mathcal{T}_{\mathrm{RVE}}
=
\left\{
\left(
\mathbf{g}^{\mathrm{RVE}}_q,
v_q
\right)
\right\}_{q=1}^{N_{\mathrm{RVE}}},
\qquad
\sum_{q=1}^{N_{\mathrm{RVE}}} v_q = 1,
\label{eq:rve_texture}
\end{equation}
where $N_{\mathrm{RVE}}$ is the number of orientations, here one per grain, $\mathbf{g}^{\mathrm{RVE}}_q$ is the $q$th orientation, and $v_q$ is its volume fraction. For a
voxel-based RVE, $v_q$ is the fraction of voxels
that carry the $q$th orientation.

For this discrete texture, the integral in Eq.~\eqref{eq:gsh_coeff_integral} becomes a weighted sum. Without the factor $(2l+1)$, it gives the normalized GSH coefficients of the target RVE texture, marked by a hat,
\begin{equation}
\widehat{C}_l^{\mu\nu,\mathrm{RVE}}
=
\sum_{q=1}^{N_{\mathrm{RVE}}}
v_q
\left[
\tridotT\left(
\mathbf{g}^{\mathrm{RVE}}_q
\right)
\right]^*.
\label{eq:rve_gsh_coeff}
\end{equation}
The coefficients of the expansion in Eq.~\eqref{eq:gsh_expansion} are then
\begin{equation}
C_l^{\mu\nu,\mathrm{RVE}}
=
(2l+1)
\widehat{C}_l^{\mu\nu,\mathrm{RVE}}.
\label{eq:rve_gsh_coeff_scaled}
\end{equation}
Retaining the independent coefficients up to the truncation
order $L$ represents the RVE texture by a finite-dimensional
vector of GSH coefficients.

Texture compaction constructs a reduced set of weighted orientations
with exactly the number of orientations required by
the ODMN. An ODMN of depth $N$ has $N_m=2^N$ material nodes,
and the compact texture, marked by the label $\mathrm c$, is written as
\begin{equation}
\mathcal{T}_{\mathrm{c}}
=
\left\{
\left(
\mathbf{g}_i,
w_i
\right)
\right\}_{i=1}^{N_m},
\qquad
\sum_{i=1}^{N_m} w_i = 1,
\label{eq:compact_texture}
\end{equation}
where $\mathbf{g}_i$ and $w_i$ are the $i$th compact orientation and its
weight.

For a trial compact texture, the normalized GSH coefficients
are
\begin{equation}
\widehat{C}_l^{\mu\nu,\mathrm{c}}
=
\sum_{i=1}^{N_m}
w_i
\left[
\tridotT\left(
\mathbf{g}_i
\right)
\right]^*.
\label{eq:compact_gsh_coeff}
\end{equation}
Following Marki and Knezevic~\cite{marki2024nonlinear}, auxiliary
nonnegative variables $\zeta_i$ are introduced to parameterize the
orientation weights as
\begin{equation}
w_i
=
\frac{\zeta_i}
{\displaystyle\sum_{j=1}^{N_m}\zeta_j},
\qquad
\zeta_i \geq 0,
\label{eq:weight_normalization}
\end{equation}
so that the unit-sum constraint on the weights is
satisfied automatically. A single index $r$ enumerates all independent components $(l,\mu,\nu)$ retained up to order $L$.
The normalized GSH coefficients of the compact texture then read
\begin{equation}
\widehat{C}_{r}^{\mathrm{c}}
=
\frac{
\displaystyle
\sum_{i=1}^{N_m}
\zeta_i
\left[
\tridotTr\left(
\mathbf{g}_i
\right)
\right]^*
}{
\displaystyle
\sum_{j=1}^{N_m}
\zeta_j
},
\label{eq:gsh_coeff_alpha}
\end{equation}
where $\tridotTr$ is the symmetry-adapted GSH basis function
of the $r$th retained component. The residual between the
compact and target textures is
\begin{equation}
G_r
=
\widehat{C}_{r}^{\mathrm{c}}
-
\widehat{C}_{r}^{\mathrm{RVE}},
\qquad
r=1,\ldots,N_{\mathrm{GSH}},
\label{eq:gsh_residual}
\end{equation}
where $N_{\mathrm{GSH}}$ is the number of independent GSH coefficients
included in the fit. The fitting error is the
Euclidean norm of the residual vector,
\begin{equation}
\mathcal{E}
=
\left[
\sum_{r=1}^{N_{\mathrm{GSH}}}
\left|G_r\right|^2
\right]^{1/2}.
\label{eq:gsh_error}
\end{equation}

Both the orientations and their weights are
optimized. The design variables are collected into the vector
\begin{equation}
\begin{aligned}
\mathbf{x}_{\mathrm{GSH}}
={}&
\left[
\varphi_1^{1},
\Phi^{1},
\varphi_2^{1},
\zeta_1,
\ldots,
\right.
\\
&\left.
\varphi_1^{N_m},
\Phi^{N_m},
\varphi_2^{N_m},
\zeta_{N_m}
\right]^{\mathsf T},
\end{aligned}
\label{eq:gsh_design_variables}
\end{equation}
where the superscripts on the Euler angles number the compact orientations. This gives $4N_m$ optimization variables. The compact texture
is obtained by solving the nonlinear optimization problem
\begin{equation}
\mathbf{x}_{\mathrm{GSH}}^{*}
=
\underset{\mathbf{x}_{\mathrm{GSH}}}{\arg\min}
\;
\mathcal{E}\left(
\mathbf{x}_{\mathrm{GSH}}
\right),
\label{eq:gsh_optimization}
\end{equation}
subject to the admissible ranges of the Bunge Euler angles and the
nonnegativity constraints $\zeta_i \geq 0$. Because the GSH coefficients
are analytical functions of the Euler angles and weights,
the analytical gradient and Hessian derived by Marki and
Knezevic~\cite{marki2024nonlinear} are used to solve the problem.

The optimized compact texture, marked by an asterisk on the orientations and weights, is
\begin{equation}
\mathcal{T}_{\mathrm{c}}^{*}
=
\left\{
\left(
\mathbf{g}_i^{*},
w_i^{*}
\right)
\right\}_{i=1}^{N_m},
\qquad
N_m=2^N,
\label{eq:optimized_compact_texture}
\end{equation}
for which the retained GSH coefficients satisfy
\begin{equation}
\widehat{C}_{r}^{\mathrm{c}}
\left(
\mathcal{T}_{\mathrm{c}}^{*}
\right)
\approx
\widehat{C}_{r}^{\mathrm{RVE}},
\qquad
r=1,\ldots,N_{\mathrm{GSH}}.
\label{eq:gsh_matching}
\end{equation}
The optimized weighted orientations are then assigned to the material nodes of the ODMN,
\begin{equation}
\mathbf{g}_i
\leftarrow
\mathbf{g}_i^{*},
\qquad
w_i
\leftarrow
w_i^{*},
\qquad
i=1,\ldots,N_m.
\label{eq:gsh_odmn_mapping}
\end{equation}
The orientations and volume fractions of the
material nodes are thus prescribed by the compact representation of the
RVE texture and are not identified through offline training.
The $i$th compact orientation is assigned to the $i$th material node, with the material nodes numbered from left to right in the binary tree, in the order in which the orientations are returned by the compaction. This assignment is arbitrary, because the compaction does not place similar orientations in neighboring material nodes. Its effect on the predictions is examined in Section~\ref{sec34}.

\subsection{Revised orientation-aware mechanism with Bunge Euler angles}
\label{sec23}

The orientation-aware mechanism associates each material node with a
crystallographic orientation and provides the transformation between the
crystal and sample reference frames. Here, the
orientations obtained from the GSH compaction of
Section~\ref{sec22} are used directly as Bunge Euler angles. The same
angles then enter the GSH representation, the rotation of the crystal-frame
stiffness during offline training, and the initialization of the
CP constitutive model during online prediction.

\subsubsection{Bunge Euler angles and crystal-to-sample rotation}

For material node $\mathcal{M}_i$, the crystallographic orientation is
parameterized by the Bunge Euler angles
\begin{equation}
\mathbf{g}_i
=
\left(
\varphi_1^i,\Phi^i,\varphi_2^i
\right),
\label{eq:bunge_triplet}
\end{equation}
where all angles are expressed in radians. To distinguish the orientation
variable $\mathbf{g}_i$ used in the GSH representation from its matrix
representation, the associated Bunge orientation matrix is denoted by
$\bm G_i$.

The right-handed active rotations by an angle $\omega$ about the $z$- and
$x$-axes are
\begin{equation}
\begin{aligned}
\bm R_z(\omega)
&=
\begin{bmatrix}
\cos\omega & -\sin\omega & 0 \\
\sin\omega &  \cos\omega & 0 \\
0          &  0          & 1
\end{bmatrix},
\\[1ex]
\bm R_x(\omega)
&=
\begin{bmatrix}
1 & 0          & 0 \\
0 & \cos\omega & -\sin\omega \\
0 & \sin\omega &  \cos\omega
\end{bmatrix}.
\end{aligned}
\label{eq:elementary_rotations}
\end{equation}
Under the Bunge convention, the orientation matrix $\bm G_i$ defines the
passive transformation of vector components from the sample frame
$(\mathrm{s})$ to the crystal frame $(\mathrm{c})$,
\begin{equation}
\bm v^{\mathrm c}
=
\bm G_i
\bm v^{\mathrm s},
\quad
\bm G_i
=
\bm R_z(-\varphi_2^i)
\bm R_x(-\Phi^i)
\bm R_z(-\varphi_1^i),
\label{eq:bunge_passive}
\end{equation}
where $\bm v^{\mathrm s}$ and $\bm v^{\mathrm c}$ collect the components of a vector in the sample and crystal frames. The resulting orientation matrix is
\begin{equation}
\bm G_i
=
\begin{bmatrix}
c_1 c_2-s_1 s_2 c_\Phi
&
s_1 c_2+c_1 s_2 c_\Phi
&
s_2 s_\Phi
\\
-c_1 s_2-s_1 c_2 c_\Phi
&
-s_1 s_2+c_1 c_2 c_\Phi
&
c_2 s_\Phi
\\
s_1 s_\Phi
&
-c_1 s_\Phi
&
c_\Phi
\end{bmatrix},
\label{eq:bunge_matrix}
\end{equation}
where
\begin{equation}
\begin{aligned}
c_1 &= \cos\varphi_1^i,
\quad
s_1 = \sin\varphi_1^i,
\quad
c_2 = \cos\varphi_2^i,
\\
s_2 &= \sin\varphi_2^i,
\quad
c_\Phi = \cos\Phi^i,
\quad
s_\Phi = \sin\Phi^i.
\end{aligned}
\end{equation}

Since $\bm G_i$ is a proper rotation, that is, an element of the special orthogonal group $SO(3)$, it satisfies
\begin{equation}
\bm G_i^{-1}
=
\bm G_i^{\mathsf T},
\end{equation}
where the superscript $\mathsf T$ denotes the transpose. The inverse transformation from the crystal frame to the
sample frame is therefore
\begin{equation}
\bm v^{\mathrm s}
=
\bm R_i
\bm v^{\mathrm c},
\qquad
\bm R_i
=
\bm G_i^{-1}
=
\bm G_i^{\mathsf T},
\label{eq:crystal_sample_rotation}
\end{equation}
where $\bm R_i$ is the crystal-to-sample rotation matrix used in the
orientation-aware formulation. From Eq.~\eqref{eq:bunge_passive},
\begin{equation}
\bm R_i
=
\bm R_z(\varphi_1^i)
\bm R_x(\Phi^i)
\bm R_z(\varphi_2^i).
\label{eq:bunge_active}
\end{equation}

The matrices $\bm G_i$ and $\bm R_i$ thus represent the same
orientation with opposite transformation directions.
The rotation $\bm R_i$ is used
for the crystal-to-sample transformation of the stiffness
during offline training, and $\bm G_i$ initializes the CP
model during online prediction.

\subsubsection{Rotation of the elastic stiffness for offline training}

During offline training, the elastic stiffness defined in the crystal
frame must be transformed to the sample frame according to the
orientation of each material node. Let
$\mathbb{C}^{\mathrm c}$ denote the fourth-order elastic stiffness tensor
in the crystal frame, with components $C_{pqrs}^{\mathrm c}$. For material node $\mathcal{M}_i$, the
sample-frame components follow from the transformation rule of
a fourth-order tensor,
\begin{equation}
C_{abcd}^{\mathrm{s},i}
=
R_{ap}^{i}
R_{bq}^{i}
R_{cr}^{i}
R_{ds}^{i}
C_{pqrs}^{\mathrm c},
\label{eq:fourth_order_rotation}
\end{equation}
where $R_{ap}^{i}$ are the components of $\bm R_i$ and summation over repeated indices is implied. Since $\bm R_i=\bm G_i^{\mathsf T}$, Eq.~\eqref{eq:fourth_order_rotation}
can equivalently be written with the components $G_{pa}^{i}$ of the Bunge orientation
matrix as
\begin{equation}
C_{abcd}^{\mathrm{s},i}
=
G_{pa}^{i}
G_{qb}^{i}
G_{rc}^{i}
G_{sd}^{i}
C_{pqrs}^{\mathrm c}.
\label{eq:fourth_order_rotation_passive}
\end{equation}

For the numerical implementation, the symmetric stress tensor $\bm\sigma$ and strain tensor $\bm\varepsilon$ are
stored in Voigt notation, marked by the superscript $\mathrm V$, as
\begin{equation}
\bm{\sigma}^{\mathrm V}
=
\begin{bmatrix}
\sigma_{11} &
\sigma_{22} &
\sigma_{33} &
\sigma_{23} &
\sigma_{13} &
\sigma_{12}
\end{bmatrix}^{\mathsf T},
\label{eq:voigt_stress}
\end{equation}
and
\begin{equation}
\bm{\varepsilon}^{\mathrm V}
=
\begin{bmatrix}
\varepsilon_{11} &
\varepsilon_{22} &
\varepsilon_{33} &
2\varepsilon_{23} &
2\varepsilon_{13} &
2\varepsilon_{12}
\end{bmatrix}^{\mathsf T}.
\label{eq:voigt_strain}
\end{equation}
With this engineering-strain convention, the tensor inner product is
preserved in Voigt form:
\begin{equation}
\bm{\sigma}:\bm{\varepsilon}
=
\left(
\bm{\sigma}^{\mathrm V}
\right)^{\mathsf T}
\bm{\varepsilon}^{\mathrm V}.
\label{eq:voigt_work_conjugacy}
\end{equation}
The crystal-to-sample transformation of a symmetric second-order tensor
is
\begin{equation}
\bm{\sigma}^{\mathrm s}
=
\bm R_i
\bm{\sigma}^{\mathrm c}
\bm R_i^{\mathsf T},
\qquad
\bm{\varepsilon}^{\mathrm s}
=
\bm R_i
\bm{\varepsilon}^{\mathrm c}
\bm R_i^{\mathsf T},
\label{eq:second_order_rotation}
\end{equation}
where the superscripts $\mathrm s$ and $\mathrm c$ refer to the sample and crystal frames. The corresponding Voigt vectors transform as
\begin{equation}
\bm{\sigma}^{\mathrm V,s}
=
\bm M_\sigma(\bm R_i)
\bm{\sigma}^{\mathrm V,c},
\qquad
\bm{\varepsilon}^{\mathrm V,s}
=
\bm M_\varepsilon(\bm R_i)
\bm{\varepsilon}^{\mathrm V,c},
\label{eq:voigt_transformations}
\end{equation}
where $\bm M_\sigma$ and $\bm M_\varepsilon$ are the $6\times6$ transformation
matrices for the Voigt stress and the engineering strain, respectively. For brevity, let
$r_{ab} \equiv R_{ab}^{i}$. With the Voigt ordering
$(11,22,33,23,13,12)$ adopted above, the stress transformation matrix is
\begin{equation}
\bm M_\sigma(\bm R_i)
=
\begin{bmatrix}
\bm M_{\sigma}^{(n)} & \bm M_{\sigma}^{(s)}
\end{bmatrix},
\label{eq:voigt_stress_transformation_matrix}
\end{equation}
where the $6\times3$ blocks $\bm M_{\sigma}^{(n)}$ and $\bm M_{\sigma}^{(s)}$ act on the normal and shear components, respectively, and are given by
\begin{equation}
\bm M_{\sigma}^{(n)}
=
\begin{bmatrix}
r_{11}^2 & r_{12}^2 & r_{13}^2 \\
r_{21}^2 & r_{22}^2 & r_{23}^2 \\
r_{31}^2 & r_{32}^2 & r_{33}^2 \\
r_{21}r_{31} & r_{22}r_{32} & r_{23}r_{33} \\
r_{11}r_{31} & r_{12}r_{32} & r_{13}r_{33} \\
r_{11}r_{21} & r_{12}r_{22} & r_{13}r_{23}
\end{bmatrix},
\end{equation}
and
\begingroup\footnotesize\setlength{\arraycolsep}{2.5pt}
\begin{equation}
\bm M_{\sigma}^{(s)}
=
\begin{bmatrix}
2r_{12}r_{13} &
2r_{11}r_{13} &
2r_{11}r_{12}
\\
2r_{22}r_{23} &
2r_{21}r_{23} &
2r_{21}r_{22}
\\
2r_{32}r_{33} &
2r_{31}r_{33} &
2r_{31}r_{32}
\\
r_{22}r_{33}+r_{23}r_{32} &
r_{21}r_{33}+r_{23}r_{31} &
r_{21}r_{32}+r_{22}r_{31}
\\
r_{12}r_{33}+r_{13}r_{32} &
r_{11}r_{33}+r_{13}r_{31} &
r_{11}r_{32}+r_{12}r_{31}
\\
r_{12}r_{23}+r_{13}r_{22} &
r_{11}r_{23}+r_{13}r_{21} &
r_{11}r_{22}+r_{12}r_{21}
\end{bmatrix}.
\end{equation}
\endgroup

The scalar contraction between stress and strain is invariant
under rotation,
\begin{equation}
\left(
\bm{\sigma}^{\mathrm V,s}
\right)^{\mathsf T}
\bm{\varepsilon}^{\mathrm V,s}
=
\left(
\bm{\sigma}^{\mathrm V,c}
\right)^{\mathsf T}
\bm{\varepsilon}^{\mathrm V,c}.
\label{eq:stress_power_invariance}
\end{equation}
Substituting Eq.~\eqref{eq:voigt_transformations} into
Eq.~\eqref{eq:stress_power_invariance} yields
\begin{equation}
\left(
\bm{\sigma}^{\mathrm V,c}
\right)^{\mathsf T}
\bm M_\sigma(\bm R_i)^{\mathsf T}
\bm M_\varepsilon(\bm R_i)
\bm{\varepsilon}^{\mathrm V,c}
=
\left(
\bm{\sigma}^{\mathrm V,c}
\right)^{\mathsf T}
\bm{\varepsilon}^{\mathrm V,c}.
\end{equation}
Since this relation must hold for arbitrary stress and strain states,
the two transformation matrices satisfy
\begin{equation}
\bm M_\sigma(\bm R_i)^{\mathsf T}
\bm M_\varepsilon(\bm R_i)
=
\bm I,
\label{eq:voigt_duality}
\end{equation}
where $\bm I$ is the identity matrix, or equivalently,
\begin{equation}
\bm M_\varepsilon(\bm R_i)^{-1}
=
\bm M_\sigma(\bm R_i)^{\mathsf T}.
\label{eq:voigt_inverse_relation}
\end{equation}
This duality follows from the work conjugacy between
the Voigt stress vector and the engineering-strain vector.

The constitutive relation in the crystal frame is
\begin{equation}
\bm{\sigma}^{\mathrm V,c}
=
\bm C^{\mathrm c}
\bm{\varepsilon}^{\mathrm V,c},
\label{eq:crystal_hooke}
\end{equation}
where $\bm C^{\mathrm c}$ is the $6\times6$ Voigt matrix of $\mathbb{C}^{\mathrm c}$. Combining this relation with Eq.~\eqref{eq:voigt_transformations} gives
\begin{align}
\bm{\sigma}^{\mathrm V,s}
&=
\bm M_\sigma(\bm R_i)
\bm C^{\mathrm c}
\bm{\varepsilon}^{\mathrm V,c}
\nonumber\\
&=
\bm M_\sigma(\bm R_i)
\bm C^{\mathrm c}
\bm M_\varepsilon(\bm R_i)^{-1}
\bm{\varepsilon}^{\mathrm V,s}.
\end{align}
Using Eq.~\eqref{eq:voigt_inverse_relation}, the elastic stiffness
associated with material node $\mathcal{M}_i$ in the sample frame is
therefore
\begin{equation}
\bm C_i^{\mathrm s}
=
\bm M_\sigma(\bm R_i)
\bm C^{\mathrm c}
\bm M_\sigma(\bm R_i)^{\mathsf T}
.
\label{eq:voigt_stiffness_rotation}
\end{equation}
The stiffness transformation thus requires only matrix products and a
transpose, and no matrix inverse has to be computed during offline training.

\subsubsection{Initialization of crystallographic orientation for online prediction}

The optimized orientations are fixed network parameters during offline training. During the nonlinear online prediction, they define only the initial lattice orientations, which then evolve according to the CP constitutive response.

At each material node $\mathcal{M}_i$, the deformation gradient is
multiplicatively decomposed as
\begin{equation}
\bm F_i
=
\bm F_i^{\mathrm e}
\bm F_i^{\mathrm i}
\bm F_i^{\mathrm p},
\label{eq:multiplicative_decomposition}
\end{equation}
where $\bm F_i^{\mathrm p}$ denotes the lattice-preserving inelastic deformation gradient associated with plastic flow, $\bm F_i^{\mathrm i}$ denotes the lattice-distorting inelastic deformation gradient associated with eigenstrains, and $\bm F_i^{\mathrm e}$ denotes the elastic deformation gradient. In the absence of an initial eigenstrain,
\begin{equation}
\bm F_i^{\mathrm i}(0)
=
\bm I,
\end{equation}
where the argument $0$ refers to the initial time $t=0$. Following the kinematic formulation of the D\"usseldorf Advanced Material Simulation Kit
(DAMASK)~\cite{roters2019damask},
the prescribed initial orientation $\bm G_i$ enters the initial multiplicative
decomposition as
\begin{equation}
\begin{aligned}
\bm F_i^{\mathrm p}(0)
&=
\bm G_i,
\\
\bm F_i^{\mathrm e}(0)
&=
\bm G_i^{-1}
=
\bm R_i.
\end{aligned}
\label{eq:cp_orientation_initialization}
\end{equation}

During subsequent loading, $\bm F_i^{\mathrm p}$ evolves according to the
CP constitutive equations. The lattice
orientation at time $t$ follows from the polar decomposition of the elastic
deformation gradient~\cite{ma2006dislocation},
\begin{equation}
\bm F_i^{\mathrm e}(t)
=
\bm R_i(t)
\bm U_i^{\mathrm e}(t),
\label{eq:elastic_polar_decomposition}
\end{equation}
where $\bm R_i(t)$ is the evolving active crystal-to-sample rotation and
$\bm U_i^{\mathrm e}(t)$ is the elastic right stretch tensor.

\subsection{Offline training}
\label{sec24}

Two offline training strategies are considered: the proposed GSH-based strategy and the conventional ODMN strategy of our previous work~\cite{wei2025orientation}. Both use the same linear-elastic dataset, objective function, network topology, and optimization settings. They differ in how the weights and orientations of the material nodes are determined and, as a result, in which quantities of the hierarchical homogenization are trainable.

\subsubsection{Parameterization of the material nodes}

In both strategies, material node $\mathcal M_i$ carries the Bunge triplet $\mathbf g_i=(\varphi_1^i,\Phi^i,\varphi_2^i)$ of Section~\ref{sec23} and a weight. In the conventional strategy, the three angles are trainable and are initialized uniformly on the rotation group, that is, $\varphi_1^i,\varphi_2^i\sim\mathcal U(0,2\pi)$ and $\cos\Phi^i\sim\mathcal U(-1,1)$, where $\mathcal U(a,b)$ denotes the uniform distribution on the interval $(a,b)$. Each weight is represented by an unconstrained scalar $z_i$ through the softplus map with sharpness parameter $\kappa$,
\begin{equation}
\tilde w_i
=
\frac{1}{\kappa}
\ln\left(1+e^{\kappa z_i}\right),
\qquad
\kappa=10,
\label{eq:softplus_weight}
\end{equation}
which keeps the unnormalized weight $\tilde w_i$ positive for any value of $z_i$. The relative volume fractions entering the homogenization building blocks, Eq.~\eqref{eq:local_volume_fractions} below, involve only ratios of the weights, so that no normalization is required during training. After training, the volume fractions used in the online stage are obtained by normalizing the trained weights,
\begin{equation}
w_i
=
\frac{\tilde w_i}{\displaystyle\sum_{j=1}^{2^N}\tilde w_j},
\qquad
\sum_{i=1}^{2^N} w_i=1.
\label{eq:weight_normalization_conv}
\end{equation}
In the GSH-based strategy, the orientations and weights are prescribed by Eq.~\eqref{eq:gsh_odmn_mapping}. The prescribed weights already satisfy the unit-sum constraint, and neither quantity is updated during training. In the implementation, the prescribed weights are stored through the inverse of Eq.~\eqref{eq:softplus_weight}, so that both strategies share the same code path. In both strategies, the angular variables $\theta_p^l$ and $\phi_p^l$ of the interaction directions are initialized with values in $(0,1)$ drawn from a fixed random seed.

\subsubsection{Linear-elastic training dataset}

For the cubic crystals considered in this work, the single-crystal stiffness
matrix expressed in the crystal frame is
\begin{equation}
\bm C^{\mathrm c}
=
\begin{bmatrix}
C_{11} & C_{12} & C_{12} & 0 & 0 & 0\\
C_{12} & C_{11} & C_{12} & 0 & 0 & 0\\
C_{12} & C_{12} & C_{11} & 0 & 0 & 0\\
0 & 0 & 0 & C_{44} & 0 & 0\\
0 & 0 & 0 & 0 & C_{44} & 0\\
0 & 0 & 0 & 0 & 0 & C_{44}
\end{bmatrix}.
\label{eq:cubic_stiffness_training}
\end{equation}
The three elastic constants $C_{11}$, $C_{12}$, and $C_{44}$ are sampled independently over a
wide range,
\begin{equation}
\begin{aligned}
&C_{11},C_{12},C_{44}
\sim
\mathcal{U}\left(10^{-3},10^{3}\right)\,\mathrm{GPa},
\\
&C_{11}-C_{12}>0,
\end{aligned}
\label{eq:elastic_sampling}
\end{equation}
following the sampling strategy adopted in our previous ODMN
work~\cite{wei2025orientation}.

For each sample $s$, with single-crystal stiffness
$\bm C^{\mathrm c,(s)}$, the effective stiffness of the
RVE, $\overline{\bm C}^{\mathrm{DNS},(s)}$, is obtained from
a full-field direct numerical simulation (DNS) with the CPFFT solver of DAMASK~\cite{roters2019damask}.
Here and below, an overline denotes a macroscopic, that is, homogenized quantity.
The offline dataset is
\begin{equation}
\mathcal{D}_{\mathrm{off}}
=
\left\{
\left(
\bm C^{\mathrm c,(s)},
\overline{\bm C}^{\mathrm{DNS},(s)}
\right)
\right\}_{s=1}^{N_{\mathrm{data}}},
\label{eq:offline_dataset}
\end{equation}
where $N_{\mathrm{data}}$ is the number of samples. The same dataset is used for both training strategies.

\subsubsection{Material-node stiffnesses}

For each training sample $s$, the sampled single-crystal stiffness is assigned
to all material nodes and rotated to the sample frame with
Eq.~\eqref{eq:voigt_stiffness_rotation},
\begin{equation}
\bm C_i^{\mathrm s,(s)}
=
\bm M_\sigma(\bm R_i)
\bm C^{\mathrm c,(s)}
\bm M_\sigma(\bm R_i)^{\mathsf T},
\;\;
i=1,\ldots,2^N,
\label{eq:training_node_stiffness}
\end{equation}
where $\bm R_i$ is the crystal-to-sample rotation of material node
$\mathcal M_i$ defined in Section~\ref{sec23}.

The material-node stiffnesses are the inputs at the terminal level of the
network. From there, they are combined recursively by the building blocks of
Section~\ref{sec21} until the effective stiffness is obtained at the root.

\subsubsection{Hierarchical homogenization}

For building block $(l,p)$, the relative volume fractions of the two child
branches, with the material-node sets $\mathcal J^0_{l,p}$ and
$\mathcal J^1_{l,p}$ of Eqs.~\eqref{eq:J0_defined}
and~\eqref{eq:J1_defined}, follow from the material-node weights as
\begin{equation}
f^0_{l,p}
=
\frac{
\displaystyle
\sum_{\mathcal M_i\in\mathcal J^0_{l,p}} w_i
}{
\displaystyle
\sum_{\mathcal M_i\in
\mathcal J^0_{l,p}\cup\mathcal J^1_{l,p}} w_i
},
\qquad
f^1_{l,p}
=
\frac{
\displaystyle
\sum_{\mathcal M_i\in\mathcal J^1_{l,p}} w_i
}{
\displaystyle
\sum_{\mathcal M_i\in
\mathcal J^0_{l,p}\cup\mathcal J^1_{l,p}} w_i
}.
\label{eq:local_volume_fractions}
\end{equation}

Let $\bm C_{l,p}^{(s)}$ denote the effective stiffness of building
block $(l,p)$ for training sample $s$. The bottom-up homogenization
is written recursively as
\begin{equation}
\begin{aligned}
\bm C_{l,p}^{(s)}
&=
\mathbb H_2\bigl(
\bm C_{l+1,2p}^{(s)},
\bm C_{l+1,2p+1}^{(s)},
f^0_{l,p},f^1_{l,p},\bm N_p^l
\bigr),
\\[-0.3ex]
&\qquad l<N-1,
\\
\bm C_{l,p}^{(s)}
&=
\mathbb H_2\bigl(
\bm C_{2p+1}^{\mathrm s,(s)},
\bm C_{2p+2}^{\mathrm s,(s)},
f^0_{l,p},f^1_{l,p},\bm N_p^l
\bigr),
\\[-0.3ex]
&\qquad l=N-1.
\end{aligned}
\label{eq:recursive_odmn_homogenization}
\end{equation}
At the terminal level $l=N-1$, the two children are the
material nodes $\mathcal M_{2p+1}$ and $\mathcal M_{2p+2}$. At all higher
levels, the child stiffnesses are the effective stiffnesses of the two
child building blocks.

For two children with stiffnesses $\bm C^0$ and $\bm C^1$, relative volume
fractions $f^0$ and $f^1$, and interaction direction $\bm N$, the two-phase
homogenization operator $\mathbb H_2$ is defined as
\begin{equation}
\begin{aligned}
&\mathbb H_2\left(\bm C^0,\bm C^1,f^0,f^1,\bm N\right)
=
f^0\bm C^0+f^1\bm C^1
\\
&\qquad-
f^0f^1
\left(\bm C^0-\bm C^1\right)
\bm Q
\left(\bm C^0-\bm C^1\right),
\end{aligned}
\label{eq:H2_function}
\end{equation}
where
\begin{equation}
\bm Q
=
\bm H(\bm N)\,
\bm Z^{-1}\,
\bm H(\bm N)^{\mathsf T},
\label{eq:Q_matrix}
\end{equation}
with
\begin{equation}
\bm Z
=
\bm H(\bm N)^{\mathsf T}
\left(
f^1\bm C^0+f^0\bm C^1
\right)
\bm H(\bm N).
\label{eq:S_matrix}
\end{equation}
The matrix $\bm H(\bm N)$ associated with the interaction direction
$\bm N=[N_1,N_2,N_3]^{\mathsf T}$ is
\begin{equation}
\bm H(\bm N)
=
\begin{bmatrix}
N_1 & 0   & 0   \\
0   & N_2 & 0   \\
0   & 0   & N_3 \\
0   & N_3 & N_2 \\
N_3 & 0   & N_1 \\
N_2 & N_1 & 0
\end{bmatrix}.
\label{eq:H_matrix}
\end{equation}
For building block $(l,p)$, the interaction direction $\bm N_p^l$ follows
from Eq.~\eqref{eq:direction_vector} with the angular variables
$\theta_p^l$ and $\phi_p^l$.
Recursive application of
Eq.~\eqref{eq:recursive_odmn_homogenization} from the terminal level to the
root yields the effective stiffness predicted by the ODMN, $\overline{\bm C}^{\mathrm{ODMN},(s)}
=
\bm C_{0,0}^{(s)}$. Equation~\eqref{eq:H2_function} also shows how the two training strategies differ. At each building block, the operator depends on the child stiffnesses $\bm C^0$ and $\bm C^1$, their relative volume fractions $f^0$ and $f^1$, and the interaction direction $\bm N$. The strategies differ in which of these quantities are prescribed and which are identified from the full-field data.

In the GSH-based strategy, the orientations
$\mathbf g_i$ and weights $w_i$ of the material nodes are prescribed by
Eq.~\eqref{eq:gsh_odmn_mapping}. The terminal stiffnesses
$\bm C_i^{\mathrm s,(s)}$ of
Eq.~\eqref{eq:training_node_stiffness} and the relative volume fractions
$f^0_{l,p}$ and $f^1_{l,p}$ of Eq.~\eqref{eq:local_volume_fractions} are
therefore fixed during offline training, and only the interaction directions
$\bm N_p^l$ are identified. The trainable
parameter set is
\begin{equation}
\boldsymbol{\Theta}_{\mathrm{GSH}}
=
\left\{
\theta_p^l,\phi_p^l
\;\middle|\;
0\le l<N,\;
0\le p<2^l
\right\}.
\label{eq:trainable_parameters_gsh}
\end{equation}

The child stiffnesses at the higher levels still depend on the trainable
interaction directions, because they result from the homogenization
operations at the lower levels. The strategy thus preserves the texture
given by the GSH compaction and learns only the interactions needed
to reproduce the full-field effective response.

In the conventional strategy, the orientations and weights of the material
nodes are identified together with the interaction directions. All
quantities entering Eq.~\eqref{eq:H2_function} then depend
on trainable parameters: the child stiffnesses through the Bunge angles
$(\varphi_1^i,\Phi^i,\varphi_2^i)$ of the material nodes, the relative
volume fractions through the weight parameters $z_i$ of
Eq.~\eqref{eq:softplus_weight}, and the interaction directions through
$(\theta_p^l,\phi_p^l)$. The trainable parameter set is
\begin{equation}
\begin{aligned}
\boldsymbol{\Theta}_{\mathrm{conv}}
={}&
\left\{
\theta_p^l,\phi_p^l
\;\middle|\;
0\le l<N,\;
0\le p<2^l
\right\}
\\
&\cup
\left\{
z_i,\varphi_1^i,\Phi^i,\varphi_2^i
\;\middle|\;
1\le i\le 2^N
\right\},
\end{aligned}
\label{eq:trainable_parameters_conventional}
\end{equation}
which contains $6\cdot2^N-2$ parameters, compared with $2(2^N-1)$ in
Eq.~\eqref{eq:trainable_parameters_gsh}.

\subsubsection{Objective function and optimization}

For both strategies, the trainable parameters are identified by minimizing
the mean relative error, measured in the Frobenius norm, between the ODMN-predicted effective stiffness
and the corresponding full-field homogenized stiffness. For a minibatch of
$N_{\mathrm{batch}}$ samples, the objective function is
\begin{equation}
\mathcal{L}(\boldsymbol{\Theta})
=
\frac{1}{N_{\mathrm{batch}}}
\sum_{s=1}^{N_{\mathrm{batch}}}
\frac{
\left\|
\overline{\bm C}^{\mathrm{ODMN},(s)}
-
\overline{\bm C}^{\mathrm{DNS},(s)}
\right\|_{\mathrm F}
}{
\left\|
\overline{\bm C}^{\mathrm{DNS},(s)}
\right\|_{\mathrm F}
}.
\label{eq:offline_loss}
\end{equation}
Here, $\|\cdot\|_{\mathrm F}$ denotes the Frobenius norm, and
$\boldsymbol{\Theta}$ stands for
$\boldsymbol{\Theta}_{\mathrm{GSH}}$ or
$\boldsymbol{\Theta}_{\mathrm{conv}}$, depending on the training
strategy. The same quantity evaluated on the validation set is reported as the validation error in Section~\ref{sec33}.

In total, $N_{\mathrm{data}}=500$ samples are generated, of which 400 are
used for training and 100 for validation.
Both strategies use the same optimization settings. Each
network is trained for 150 epochs with the AdamW optimizer, zero weight
decay, and a minibatch size of $N_{\mathrm{batch}}=20$. The learning rate starts at
$10^{-2}$ and decreases to $10^{-4}$ along a cosine-annealing schedule.

\subsection{Nonlinear online prediction}
\label{sec25}

After offline training, the interaction directions $\bm N_p^l$ and the volume fractions $w_i$ are fixed. For the conventional strategy, $w_i$ denotes the normalized weights of Eq.~\eqref{eq:weight_normalization_conv}. The online prediction is formulated as a localization problem under a prescribed macroscopic deformation gradient and temperature. Its unknowns are the interaction vectors of all building blocks,
\begin{equation} 
\mathcal{A} = \left\{ \bm a_p^l \;\middle|\; l=0,\ldots,N-1,\; p=0,\ldots,2^l-1 \right\}, \label{eq:online_interaction_vectors}
\end{equation} 
where $\bm a_p^l$ is the interaction vector of building block $(l,p)$. The interaction vectors describe the fluctuations of the deformation gradient in the network. At each loading increment, they are obtained with the Newton--Raphson method from the equilibrium equations that follow from the Hill--Mandel macrohomogeneity condition, with the nonlinear constitutive response evaluated at the material nodes.

The RVE is assumed to be isothermal, so that the temperature requires no localization and the macroscopic temperature $\overline{T}$ is assigned to all material nodes,
\begin{equation}
T_i = \overline{T}, \qquad i=1,\ldots,2^N, \label{eq:isothermal_temperature}
\end{equation}
where $T_i$ is the temperature of material node $\mathcal{M}_i$.

For a prescribed macroscopic deformation gradient $\overline{\bm F}$, the local deformation gradient $\bm F_i$ at material node $\mathcal{M}_i$ is obtained by hierarchical downscaling,
\begin{equation}
\begin{aligned}
\bm F_i &= \overline{\bm F} + \sum_{l=0}^{N-1} \sum_{p=0}^{2^l-1} \alpha_{l,p}^{i}\, \bm a_p^l \otimes \bm N_p^l \\
&= \overline{\bm F} + \sum_{l=0}^{N-1} \alpha_{l,p_l(i)}^{i}\, \bm a_{p_l(i)}^l \otimes \bm N_{p_l(i)}^l .
\end{aligned}
\label{eq:deformation_gradient_localization}
\end{equation}
Here, $\otimes$ denotes the dyadic product, and $\alpha_{l,p}^{i}$ is the localization coefficient that weights the contribution of building block $(l,p)$ to material node $\mathcal{M}_i$. Unlike $\bm a_p^l$, these coefficients are not unknowns of the localization problem. They are determined by the network topology and the fixed volume fractions. Owing to the binary-tree structure, building block $(l,p)$ contributes to $\mathcal{M}_i$ only if it lies on the unique path from the root to that material node, so that $\alpha_{l,p}^{i}=0$ otherwise. This path contains exactly one building block on each level $l$, namely $(l,p_l(i))$ with $p_l(i)=\lfloor (i-1)/2^{N-l}\rfloor$, where $\lfloor\cdot\rfloor$ denotes the floor function. Retaining only these contributions yields the second form of Eq.~\eqref{eq:deformation_gradient_localization}. Each material node is thus connected to the root through $N$ nested building blocks, and its local deformation gradient is obtained from the macroscopic one through the $N$ interaction vectors along this path.

Using the material-node sets $\mathcal{J}_{l,p}^{0}$ and $\mathcal{J}_{l,p}^{1}$ introduced in Section~\ref{sec21}, the localization coefficients are defined as \begin{equation} \alpha_{l,p}^{i} = \begin{cases} \displaystyle \frac{1}{ \sum\limits_{\mathcal{M}_k\in\mathcal{J}_{l,p}^{0}} w_k }, & \mathcal{M}_i\in\mathcal{J}_{l,p}^{0}, \\[1.5ex] \displaystyle -\frac{1}{ \sum\limits_{\mathcal{M}_k\in\mathcal{J}_{l,p}^{1}} w_k }, & \mathcal{M}_i\in\mathcal{J}_{l,p}^{1}, \\[1.5ex] 0, & \text{otherwise}. \end{cases} \label{eq:alpha_defined} \end{equation} 
The localization mapping and its coefficients are derived in our previous work on the piezoelectric DMN~\cite{wei2027deep}. The online prediction has three ingredients: the CP constitutive update at each material node (Section~\ref{sec251}), the enforcement of the Hill--Mandel condition with the Newton--Raphson solution for the interaction vectors (Section~\ref{sec252}), and the homogenized stress with its consistent macroscopic tangent (Section~\ref{sec253}).

\subsubsection{Crystal-plasticity constitutive update at the material nodes}
\label{sec251}

Given the local deformation gradient $\bm F_i$ of Eq.~\eqref{eq:deformation_gradient_localization}, the first Piola--Kirchhoff stress $\bm P_i$ and its consistent tangent $\partial\bm P_i/\partial\bm F_i$ at material node $\mathcal{M}_i$ are evaluated by a finite-strain CP constitutive update. The update is performed with the constitutive framework of DAMASK~\cite{roters2019damask}, which is embedded as the local material law of the ODMN. The same CP formulation and material parameters are used in the full-field CPFFT reference simulations, so that the ODMN and the DNS differ only in the homogenization scheme. In the remainder of this subsection, the node index $i$ is omitted. All quantities refer to a single material node at temperature $T=\overline{T}$, and the internal variables of each node are stored and evolved independently. The time integration and the consistent tangent follow the DAMASK implementation and are summarized in Appendix~\ref{app:consistent_tangent}.

\paragraph{Kinematics}

Following Eq.~\eqref{eq:multiplicative_decomposition}, the deformation gradient is decomposed multiplicatively as $\bm F=\bm F^{\mathrm e}\bm F^{\mathrm i}\bm F^{\mathrm p}$. The plastic part $\bm F^{\mathrm p}$ maps the reference configuration onto the plastic configuration, which is also referred to as the lattice configuration. The eigen part $\bm F^{\mathrm i}$ maps the plastic configuration onto a stress-free intermediate configuration and carries all inelastic deformation other than slip, which here is thermal expansion. The elastic part $\bm F^{\mathrm e}$ maps the intermediate configuration onto the current configuration and generates stress. The plastic and eigen deformation gradients evolve according to the flow rules
\begin{equation}
\dot{\bm F}^{\mathrm p}
=
\bm L^{\mathrm p}\bm F^{\mathrm p},
\qquad
\dot{\bm F}^{\mathrm i}
=
\bm L^{\mathrm i}\bm F^{\mathrm i},
\label{eq:cp_flow_rules}
\end{equation}
where a superposed dot denotes the time derivative, $\bm L^{\mathrm p}$ is the plastic velocity gradient defined in the lattice configuration, and $\bm L^{\mathrm i}$ is the eigen velocity gradient defined in the intermediate configuration~\cite{roters2019damask}. Both velocity gradients are given by constitutive laws of the general form
\begin{equation}
\bm L^{\mathrm p}
=
\bm L^{\mathrm p}\!\left(\bm M^{\mathrm p},\bm\xi\right),
\qquad
\bm L^{\mathrm i}
=
\bm L^{\mathrm i}\!\left(\bm M^{\mathrm i},\bm F^{\mathrm i},T,\dot T\right),
\label{eq:cp_velocity_gradient_laws}
\end{equation}
which are driven by the Mandel stresses $\bm M^{\mathrm p}$ and $\bm M^{\mathrm i}$ that are work-conjugate to $\bm L^{\mathrm p}$ and $\bm L^{\mathrm i}$ in their respective configurations, by the vector of internal state variables $\bm\xi$, here the slip resistances, and by the temperature $T$ and its rate $\dot T$.

\paragraph{Thermoelastic law}

The elastic response is formulated in terms of the second
Piola--Kirchhoff stress $\bm S$ defined in the lattice configuration.
The constitutive relation is given by the generalized Hooke's law,
\begin{equation}
\begin{aligned}
\bm S
&=
\mathbb{C}^{\mathrm c}(T):\bm E^{\mathrm e},
\\
\bm E^{\mathrm e}
&=
\left(\bm F^{\mathrm i}\right)^{\mathsf T}
\left[
\tfrac{1}{2}
\left(
\left(\bm F^{\mathrm e}\right)^{\mathsf T}
\bm F^{\mathrm e}
-
\bm I
\right)
\right]
\bm F^{\mathrm i}
\\
&=
\tfrac{1}{2}
\left[
\left(\bm F^{\mathrm i}\right)^{\mathsf T}
\left(\bm F^{\mathrm e}\right)^{\mathsf T}
\bm F^{\mathrm e}
\bm F^{\mathrm i}
-
\left(\bm F^{\mathrm i}\right)^{\mathsf T}
\bm F^{\mathrm i}
\right].
\end{aligned}
\label{eq:cp_hooke}
\end{equation}
Here, the colon denotes the double contraction, $\bm I$ is the second-order identity tensor, and $\bm E^{\mathrm e}$ is the elastic Green--Lagrange strain pulled back to the lattice configuration by $\bm F^{\mathrm i}$~\cite{roters2019damask}.

The fourth-order tensor $\mathbb{C}^{\mathrm c}(T)$ is the
temperature-dependent elastic stiffness in the crystal frame.
For the cubic crystal considered here, it has the form of
Eq.~\eqref{eq:cubic_stiffness_training} with
the three elastic constants
$C_{11}$, $C_{12}$, and $C_{44}$.
Their temperature dependence is described by the quadratic expansion
\begin{equation}
C_{ij}(T)
=
C_{ij}^{(0)}
+
C_{ij}^{(1)}
\left(T-T_{\mathrm{ref}}\right)
+
C_{ij}^{(2)}
\left(T-T_{\mathrm{ref}}\right)^2,
\label{eq:cp_elastic_T}
\end{equation}
for $ij\in\{11,12,44\}$, where $T_{\mathrm{ref}}$ is the reference
temperature and $C_{ij}^{(0)}$, $C_{ij}^{(1)}$, and $C_{ij}^{(2)}$
are the constant, linear, and quadratic coefficients.

\paragraph{Plastic flow}

Plastic deformation is assumed to arise from crystallographic slip on
$N_{\mathrm s}$ slip systems. The plastic velocity gradient is expressed
as the superposition of the slip-system shear rates,
\begin{equation}
\bm L^{\mathrm p}
=
\sum_{\alpha=1}^{N_{\mathrm s}}
\dot\gamma^{\alpha}\,
\bm P^{\alpha},
\qquad
\bm P^{\alpha}
=
\bm s^{\alpha}\otimes\bm n^{\alpha},
\label{eq:cp_Lp}
\end{equation}
where $\dot\gamma^{\alpha}$ is the shear rate of slip system $\alpha$,
$\bm s^{\alpha}$ and $\bm n^{\alpha}$ are its unit slip direction
and slip-plane normal in the
crystal frame, and $\bm P^{\alpha}$ is the corresponding Schmid tensor.
The Schmid tensor always carries a slip-system superscript and is thereby
distinguished from the first Piola--Kirchhoff stress $\bm P$.

Following the DAMASK formulation~\cite{roters2019damask}, the driving
stress for plastic slip is approximated by the Mandel stress
\begin{equation}
\bm M^{\mathrm p}
=
\left(\bm F^{\mathrm i}\right)^{\mathsf T}
\bm F^{\mathrm i}
\bm S,
\label{eq:cp_mandel}
\end{equation}
where the contribution of the elastic stretch is neglected. A derivation
of Eq.~\eqref{eq:cp_mandel} is given in
Appendix~\ref{app:mandel_stress}. In the absence of lattice-distorting
inelastic deformation, $\bm F^{\mathrm i}=\bm I$, and
$\bm M^{\mathrm p}$ reduces to $\bm S$.

The resolved shear stress on slip system $\alpha$ is obtained from
Schmid's law as
\begin{equation}
\tau^{\alpha}
=
\bm M^{\mathrm p}:\bm P^{\alpha}.
\label{eq:cp_rss}
\end{equation}
The corresponding shear rate is governed by the phenomenological
power-law relation~\cite{PEIRCE19831951,roters2019damask},
\begin{equation}
\dot\gamma^{\alpha}
=
\dot\gamma_{0}
\left|
\frac{\tau^{\alpha}}{\xi^{\alpha}}
\right|^{n}
\operatorname{sgn}\!\left(\tau^{\alpha}\right),
\label{eq:cp_flow_rule}
\end{equation}
where $\dot\gamma_{0}$ is the reference shear rate, $n$ is the stress
exponent, $\xi^{\alpha}$ is the slip resistance of system
$\alpha$, and $\operatorname{sgn}$ is the sign function.

The slip resistances
$\bm\xi=\{\xi^{\alpha}\}_{\alpha=1}^{N_{\mathrm s}}$
constitute the hardening state variables and evolve from the initial
value $\xi_{0}$ toward the saturation value $\xi_{\infty}$ according to
\begin{equation}
\dot\xi^{\alpha}
=
h_{0}
\left|
1-\frac{\xi^{\alpha}}{\xi_{\infty}}
\right|^{a}
\operatorname{sgn}\!\left(
1-\frac{\xi^{\alpha}}{\xi_{\infty}}
\right)
\sum_{\beta=1}^{N_{\mathrm s}}
h^{\alpha\beta}
\left|\dot\gamma^{\beta}\right|,
\label{eq:cp_hardening}
\end{equation}
where $h_{0}$ is the reference hardening modulus, $a$ is the hardening
exponent, and $h^{\alpha\beta}$ denotes the slip--slip interaction
coefficient. For the octahedral slip systems considered here,
$h^{\alpha\beta}$ accounts for self-hardening and latent hardening
associated with coplanar, collinear, Hirth-lock, glissile-junction (two
types), and Lomer-lock interactions~\cite{madec2017dislocation, roters2019damask, prabhu2026incorporation}.

\paragraph{Thermal eigen deformation}
Thermal expansion is introduced as an eigen deformation that evolves $\bm F^{\mathrm i}$ through the rate law
\begin{equation}
\bm L^{\mathrm i}
=
\det\!\left(\bm F^{\mathrm i}\right)
\bm F^{\mathrm i}
\left[
\dot T\,
\bm\alpha(T)
\right]
\left(\bm F^{\mathrm i}\right)^{-1},
\label{eq:cp_Li_thermal}
\end{equation}
where $\bm\alpha(T)$ is the second-order tensor of thermal expansion coefficients defined in the crystal frame, and the outer transformation pushes the rate $\dot T\bm\alpha(T)$ from the lattice configuration forward to the intermediate configuration in which $\bm L^{\mathrm i}$ is defined~\cite{roters2019damask}. This mapping is the one implemented in DAMASK, and its form is written out in Appendix~\ref{app:eigen_rate_mapping}. For the cubic crystal considered here, $\bm\alpha(T)=\alpha_{11}\bm I$ is isotropic with the linear thermal expansion coefficient $\alpha_{11}$, and Eq.~\eqref{eq:cp_Li_thermal} reduces to $\bm L^{\mathrm i}=\det(\bm F^{\mathrm i})\,\alpha_{11}\dot T\,\bm I$. The thermal expansion tensor $\bm\alpha$ and its component $\alpha_{11}$ are unrelated to the localization coefficients $\alpha_{l,p}^{i}$ and to the slip-system index $\alpha$. Thermal expansion is stress independent, so that $\partial\bm L^{\mathrm i}/\partial\bm M^{\mathrm i}=\bm 0$. Equation~\eqref{eq:cp_Li_thermal} is a rate form. The thermal eigenstrain is therefore accumulated along the actual temperature history from the stress-free initial configuration and is not prescribed as a total strain $\alpha_{11}(T-T_{\mathrm{ref}})$. Under the isothermal-RVE assumption of Eq.~\eqref{eq:isothermal_temperature}, all material nodes experience the same temperature rate within a loading increment, $\dot T=\Delta\overline{T}/\Delta t$, where $\Delta\overline{T}$ and $\Delta t$ are the increments of the macroscopic temperature and of time.

\subsubsection{Hill--Mandel equilibrium and Newton--Raphson solution}
\label{sec252}

The interaction vectors $\mathcal{A}$ in
Eq.~\eqref{eq:online_interaction_vectors} are determined by enforcing the
Hill--Mandel macrohomogeneity condition. In terms of the first
Piola--Kirchhoff stress and deformation gradient, this condition reads
\begin{equation}
\begin{aligned}
\overline{\bm P}:\delta\overline{\bm F}
&=
\sum_{i=1}^{2^N}
w_i\,
\bm P_i:\delta\bm F_i
\\
&=
\sum_{i=1}^{2^N}
w_i\,
\bm P_i:
\Big(
\delta\overline{\bm F}
\\
&\qquad
+
\sum_{l=0}^{N-1}
\sum_{p=0}^{2^l-1}
\alpha_{l,p}^{i}\,
\delta\bm a_p^l
\otimes
\bm N_p^l
\Big),
\end{aligned}
\label{eq:hill_mandel}
\end{equation}
where $\overline{\bm P}$ is the macroscopic first Piola--Kirchhoff stress,
$\delta$ denotes a virtual variation, and
Eq.~\eqref{eq:deformation_gradient_localization} has been used for
$\delta\bm F_i$.

Since the variations
$\delta\bm a_p^l$ are arbitrary and independent, their
coefficients in Eq.~\eqref{eq:hill_mandel} must vanish.
The equilibrium condition of building block $(l,p)$, with residual
$\bm r_p^l$, is therefore
\begin{equation}
\bm r_p^l
\equiv
\sum_{i=1}^{2^N}
w_i\,
\alpha_{l,p}^{i}\,
\bm P_i \cdot \bm N_p^l
=
\bm 0,
\label{eq:building_block_equilibrium}
\end{equation}
for $l=0,\ldots,N-1$ and $p=0,\ldots,2^l-1$.

For the numerical solution, all interaction vectors are collected into
the global vector
\begin{equation}
\mathbf A
=
\begin{bmatrix}
\bm a_0^0 \\
\bm a_0^1 \\
\bm a_1^1 \\
\vdots \\
\bm a_{2^{N-1}-1}^{N-1}
\end{bmatrix}
\in
\mathbb{R}^{3(2^N-1)}.
\label{eq:A_vector}
\end{equation}
Let $\operatorname{vec}(\cdot)$ denote row-wise vectorization,
\begin{equation}
\begin{aligned}
\operatorname{vec}(\bm F)
={}&
[F_{11},F_{12},F_{13},F_{21},F_{22},
\\
&\;F_{23},F_{31},F_{32},F_{33}]^{\mathsf T}.
\end{aligned}
\end{equation}
For an interaction direction
$\bm N=[N_1,N_2,N_3]^{\mathsf T}$, define
\begin{equation}
\bm V(\bm N)
=
\begin{bmatrix}
N_1 & 0   & 0 \\
N_2 & 0   & 0 \\
N_3 & 0   & 0 \\
0   & N_1 & 0 \\
0   & N_2 & 0 \\
0   & N_3 & 0 \\
0   & 0   & N_1 \\
0   & 0   & N_2 \\
0   & 0   & N_3
\end{bmatrix},
\label{eq:R_matrix}
\end{equation}
such that
$\operatorname{vec}(\bm a\otimes\bm N)
=
\bm V(\bm N)\bm a$.
The localization relation of
Eq.~\eqref{eq:deformation_gradient_localization} can then be expressed as
\begin{equation}
\operatorname{vec}(\bm F_i)
=
\operatorname{vec}(\overline{\bm F})
+
\bm D_i\mathbf A,
\label{eq:D_matrix}
\end{equation}
where
\begin{equation}
\bm D_i
=
\operatorname{row}_{(l,p)}
\left(
\alpha_{l,p}^{i}\,
\bm V(\bm N_p^l)
\right)
\in
\mathbb{R}^{9\times3(2^N-1)}.
\label{eq:D_definition}
\end{equation}
Here, $\operatorname{row}_{(l,p)}(\cdot)$ denotes horizontal block
concatenation over all building blocks $(l,p)$.
By the second form of Eq.~\eqref{eq:deformation_gradient_localization},
only the $N$ blocks of $\bm D_i$ that belong to the building blocks
$(l,p_l(i))$ on the path from the root to $\mathcal{M}_i$ are nonzero.
Using this representation, the equilibrium equations of all building
blocks are assembled into the global residual
\begin{equation}
\bm r(\mathbf A)
=
\sum_{i=1}^{2^N}
w_i\,
\bm D_i^{\mathsf T}
\operatorname{vec}(\bm P_i)
=
\bm 0,
\qquad
\bm r\in\mathbb{R}^{3(2^N-1)}.
\label{eq:residual_vector}
\end{equation}
Here,
$\bm P_i= \bm P_i(\bm F_i(\mathbf A),\overline T,\text{history})$
is obtained from the CP constitutive update described in
Section~\ref{sec251}.

The nonlinear system in Eq.~\eqref{eq:residual_vector} is solved using
the Newton--Raphson method. The consistent material tangent at material
node $\mathcal M_i$ is written in matrix form as
\begin{equation}
\bm{\mathsf L}_i
\equiv
\frac{
\partial\operatorname{vec}(\bm P_i)
}{
\partial\operatorname{vec}(\bm F_i)
}
\in
\mathbb{R}^{9\times9}.
\label{eq:local_tangent_matrix}
\end{equation}
Linearization of the global residual with respect to $\mathbf A$ then
gives the Newton Jacobian
\begin{equation}
\bm K
\equiv
\frac{\partial\bm r}{\partial\mathbf A}
=
\sum_{i=1}^{2^N}
w_i\,
\bm D_i^{\mathsf T}
\bm{\mathsf L}_i
\bm D_i.
\label{eq:newton_K}
\end{equation}
Because each $\bm D_i$ has only $N$ nonzero blocks, the $3\times3$ block
of $\bm K$ that couples building blocks $(l,p)$ and $(l',p')$ vanishes
unless both lie on a common root-to-node path, so that $\bm K$ is
assembled from the $N^2$ block pairs of each material node rather than
from dense products $\bm D_i^{\mathsf T}\bm{\mathsf L}_i\bm D_i$.

At Newton iteration $k$, the interaction-vector correction is obtained
from
\begin{equation}
\bm K_{\mathrm{reg}}^{(k)}
\delta\mathbf A
=
-\bm r^{(k)},
\qquad
\mathbf A^{(k+1)}
=
\mathbf A^{(k)}
+
\eta^{(k)}
\delta\mathbf A,
\label{eq:newton_update}
\end{equation}
where $\delta\mathbf A$ is the correction, $\bm K_{\mathrm{reg}}$ is the
regularized Jacobian defined below, and the step length
$\eta^{(k)}\in(0,1]$ is determined by a backtracking line search.
Starting from $\eta^{(k)}=1$, the step length is halved
until the sufficient-decrease condition
\begin{equation}
\left\|
\bm r
\left(
\mathbf A^{(k)}
+
\eta^{(k)}\delta\mathbf A
\right)
\right\|_{\infty}
\le
\left(
1-c\,\eta^{(k)}
\right)
\left\|
\bm r^{(k)}
\right\|_{\infty}
\label{eq:line_search}
\end{equation}
is satisfied, where $\|\cdot\|_{\infty}$ denotes the maximum norm and $c=10^{-4}$. At most $q_{\max}=8$ halvings are performed per Newton step, and the Newton loop is limited to $k_{\max}=30$ iterations.
For each trial step, the local deformation gradients are recomputed
with Eq.~\eqref{eq:D_matrix} and the constitutive update is repeated at all
material nodes.

To stabilize the Newton solve, a diagonal regularization
is applied to the Jacobian,
\begin{equation}
\bm K_{\mathrm{reg}}
=
\bm K
+
\lambda
\max_j |K_{jj}|
\bm I,
\label{eq:newton_regularization}
\end{equation}
where $K_{jj}$ are the diagonal entries of $\bm K$ and $\lambda$ is the
regularization parameter, with the baseline value $\lambda_0=10^{-10}$.
If the line search is exhausted, the trial step with the smallest
residual is retained and $\lambda$ is increased by a factor of $100$,
with an activation floor of $10^{-8}$ and an upper bound of $10^{-2}$.
After each full Newton step, that is, when
$\eta^{(k)}=1$, $\lambda$ is reduced by a factor of $10$
toward $\lambda_0$.
Upon convergence, $\lambda$ is reset to $\lambda_0$ before reassembling
the Jacobian used to evaluate the homogenized tangent.

The Newton iterations are terminated when
\begin{equation}
\bigl\|
\bm r^{(k)}
\bigr\|_{\infty}
\le
\max\bigl[
\epsilon_{\mathrm{abs}},
\epsilon_{\mathrm{rel}}
d_{\max}
\max\bigl(
p_{\max}^{(k)},
p_{\mathrm{floor}}
\bigr)
\bigr],
\label{eq:newton_convergence}
\end{equation}
where $d_{\max}=\max_i\|\bm D_i\|_{\max}$ and
$p_{\max}^{(k)}=\max_i\|\bm P_i^{(k)}\|_{\max}$ characterize the
maximum absolute components of the localization matrices and the
first Piola--Kirchhoff stresses, respectively, at Newton iteration $k$.
Here, $\|\cdot\|_{\max}$ denotes the maximum absolute component, and
the stress floor is set to $p_{\mathrm{floor}}=1\,\mathrm{Pa}$.
The tolerances $\epsilon_{\mathrm{abs}}=10^{-4}\,\mathrm{Pa}$ and
$\epsilon_{\mathrm{rel}}=10^{-7}$ are used throughout this work.

\paragraph{Tree-structured LU factorization}
Let $b=(l,p)$ index the building blocks, with
$\alpha_b^i\equiv\alpha_{l,p}^i$ and $\bm N_b\equiv\bm N_p^l$. Let
$\bm K_{bu}$ denote the $3\times3$ block of $\bm K$ that couples
building blocks $b$ and $u$, and let $\bm r_b$ denote the part of the
residual associated with $b$. From Eqs.~\eqref{eq:D_definition} and
\eqref{eq:newton_K},
\begin{equation}
\bm K_{bu}
=
\sum_{i=1}^{2^N}
w_i\,
\alpha_b^i\,\alpha_u^i\,
\bm V(\bm N_b)^{\mathsf T}
\bm{\mathsf L}_i
\bm V(\bm N_u),
\label{eq:K_block}
\end{equation}
which is nonzero only if $b=u$ or if one of the two building blocks is
an ancestor of the other in the binary tree. Let $\mathcal P(b)$ denote
the set of the $l$ ancestors of $b=(l,p)$, that is, the building blocks
$(l',\lfloor p/2^{\,l-l'}\rfloor)$ with $l'<l$.

The linear system of Eq.~\eqref{eq:newton_update} is solved by block
Gaussian elimination from the terminal level toward the root. For
$l=N-1,\ldots,1$ and for every building block $b$ on level $l$, the
unknown $\delta\bm a_b$ is eliminated through the Schur-complement
update
\begin{equation}
\begin{aligned}
\bm K_{uv}
&\leftarrow
\bm K_{uv}
-
\bm K_{ub}\,
\bm K_{bb}^{-1}\,
\bm K_{bv},
\\
\bm r_u
&\leftarrow
\bm r_u
-
\bm K_{ub}\,
\bm K_{bb}^{-1}\,
\bm r_b,
\end{aligned}
\qquad
u,v\in\mathcal P(b).
\label{eq:tree_schur}
\end{equation}
All ancestors of $b$ lie on one root-to-node path and are therefore
already coupled to each other. The update thus creates no new nonzero
blocks, so the factorization proceeds without fill-in. Building blocks
on the same level are not coupled and can be eliminated independently.
After this forward sweep, the corrections follow by back substitution
from the root to the terminal level. For $l=0,\ldots,N-1$ and every
building block $b$ on level $l$,
\begin{equation}
\delta\bm a_b
=
-\bm K_{bb}^{-1}
\Big(
\bm r_b
+
\sum_{u\in\mathcal P(b)}
\bm K_{bu}\,
\delta\bm a_u
\Big),
\label{eq:tree_backsub}
\end{equation}
where $\bm K$ and $\bm r$ denote the updated quantities and the sum is
empty for the root. In the implementation, the regularized Jacobian
$\bm K_{\mathrm{reg}}$ of Eq.~\eqref{eq:newton_regularization} takes the
place of $\bm K$.

The CP tangent $\bm{\mathsf L}_i$ does not possess major
symmetry in general, so that $\bm K$ is not symmetric. Both
$\bm K_{bu}$ and $\bm K_{ub}$ are therefore stored, and each $3\times3$
pivot block $\bm K_{bb}$ is factorized by an LU decomposition, not by
a Cholesky decomposition.

\paragraph{Computational complexity}
Each Newton iteration requires one CP constitutive update at every
material node, whose cost is independent of the network depth $N$,
followed by the assembly and solution of Eq.~\eqref{eq:newton_update}.
Exploiting the block sparsity described above, $\bm K$ is assembled from
the $N^2$ block pairs of each material node, at a total cost of
$O(2^N N^2)$ operations. The tree-structured LU factorization performs
$l^2$ block updates of Eq.~\eqref{eq:tree_schur} for each of the $2^l$
building blocks on level $l$ and therefore also requires $O(2^N N^2)$
operations. The same procedure is used to solve
Eq.~\eqref{eq:dA_dF} for the consistent tangent of
Section~\ref{sec253}. The cost of each Newton iteration therefore
consists of $O(2^N)$ operations for the constitutive updates and
$O(2^N N^2)$ operations for the network algebra. Since the network
contains $2^N$ material nodes, the latter scales as the number of
material nodes multiplied by the square of the network depth. For the
depths considered here, the network algebra remains a minor fraction of
the iteration cost (about 30\,\% at $N=8$), so that the cost per
iteration grows essentially in proportion to the number of material
nodes. If $\bm D_i$ and $\bm K$ were treated as dense matrices,
each iteration would require $O(2^N[3(2^N-1)]^2)=O(8^N)$ operations, and
the network algebra would dominate the cost for
$N\ge\nobreak6$. In the original DMN, the corresponding linear problem is solved by a recursive bottom-up homogenization and top-down de-homogenization over the building blocks, at a cost per Newton iteration that is linear in the number of nodes~\cite{liu2019deep}. The tree-structured LU factorization is the counterpart of this recursion for the formulation adopted here, in which the unknowns are the interface vectors of all building blocks~\cite{gajek2020micromechanics,noels2022micromechanics}. In this formulation, each material node is coupled to all $N$ of its ancestors, which gives rise to the factor $N^2$, and the Newton system is a single global linear system. The $O(8^N)$ estimate refers to a dense solution of that system.

\subsubsection{Homogenized stress, consistent macroscopic tangent, and Abaqus implementation}
\label{sec253}

Once the localization problem has converged, the coefficient of
$\delta\overline{\bm F}$ in Eq.~\eqref{eq:hill_mandel} gives the
homogenized first Piola--Kirchhoff stress,
\begin{equation}
\overline{\bm P}
=
\sum_{i=1}^{2^N}
w_i\,\bm P_i.
\label{eq:homogenized_stress}
\end{equation}
Since $\sum_i w_i=1$, Eq.~\eqref{eq:homogenized_stress} is the
volume average of the first Piola--Kirchhoff stress over the material
nodes.

The material tangent for the macroscopic finite-element problem must
include both the constitutive response of the individual
material nodes and the redistribution of the interaction
vectors that maintains equilibrium after a perturbation of the
macroscopic deformation gradient. At the converged state,
\begin{equation}
\bm r
\left(
\mathbf A(\overline{\bm F}),
\overline{\bm F}
\right)
=
\bm 0.
\end{equation}
Differentiating the equilibrium equation with respect to
$\operatorname{vec}(\overline{\bm F})$ gives
\begin{equation}
\bm K
\frac{
\partial\mathbf A
}{
\partial\operatorname{vec}(\overline{\bm F})
}
+
\sum_{i=1}^{2^N}
w_i\,
\bm D_i^{\mathsf T}
\bm{\mathsf L}_i
=
\bm 0,
\end{equation}
where $\bm K$ is the Newton Jacobian defined in
Eq.~\eqref{eq:newton_K}. Hence,
\begin{equation}
\frac{
\partial\mathbf A
}{
\partial\operatorname{vec}(\overline{\bm F})
}
=
-
\bm K^{-1}
\sum_{i=1}^{2^N}
w_i\,
\bm D_i^{\mathsf T}
\bm{\mathsf L}_i.
\label{eq:dA_dF}
\end{equation}

Using Eq.~\eqref{eq:D_matrix}, the sensitivity of the local deformation
gradient to the macroscopic deformation gradient follows as
\begin{equation}
\frac{
\partial\operatorname{vec}(\bm F_i)
}{
\partial\operatorname{vec}(\overline{\bm F})
}
=
\bm I_9
+
\bm D_i
\frac{
\partial\mathbf A
}{
\partial\operatorname{vec}(\overline{\bm F})
},
\label{eq:dFi_dFbar}
\end{equation}
where $\bm I_9$ is the $9\times9$ identity matrix.
The consistent macroscopic tangent in terms of the first
Piola--Kirchhoff stress and the deformation gradient is
therefore
\begin{equation}
\begin{aligned}
\overline{\bm{\mathsf L}}
&\equiv
\frac{
\partial\operatorname{vec}(\overline{\bm P})
}{
\partial\operatorname{vec}(\overline{\bm F})
}
\\
&=
\sum_{i=1}^{2^N}
w_i\,
\bm{\mathsf L}_i
\\
&\quad-
\Big(
\sum_{i=1}^{2^N}
w_i\,
\bm{\mathsf L}_i\bm D_i
\Big)
\bm K^{-1}
\Big(
\sum_{j=1}^{2^N}
w_j\,
\bm D_j^{\mathsf T}\bm{\mathsf L}_j
\Big).
\end{aligned}
\label{eq:homogenized_tangent}
\end{equation}
The first term in Eq.~\eqref{eq:homogenized_tangent} is the
constitutive response at fixed interaction vectors, and the
second term accounts for their redistribution. The second term is
needed for a consistent linearization of the converged
localization problem.

Equations~\eqref{eq:dA_dF} and
\eqref{eq:homogenized_tangent} are written in terms of the exact Newton
Jacobian $\bm K$. In the implementation, the Jacobian used in
the final sensitivity solve is reassembled after convergence with the
regularization parameter returned to its baseline value
$\lambda_0=10^{-10}$. This keeps the temporary damping of
difficult Newton iterations out of the homogenized
tangent and retains only a small regularization of the final
linear solve.

The ODMN is implemented at
each macroscopic integration point in Abaqus/Standard as a
user material subroutine (\texttt{UMAT}). The converged localization solution provides
$\overline{\bm P}$ and the fourth-order tangent
$\overline{\mathbb L}
=
\partial\overline{\bm P}/\partial\overline{\bm F}$,
which is the tensor form of $\overline{\bm{\mathsf L}}$. Both
must be converted to the stress measure and
constitutive Jacobian required by Abaqus.

The Kirchhoff stress $\overline{\bm\tau}$ and the Cauchy stress $\overline{\bm\sigma}$ are obtained as
\begin{equation}
\overline{\bm\tau}
=
\overline{\bm P}\,
\overline{\bm F}^{\mathsf T},
\qquad
\overline{\bm\sigma}
=
\frac{1}{\overline J}\,
\overline{\bm\tau},
\qquad
\overline J
=
\det\overline{\bm F},
\label{eq:cauchy_stress_abaqus}
\end{equation}
and the Cauchy stress $\overline{\bm\sigma}$ is returned through the
\texttt{STRESS} array.

For finite-deformation continuum elements, the constitutive Jacobian
required by Abaqus/Standard is associated with the Jaumann objective rate
of the Kirchhoff stress. In incremental form,
\begin{equation}
\delta^{\nabla J}\overline{\bm\tau}
=
\overline J\,
\overline{\mathbb C}^{\mathrm{ABQ}}
:
\delta\overline{\bm D},
\label{eq:abaqus_jaumann_relation}
\end{equation}
where
\begin{equation}
\delta^{\nabla J}\overline{\bm\tau}
=
\delta\overline{\bm\tau}
-
\delta\overline{\bm W}\,
\overline{\bm\tau}
+
\overline{\bm\tau}\,
\delta\overline{\bm W},
\label{eq:jaumann_increment}
\end{equation}
and
\begin{equation}
\delta\overline{\bm H}
=
\delta\overline{\bm F}\,
\overline{\bm F}^{-1}
=
\delta\overline{\bm D}
+
\delta\overline{\bm W}.
\label{eq:incremental_velocity_gradient}
\end{equation}
Here, $\overline{\mathbb C}^{\mathrm{ABQ}}$ is the constitutive Jacobian
required by Abaqus, $\delta^{\nabla J}\overline{\bm\tau}$ is the Jaumann
increment of the Kirchhoff stress, and $\delta\overline{\bm D}$ and
$\delta\overline{\bm W}$ are the symmetric and skew-symmetric parts of the
incremental displacement gradient $\delta\overline{\bm H}$. These symbols
are unrelated to the localization matrices $\bm D_i$ and to the matrix
$\bm H(\bm N)$ of Eq.~\eqref{eq:H_matrix}.

The linearization of the Kirchhoff stress is
\begin{equation}
\delta\overline{\bm\tau}
=
\delta\overline{\bm P}\,
\overline{\bm F}^{\mathsf T}
+
\overline{\bm P}\,
\delta\overline{\bm F}^{\mathsf T}.
\label{eq:kirchhoff_linearization}
\end{equation}
Writing the fourth-order ODMN tangent in components as
\begin{equation}
\overline{\mathbb L}_{iA kB}
=
\frac{
\partial\overline P_{iA}
}{
\partial\overline F_{kB}
},
\label{eq:dPdF_tensor}
\end{equation}
where lowercase and uppercase indices refer to the current and reference
configurations and summation over repeated indices is implied, and using
$\delta\overline{\bm F}
=
\delta\overline{\bm H}\,\overline{\bm F}$,
Eq.~\eqref{eq:kirchhoff_linearization} becomes
\begin{equation}
\delta\overline{\tau}_{ij}
=
\overline{\mathbb G}_{ijkl}\,
\delta\overline H_{kl}
+
\overline{\tau}_{il}\,
\delta\overline H_{jl},
\label{eq:kirchhoff_linearization_spatial}
\end{equation}
where
\begin{equation}
\overline{\mathbb G}_{ijkl}
=
\overline F_{jA}\,
\overline{\mathbb L}_{iA kB}\,
\overline F_{lB}
\label{eq:dPdF_push_forward}
\end{equation}
is the spatial push-forward of the consistent ODMN tangent.

Substituting
$\delta\overline{\bm H}
=
\delta\overline{\bm D}
+
\delta\overline{\bm W}$
into Eq.~\eqref{eq:kirchhoff_linearization_spatial} and inserting the
result into Eq.~\eqref{eq:jaumann_increment} leaves the spin-dependent
term
$\overline{\mathbb G}:\delta\overline{\bm W}
-\delta\overline{\bm W}\,\overline{\bm\tau}$.
This term vanishes because the homogenized response is objective, so that
a pure spin increment gives
$\delta\overline{\bm P}=\delta\overline{\bm W}\,\overline{\bm P}$.
The remaining terms give
\begin{equation}
\delta^{\nabla J}\overline{\tau}_{ij}
=
\left(
\overline{\mathbb G}_{ijkl}
+
\overline{\tau}_{ik}\delta_{jl}
\right)
\delta\overline D_{kl},
\label{eq:jaumann_from_dpdf}
\end{equation}
where $\delta_{jl}$ is the Kronecker delta.
Because both
$\delta^{\nabla J}\overline{\bm\tau}$
and
$\delta\overline{\bm D}$
are symmetric, only the minor-symmetric part of the fourth-order tensor
in Eq.~\eqref{eq:jaumann_from_dpdf} contributes. With
$\operatorname{sym}_{ij}$ denoting symmetrization with respect to the
index pair $(i,j)$,
\begin{equation}
\overline J\,
\overline{\mathbb C}^{\mathrm{ABQ}}_{ijkl}
=
\operatorname{sym}_{ij}
\operatorname{sym}_{kl}
\left(
\overline{\mathbb G}_{ijkl}
+
\overline{\tau}_{ik}\delta_{jl}
\right),
\label{eq:abaqus_consistent_tangent_compact}
\end{equation}
or, equivalently,
\begin{equation}
\begin{aligned}
\overline{\mathbb C}^{\mathrm{ABQ}}_{ijkl}
=
\frac{1}{4\overline J}
\Big(
&
\overline{\mathbb G}_{ijkl}
+
\overline{\mathbb G}_{jikl}
+
\overline{\mathbb G}_{ijlk}
+
\overline{\mathbb G}_{jilk}
\\
&
+
\delta_{ik}\overline{\tau}_{jl}
+
\delta_{il}\overline{\tau}_{jk}
+
\delta_{jk}\overline{\tau}_{il}
+
\delta_{jl}\overline{\tau}_{ik}
\Big).
\end{aligned}
\label{eq:abaqus_consistent_tangent}
\end{equation}

The tensor
$\overline{\mathbb C}^{\mathrm{ABQ}}$
is mapped to the Abaqus Voigt representation and returned through the
\texttt{DDSDDE} array. The \texttt{STRESS} and \texttt{DDSDDE} quantities
are thus obtained from the
same converged localization state, and the
Jacobian is consistent with both the ODMN
linearization and the objective stress rate of
Abaqus/Standard. This consistency is required for the quadratic
convergence of the global Newton iteration.

Algorithm~\ref{alg:odmn_localization} summarizes the localization solve at a prescribed macroscopic state. The implementation avoids a global cutback of the finite-element time increment when convergence difficulties are confined to a single integration point, because a global cutback repeats the entire macroscopic increment, including the equilibrium iterations and the constitutive evaluations at all integration points. If the localization solve for the full increment fails at an integration point, convergence is first recovered locally by adaptively subdividing the prescribed loading path. Only if this local recovery fails is a global cutback requested from Abaqus/Standard.

For the local substepping, let $s\in[0,1]$ denote the fraction of the prescribed macroscopic increment that has been completed, and let $\Delta s$ denote the size of the next local subincrement. Within each subincrement, the deformation gradient and the temperature are interpolated with the same path fraction, and the time increment is scaled by $\Delta s$ so that the imposed loading rate is preserved. Algorithm~\ref{alg:odmn_umat} summarizes the complete constitutive update at an integration point, including the local substepping and the conversion to the stress and Jacobian required by Abaqus/Standard. The cutback factor, growth factor, and minimum subincrement are $\beta_{\mathrm{cut}}=0.25$, $\beta_{\mathrm{grow}}=1.5$, and $\Delta s_{\min}=10^{-3}$.

\begin{algorithm*}[t]
\footnotesize
\caption{Newton--Raphson solution of the nonlinear ODMN localization problem}
\label{alg:odmn_localization}
\begin{algorithmic}[1]

\algrenewcommand\algorithmicrequire{\textbf{Input:}}
\algrenewcommand\algorithmicensure{\textbf{Output:}}

\Require
Prescribed macroscopic state
$\overline{\bm F}^{\,*}$,
$\overline T^{\,*}$,
$\Delta t^{*}$;
reference material-node states
$\{\mathcal H_i^{\mathrm{ref}}\}$;
initial interaction-vector estimate $\mathbf A^{(0)}$

\Ensure
Convergence flag;
$\overline{\bm P}$;
$\overline{\bm{\mathsf L}}$;
converged interaction vector $\mathbf A$;
material-node trial states
$\{\mathcal H_i^{\mathrm{trial}}\}$

\State
$\mathbf A\gets\mathbf A^{(0)}$,
$\lambda\gets\lambda_0$
\Comment{Initialize the localization solve}

\State
$\mathbf A_{\mathrm{best}}\gets\mathbf A$,
$r_{\mathrm{best}}\gets+\infty$

\For{$k=0,1,\ldots,k_{\max}$}

    \For{$i=1,\ldots,2^N$}

        \State
        $\operatorname{vec}(\bm F_i)
        \gets
        \operatorname{vec}(\overline{\bm F}^{\,*})
        +\bm D_i\mathbf A$
        \Comment{Eq.~\eqref{eq:D_matrix}}

        \State
        $(\bm P_i,\bm{\mathsf L}_i,\mathcal H_i^{\mathrm{trial}})
        \gets
        \textsc{MaterialUpdate}
        (\bm F_i,\overline T^{\,*},\Delta t^{*},
        \mathcal H_i^{\mathrm{ref}})$
        \Comment{Section~\ref{sec251}}

    \EndFor

    \State
    $\bm r
    \gets
    \displaystyle
    \sum_{i=1}^{2^N}
    w_i\bm D_i^{\mathsf T}
    \operatorname{vec}(\bm P_i)$
    \Comment{Eq.~\eqref{eq:residual_vector}}

    \State
    $p_{\max}
    \gets
    \displaystyle
    \max_i\|\bm P_i\|_{\max}$

    \State
    $r_{\mathrm{tol}}
    \gets
    \max\!\left[
    \epsilon_{\mathrm{abs}},
    \epsilon_{\mathrm{rel}}\,
    d_{\max}\,
    \max(p_{\max},p_{\mathrm{floor}})
    \right]$
    \Comment{Eq.~\eqref{eq:newton_convergence}}

    \If{$\|\bm r\|_{\infty}<r_{\mathrm{best}}$}
        \State
        $\mathbf A_{\mathrm{best}}\gets\mathbf A$,
        $r_{\mathrm{best}}\gets\|\bm r\|_{\infty}$
    \EndIf

    \State
    $\texttt{failed}
    \gets
    (\text{local constitutive failure})
    \lor
    (\text{divergence})
    \lor
    (\text{iterative stagnation})$

    \If{\texttt{failed}}
        \State
        $\mathbf A\gets\mathbf A_{\mathrm{best}}$
        \Comment{Restore the best admissible localization state}
        \State
        \Return \textsc{Failed}
    \EndIf

    \If{$\|\bm r\|_{\infty}\le r_{\mathrm{tol}}$}

        \State
        $\lambda\gets\lambda_0$

        \State
        $\bm K
        \gets
        \displaystyle
        \sum_{i=1}^{2^N}
        w_i
        \bm D_i^{\mathsf T}
        \bm{\mathsf L}_i
        \bm D_i$
        \Comment{Eq.~\eqref{eq:newton_K}}

        \State
        $\bm K_{\mathrm{reg}}
        \gets
        \bm K
        +
        \lambda_0\max_j|K_{jj}|\bm I$
        \Comment{Eq.~\eqref{eq:newton_regularization}}

        \State
        $\displaystyle
        \overline{\bm P}
        \gets
        \sum_{i=1}^{2^N}w_i\bm P_i$
        \Comment{Eq.~\eqref{eq:homogenized_stress}}

        \State
        Compute $\overline{\bm{\mathsf L}}$
        \Comment{Eq.~\eqref{eq:homogenized_tangent}}

        \State
        \Return \textsc{Converged}

    \EndIf

    \State
    $\bm K
    \gets
    \displaystyle
    \sum_{i=1}^{2^N}
    w_i
    \bm D_i^{\mathsf T}
    \bm{\mathsf L}_i
    \bm D_i$
    \Comment{Eq.~\eqref{eq:newton_K}}

    \State
    $\bm K_{\mathrm{reg}}
    \gets
    \bm K
    +
    \lambda\max_j|K_{jj}|\bm I$
    \Comment{Eq.~\eqref{eq:newton_regularization}}

    \State
    Solve
    $\bm K_{\mathrm{reg}}\delta\mathbf A=-\bm r$
    \Comment{Eq.~\eqref{eq:newton_update}}

    \State
    $\mathbf A^{(k)}\gets\mathbf A$,
    $\eta\gets1$,
    $r_{\mathrm{LS}}\gets+\infty$,
    $\texttt{accepted}\gets\texttt{false}$

    \For{$q=0,1,\ldots,q_{\max}$}

        \State
        $\mathbf A^{\mathrm{trial}}
        \gets
        \mathbf A^{(k)}+\eta\delta\mathbf A$

        \State
        Re-evaluate
        $\bm r^{\mathrm{trial}}$
        at $\mathbf A^{\mathrm{trial}}$

        \If{$\|\bm r^{\mathrm{trial}}\|_{\infty}<r_{\mathrm{LS}}$}
            \State
            $\mathbf A_{\mathrm{LS}}\gets\mathbf A^{\mathrm{trial}}$,
            $r_{\mathrm{LS}}
            \gets
            \|\bm r^{\mathrm{trial}}\|_{\infty}$
            \Comment{Retain the best tested step}
        \EndIf

        \If{$
        \|\bm r^{\mathrm{trial}}\|_{\infty}
        \le
        (1-c\eta)\|\bm r\|_{\infty}
        $}

            \State
            $\mathbf A\gets\mathbf A^{\mathrm{trial}}$,
            $\texttt{accepted}\gets\texttt{true}$
            \Comment{Eq.~\eqref{eq:line_search}}

            \State
            \textbf{break}

        \EndIf

        \State
        $\eta\gets\eta/2$
        \Comment{Backtracking line search}

    \EndFor

    \If{\textbf{not} \texttt{accepted}}

        \State
        $\mathbf A\gets\mathbf A_{\mathrm{LS}}$

        \State
        $\lambda
        \gets
        \min\!\left[
        \max(100\lambda,10^{-8}),
        10^{-2}
        \right]$
        \Comment{Increase Newton damping}

    \ElsIf{$\eta=1$}

        \State
        $\lambda
        \gets
        \max(\lambda_0,\lambda/10)$
        \Comment{Reduce damping after a full Newton step}

    \EndIf

\EndFor

\State
$\mathbf A\gets\mathbf A_{\mathrm{best}}$

\State
\Return \textsc{Failed}

\end{algorithmic}
\end{algorithm*}

\begin{algorithm*}[t]
\small
\caption{Integration-point ODMN constitutive update with local adaptive substepping}
\label{alg:odmn_umat}
\begin{algorithmic}[1]

\algrenewcommand\algorithmicrequire{\textbf{Input:}}
\algrenewcommand\algorithmicensure{\textbf{Output:}}

\Require
Macroscopic increment
$\overline{\bm F}_n\rightarrow\overline{\bm F}_{n+1}$,
$\overline T_n\rightarrow\overline T_{n+1}$,
$\Delta t$;
converged material-node history $\mathcal H_n$;
converged interaction vector $\mathbf A_n$

\Ensure
Abaqus \texttt{STRESS} and \texttt{DDSDDE};
trial ODMN state $\mathcal H_{n+1}^{\mathrm{trial}}$;
global cutback request if local recovery fails

\State
Load the converged material-node history $\mathcal H_n$

\State
$\mathbf A^{(0)}\gets\mathbf A_n$
\Comment{Initialize from the preceding converged interaction vector}

\State
$\texttt{converged}
\gets
\textsc{ODMNSolve}
(\overline{\bm F}_{n+1},
 \overline T_{n+1},
 \Delta t,
 \mathcal H_n,
 \mathbf A^{(0)})$
\Comment{Algorithm~\ref{alg:odmn_localization}}

\If{\textbf{not} \texttt{converged}}

    \State
    $\mathcal H^{\mathrm{ref}}\gets\mathcal H_n$,
    $\mathbf A^{\mathrm{ref}}\gets\mathbf A_n$
    \Comment{Restore the increment-start converged state}

    \State
    $s\gets0$,
    $\Delta s\gets\beta_{\mathrm{cut}}$
    \Comment{Initialize local path subdivision}

    \While{$s<1$}

        \State
        $\Delta s\gets\min(\Delta s,1-s)$
        \Comment{Restrict the trial point to the prescribed increment}

        \State
        $s^{*}\gets s+\Delta s$
        \Comment{Trial loading-path fraction}

        \State
        $\overline{\bm F}^{\,*}
        \gets
        \overline{\bm F}_n
        +
        s^{*}
        \left(
        \overline{\bm F}_{n+1}
        -
        \overline{\bm F}_n
        \right)$
        \Comment{Interpolate the deformation-gradient path}

        \State
        $\overline T^{\,*}
        \gets
        \overline T_n
        +
        s^{*}
        \left(
        \overline T_{n+1}
        -
        \overline T_n
        \right)$
        \Comment{Interpolate the prescribed temperature path}

        \State
        $\Delta t^{*}
        \gets
        \Delta s\,\Delta t$
        \Comment{Preserve the imposed loading rate}

        \State
        $\texttt{converged}
        \gets
        \textsc{ODMNSolve}
        (\overline{\bm F}^{\,*},
         \overline T^{\,*},
         \Delta t^{*},
         \mathcal H^{\mathrm{ref}},
         \mathbf A^{\mathrm{ref}})$
        \Comment{Algorithm~\ref{alg:odmn_localization}}
        
        \If{\texttt{converged}}

            \State
            $s\gets s^{*}$
            \Comment{Advance the converged path fraction}

            \State
            $\mathcal H^{\mathrm{ref}}
            \gets
            \mathcal H^{\mathrm{trial}}$,
            $\mathbf A^{\mathrm{ref}}\gets\mathbf A$
            \Comment{Advance the local reference state}

            \If{$s<1$}
                \State
                $\Delta s
                \gets
                \min\!\left(
                1-s,\,
                \beta_{\mathrm{grow}}\Delta s
                \right)$
                \Comment{Adaptive substep growth}
            \EndIf

        \Else

            \State
            Restore
            $\mathcal H^{\mathrm{ref}}$ and $\mathbf A^{\mathrm{ref}}$
            \Comment{Rollback to the last converged local state}

            \State
            $\Delta s
            \gets
            \beta_{\mathrm{cut}}\Delta s$
            \Comment{Local substep cutback}

            \If{$\Delta s<\Delta s_{\min}$}

                \State
                Request a global Abaqus time-increment cutback
                \Comment{Local recovery exhausted}

                \State
                \Return \textsc{Failed}

            \EndIf

        \EndIf

    \EndWhile

\EndIf

\State
Convert
$(\overline{\bm P},\overline{\mathbb L})$
to
$(\overline{\bm\sigma},
\overline{\mathbb C}^{\mathrm{ABQ}})$
\Comment{Eqs.~\eqref{eq:cauchy_stress_abaqus} and \eqref{eq:abaqus_consistent_tangent}}

\State
\texttt{STRESS}
$\gets\overline{\bm\sigma}$,
\qquad
\texttt{DDSDDE}
$\gets\overline{\mathbb C}^{\mathrm{ABQ}}$
\Comment{Abaqus UMAT output}

\State
$\mathcal H_{n+1}^{\mathrm{trial}}
\gets
\mathcal H^{\mathrm{trial}}$
\Comment{Store the final converged state}

\State
\Return \textsc{Converged}

\end{algorithmic}
\end{algorithm*}

\section{Verification of the proposed framework}\label{sec3}

This section verifies the GSH-ODMN against full-field CPFFT simulations of two Inconel 625 (IN625) RVEs. Section~\ref{sec31} describes the RVEs, the constitutive model shared by the DNS and the material nodes, the loading protocols, and the error measures. Section~\ref{sec32} examines the effect of the GSH truncation order $L$ on the compact texture. Section~\ref{sec33} compares the two offline training strategies. Sections~\ref{sec34} and \ref{sec35} assess the online predictions of the macroscopic stress response and of the texture evolution, and Section~\ref{sec36} reports the computational cost.

\subsection{Reference RVEs, constitutive model, and loading protocols}
\label{sec31}

\begin{figure*}[!t]
\centering
\includegraphics[width=\textwidth]{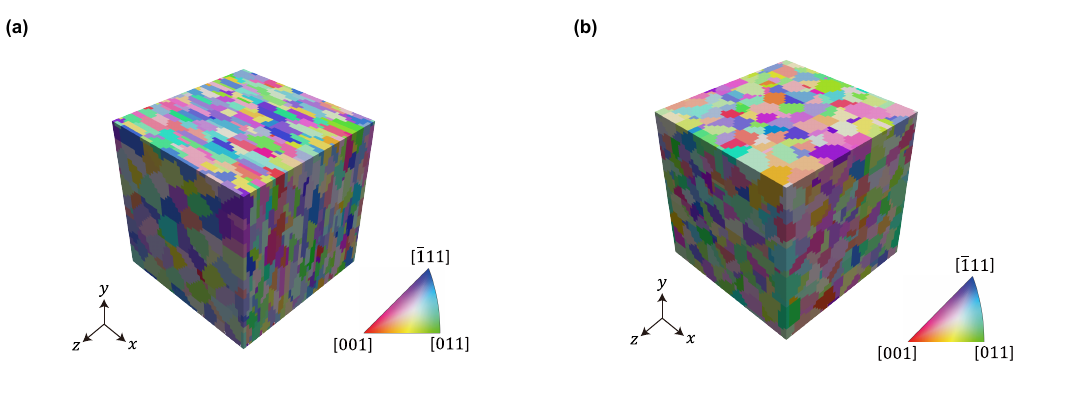}
\caption{Initial microstructures of (a) the rolled RVE and (b) the equiaxed RVE, each consisting of $48^3$ voxels and 1000 grains. Colors follow the inverse pole figure of the loading direction, the sample axis $x_1$, in the crystal reference frame.}
\label{fig:rve_ipf}
\end{figure*}

Two periodic RVEs of IN625 are taken from Prabhu and Diehl~\cite{prabhu2026incorporation}. Both contain 1000 grains discretized on a $48\times48\times48$ voxel grid and were generated by Voronoi tessellation of random seed points with DAMASK~\cite{roters2019damask}. The equiaxed RVE has globular grains and orientations drawn uniformly from the rotation group, so that its texture is nearly random. The deviation from a uniform ODF is due to the sampling fluctuations of 1000 grains. The rolled RVE has grains flattened along the normal direction, with an aspect ratio of about 3, and an idealized rolling texture of a face-centered cubic metal, composed of S, copper, and brass components superimposed on a random background. Fig.~\ref{fig:rve_ipf} shows both microstructures. The grain volumes are far from uniform. The number of voxels per grain ranges from 17 to 350 in the equiaxed RVE and from 13 to 385 in the rolled RVE, with a coefficient of variation of about 0.4 in both cases. The volume fractions \(v_q\) in Eq.~\eqref{eq:rve_texture} are therefore taken proportional to the voxel counts of the grains, so that the compaction target is the volume-weighted ODF realized on the grid.

The DNS and the ODMN material nodes use the constitutive model of Section~\ref{sec251} with the parameters in Table~\ref{tab:material}. These parameters are the fit of Prabhu and Diehl~\cite{prabhu2026incorporation} for IN625 with physics-based interaction coefficients, and the temperature-dependent elastic constants are taken from the same work. Plastic slip occurs on the twelve $\{111\}\langle1\bar{1}0\rangle$ systems. Thermal expansion is isotropic with $\alpha_{11}=1.28\times10^{-5}\,\mathrm{K}^{-1}$. Between 293.15\,K and 593.15\,K, the elastic constants $C_{11}$ and $C_{44}$ decrease by about 7\,\% and 9\,\%, respectively, whereas the slip resistances are independent of temperature in this model.

\begin{table*}[tp]
\caption{Constitutive parameters of IN625 used in the DNS and at the ODMN material nodes (fit of Prabhu and Diehl~\cite{prabhu2026incorporation}).}
\label{tab:material}
\begin{threeparttable}
\begin{tabular}{lll}
\toprule
parameter & symbol & value \\
\midrule
elastic constants at $T_{\mathrm{ref}}=293.15$ K
& $C_{11}^{(0)}, C_{12}^{(0)}, C_{44}^{(0)}$
& 251.0, 150.0, 123.7 GPa \\

linear temperature coefficients
& $C_{11}^{(1)}, C_{12}^{(1)}, C_{44}^{(1)}$
& $-49.67$, $-3.714$, $-36.21$ MPa/K \\

quadratic temperature coefficients
& $C_{11}^{(2)}, C_{12}^{(2)}, C_{44}^{(2)}$
& $-33.27$, $-7.414$, $-8.017$ kPa/K$^2$ \\

slip systems
& $N_{\mathrm{s}}$
& 12, $\{111\}\langle 1\bar{1}0\rangle$ \\

reference shear rate
& $\dot{\gamma}_0$
& $10^{-3}$ s$^{-1}$ \\

stress exponent
& $n$
& 40 \\

initial slip resistance
& $\xi_0$
& 152.4 MPa \\

saturation slip resistance
& $\xi_{\infty}$
& 1200 MPa \\

reference hardening modulus
& $h_0$
& 1.328 GPa \\

hardening exponent
& $a$
& 1.500 \\

interaction coefficients\tnote{a}
& $h^{\alpha\beta}$
& 1.000, 1.000, 5.508, 0.361, 0.836, 0.738, 1.377 \\

thermal expansion coefficient
& $\alpha_{11}$
& $1.28\times10^{-5}$ K$^{-1}$ \\
\bottomrule
\end{tabular}

\begin{tablenotes}[flushleft]
\footnotesize
\item[a] Coefficients of Madec and Kubin~\cite{madec2017dislocation}, in the order self, coplanar, collinear, Hirth lock, glissile junction I, glissile junction II, Lomer lock.
\end{tablenotes}
\end{threeparttable}
\end{table*}

The reference solutions are computed with the spectral solver of DAMASK (version 3.0.2) on the full $48^3$ grid, using the fixed-point scheme~\cite{eisenlohr2013spectral,shanthraj2015numerically} and Open Multi-Processing (OpenMP) threads on a desktop workstation (Intel Core i9-13900K). In the temperature-cycle protocol described below, the temperature is raised and lowered by a uniform volumetric heat source in the staggered thermal solver of DAMASK, so that the temperature of the RVE remains spatially uniform, consistent with Eq.~\eqref{eq:isothermal_temperature}. The linear-elastic training data of Section~\ref{sec24} are generated with the same solver by probing each sampled single-crystal stiffness with six independent load cases and assembling the homogenized stiffness from the volume-averaged stresses and strains.

The ODMN predictions are obtained with the user-subroutine implementation of Section~\ref{sec253} in Abaqus/Standard on a single reduced-integration hexahedral element whose faces are constrained to remain plane, so that the element experiences the same homogeneous deformation gradient as the DNS. The ODMN runs on one central processing unit (CPU) core of the same workstation.

Two loading protocols are considered. Both apply strain-controlled uniaxial tension along the sample axis $x_1$, with an LUR path of the axial component $\overline F_{11}$ of the macroscopic deformation gradient,
\begin{equation}
\begin{aligned}
&\overline F_{11}:\ 1.00\rightarrow1.10\rightarrow1.05\rightarrow1.15,
\\
&\left|\dot{\overline F}_{11}\right|=10^{-3}\,\mathrm{s}^{-1}.
\end{aligned}
\label{eq:lur_path}
\end{equation}
In the isothermal protocol, the path is traversed at 293.15\,K in 250\,s, with loading from 0 to 100\,s, partial unloading from 100 to 150\,s, and reloading from 150 to 250\,s. In the temperature-cycle protocol, the same three deformation legs are separated by two holds at fixed $\overline F_{11}$, during which the temperature changes at 3\,K/s. The sequence is loading at 293.15\,K (0 to 100\,s), heating to 593.15\,K (100 to 200\,s), unloading at 593.15\,K (200 to 250\,s), cooling to 293.15\,K (250 to 350\,s), and reloading at 293.15\,K (350 to 450\,s). Every deformation leg is therefore isothermal, and every temperature change takes place at fixed deformation. The partial unloading to $\overline F_{11}=1.05$ drives the RVE into reverse yielding, so that reverse slip on the active systems and the hardening state accumulated during loading are also tested.

The macroscopic response is assessed by the relative error of the axial stress $\overline P_{11}$ at time $t$,
\begin{equation}
e(t)
=
\frac{
\left|\overline P_{11}^{\mathrm{ODMN}}(t)-\overline P_{11}^{\mathrm{DNS}}(t)\right|
}{
\max_{t}\left|\overline P_{11}^{\mathrm{DNS}}(t)\right|
},
\label{eq:stress_error}
\end{equation}
whose mean and maximum over the loading history are reported. Textures are compared with the normalized texture-difference index~\cite{li2005texture}
\begin{equation}
\hat{T}^{d}
=
\frac{
\left\|f-f_{\mathrm{ref}}\right\|^{2}
}{
\left\|f_{\mathrm{ref}}\right\|^{2}
},
\label{eq:Td}
\end{equation}
where $f$ is the ODF under assessment, $f_{\mathrm{ref}}$ is the reference ODF, both reconstructed from weighted sets of orientations, and $\|\cdot\|$ is the $L^2$ norm over the orientation space. Both ODFs are estimated in MTEX 5.11.1~\cite{bachmann2010texture} with a de la Vall\'ee Poussin kernel of $10^\circ$ half-width. In Sections~\ref{sec33}--\ref{sec35}, the ODMN texture is compared with the corresponding DNS texture. In Section~\ref{sec32}, the reference is the volume-weighted texture of the 1000-grain RVE. For the DNS, the ODF is reconstructed from the $48^3$ voxel orientations with equal weights. For the ODMN, it is reconstructed from the $2^N$ material-node orientations with the weights $w_i$. The current orientation of each voxel or material node is given by the rotation $\bm R_i(t)$ in Eq.~\eqref{eq:elastic_polar_decomposition}. A value of $\hat{T}^{d}=0$ indicates identical ODFs, whereas $\hat{T}^{d}=1$ indicates that the $L^2$ norm of their difference equals that of the reference ODF.

\subsection{Effect of the GSH truncation order}
\label{sec32}

\begin{figure*}[tp]
\centering
\includegraphics[width=\textwidth]{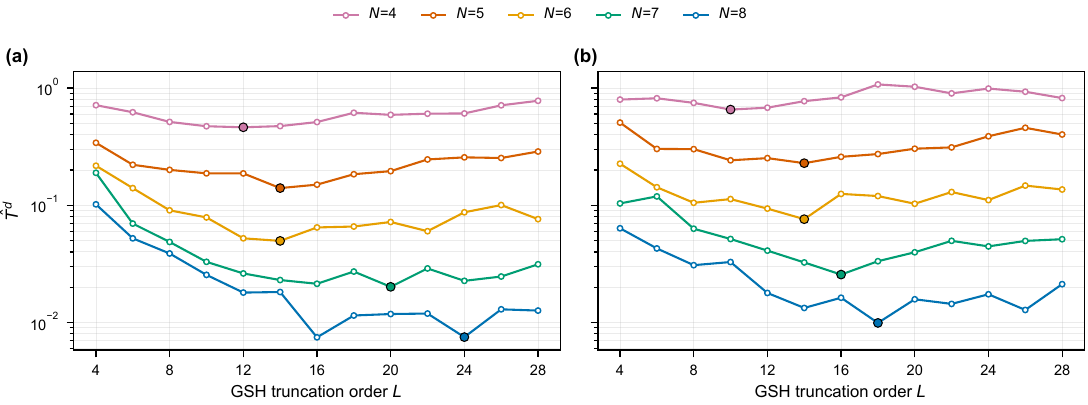}
\caption{Texture-difference index $\hat{T}^{d}$ of the compact texture with respect to the volume-weighted 1000-grain reference as a function of the GSH truncation order $L$ for $2^N$ orientations: (a) rolled RVE, (b) equiaxed RVE. Filled symbols mark the adopted orders of Table~\ref{tab:L_selection}.}
\label{fig:Td_vs_L}
\end{figure*}

The volume-weighted textures of the 1000-grain RVEs are compacted with the procedure of Section~\ref{sec22} into $N_m=2^N$ orientations for $N=4,\ldots,8$ and for all even truncation orders from $L=4$ to $L=28$. Fig.~\ref{fig:Td_vs_L} shows the texture-difference index of each compact texture relative to its 1000-grain reference texture.

Two trends emerge. First, the minimum $\hat{T}^{d}$ at each depth, reported in Table~\ref{tab:L_selection}, decreases by a factor of 2.5--3.3 with each increase in $N$. Second, $\hat{T}^{d}$ varies non-monotonically with $L$. For each $N$, an intermediate order minimizes $\hat{T}^{d}$, and retaining higher-order coefficients increases the error. The reason is that, at higher orders, the fitting residual of Eq.~\eqref{eq:gsh_error} is increasingly dominated by high-order coefficients that a limited number of orientations cannot reproduce, which degrades the fit of the low-order coefficients. The low-order coefficients describe the dominant texture components and largely determine the homogenized elastic response. As $N$ increases, more orientations are available to represent finer features of the ODF. The optimal order therefore shifts from $L=10$ or 12 at $N=4$ to $L=16$ or 18 at $N=8$, depending on the RVE.

\begin{table}[tp]
\caption{Adopted GSH truncation order $L$ and texture-difference index $\hat{T}^{d}$ of the compact texture with respect to the volume-weighted 1000-grain reference.}
\label{tab:L_selection}
\begin{tabular}{crrrr}
\toprule
& \multicolumn{2}{c}{rolled RVE} & \multicolumn{2}{c}{equiaxed RVE} \\
\cmidrule(lr){2-3}\cmidrule(lr){4-5}
$N$ & $L$ & $\hat{T}^{d}$ & $L$ & $\hat{T}^{d}$ \\
\midrule
4 & 12 & 0.4623 & 10 & 0.6549 \\
5 & 14 & 0.1401 & 14 & 0.2285 \\
6 & 14 & 0.0496 & 14 & 0.0761 \\
7 & 20 & 0.0201 & 16 & 0.0256 \\
8 & 24 & 0.0075 & 18 & 0.0099 \\
\bottomrule
\end{tabular}
\end{table}

Table~\ref{tab:L_selection} lists the adopted truncation orders and the corresponding texture-difference indices. The adopted orders are the minima in Fig.~\ref{fig:Td_vs_L}. For the rolled RVE at $N=8$, two equal minima occur at $L=16$ and $L=24$, both with $\hat{T}^{d}=0.0075$, and the higher order is adopted to retain more of the high-order content of the sharp rolling texture. The compact textures improve markedly with the number of material nodes. At $N=4$ (16 orientations), $\hat{T}^{d}$ is 0.46 for the rolled RVE and 0.65 for the equiaxed RVE. At $N=8$ (256 orientations), it is below 0.01 for both RVEs. For the adopted orders, each compaction takes less than 2\,s, and the longest compaction in the entire sweep takes 35\,s. A complete sweep over $L$, evaluated with the same texture-difference index that is used in the verification below, therefore takes only a few minutes. Such a sweep is most useful when the number of material nodes is small, because the best truncation order is then well below the highest order considered.

\subsection{Offline training results}
\label{sec33}

\begin{figure*}[tp]
\centering
\includegraphics[width=\textwidth]{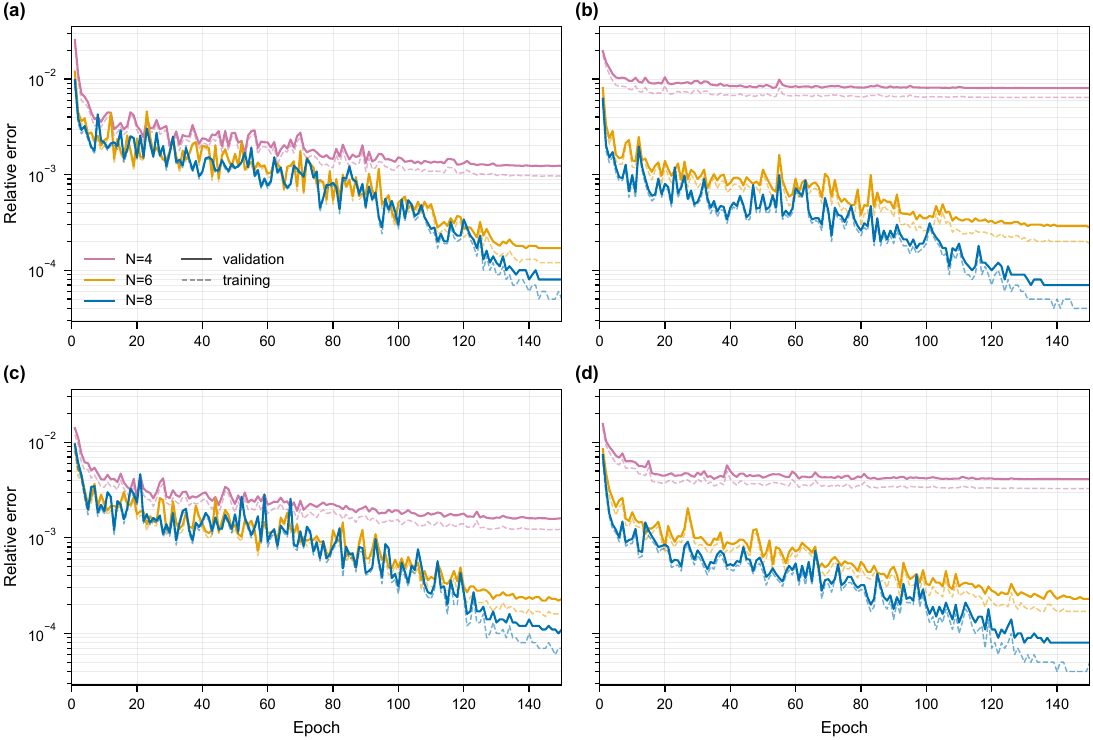}
\caption{Training (dashed) and validation (solid) errors in the homogenized stiffness during offline training for the rolled and equiaxed RVEs: (a, c) conventional ODMN and (b, d) GSH-ODMN, with (a, b) corresponding to the rolled RVE and (c, d) to the equiaxed RVE.}
\label{fig:training_curves}
\end{figure*}

Both strategies are trained for each RVE at $N=4$, 6, and 8 with the dataset and settings of Section~\ref{sec24}. The GSH-ODMN uses the compact textures of Table~\ref{tab:L_selection} as the fixed orientations and weights of its material nodes. Fig.~\ref{fig:training_curves} shows the training and validation histories. Table~\ref{tab:offline} lists the number of trainable parameters $n_{\Theta}$, the best validation error, and the training wall-clock time on the desktop workstation (Intel Core i9-13900K). Generating the dataset for one RVE takes approximately 4\,h on the same workstation, comprising 2.7\,h for CPFFT simulations (six load cases per sample, with six four-thread simulations running concurrently) and 1.4\,h for extracting the effective stiffness tensors. It also lists the texture-difference index $\hat{T}^{d}$ for the trained network at $t=0$ relative to the DNS texture.

\begin{table*}[tp]
\caption{Offline training results: number of trainable parameters $n_{\Theta}$, best validation error of the homogenized stiffness, training wall-clock time, and texture-difference index $\hat{T}^{d}$ of the trained network at $t=0$ with respect to the DNS texture.}
\label{tab:offline}
\begin{tabular}{llrrrrrrrr}
\toprule
& & \multicolumn{4}{c}{conventional ODMN} & \multicolumn{4}{c}{GSH-ODMN} \\
\cmidrule(lr){3-6}\cmidrule(lr){7-10}
RVE & $N$ & $n_{\Theta}$ & error [\%] & time & $\hat{T}^{d}$ & $n_{\Theta}$ & error [\%] & time & $\hat{T}^{d}$ \\
\midrule
rolled & 4 & 94 & 0.1229 & 5\,min 4\,s & 1.0564 & 30 & 0.8015 & 2\,min 35\,s & 0.4623 \\
rolled & 6 & 382 & 0.0167 & 22\,min 13\,s & 0.2551 & 126 & 0.0284 & 11\,min 0\,s & 0.0496 \\
rolled & 8 & 1534 & 0.0077 & 3\,h 0\,min 18\,s & 0.0964 & 510 & 0.0067 & 1\,h 0\,min 53\,s & 0.0075 \\
\midrule
equiaxed & 4 & 94 & 0.1579 & 5\,min 4\,s & 1.3525 & 30 & 0.4102 & 2\,min 37\,s & 0.6549 \\
equiaxed & 6 & 382 & 0.0220 & 22\,min 5\,s & 0.2536 & 126 & 0.0232 & 11\,min 4\,s & 0.0761 \\
equiaxed & 8 & 1534 & 0.0104 & 3\,h 0\,min 21\,s & 0.0785 & 510 & 0.0076 & 1\,h 16\,min 3\,s & 0.0099 \\
\bottomrule
\end{tabular}
\end{table*}

The offline error decreases with $N$ for both strategies. At $N=4$, the GSH-ODMN has a higher validation error than the conventional ODMN (0.80\,\% versus 0.12\,\% for the rolled RVE and 0.41\,\% versus 0.16\,\% for the equiaxed RVE). With 16 prescribed orientations, only the 15 interaction directions (30 parameters) are trainable. The network therefore has less freedom to match the effective stiffness than the conventional network, which has 94 free parameters. The gap narrows at $N=6$ (0.028\,\% versus 0.017\,\% and 0.023\,\% versus 0.022\,\%), and at $N=8$ the GSH-ODMN is slightly more accurate (0.0067\,\% versus 0.0077\,\% and 0.0076\,\% versus 0.0104\,\%). Since only one-third of the parameters are trained, the GSH-ODMN trains in one-third to one-half of the time.

\begin{figure*}[tp]
\centering
\includegraphics[width=\textwidth]{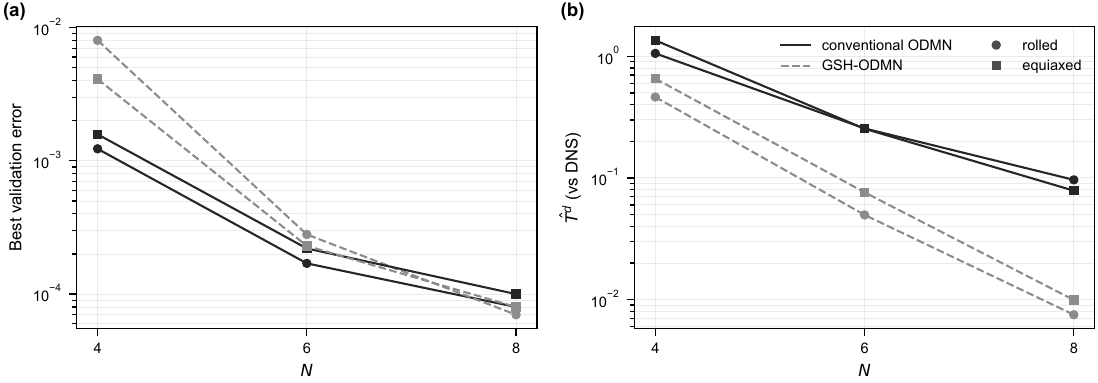}
\caption{Offline training results versus network depth $N$: (a) best validation error of the homogenized stiffness, (b) texture-difference index $\hat{T}^{d}$ of the trained network at $t=0$ with respect to the DNS texture. Line style distinguishes the training strategies, markers the RVEs.}
\label{fig:offline_summary}
\end{figure*}

The two strategies differ most in the textures they represent (Fig.~\ref{fig:offline_summary}). For the GSH-ODMN, the initial texture is the compact texture by construction, and its $\hat{T}^{d}$ coincides with the value in Table~\ref{tab:L_selection}. The conventional ODMN learns its orientations and weights from elastic data alone, and its texture-difference indices are two to thirteen times larger. Its texture error also decreases more slowly with $N$ and stays close to 0.1 at $N=8$, although the elastic validation error is only about 0.01\,\%. This is expected, because the orientation averages that govern the homogenized elastic stiffness of a cubic polycrystal depend on the ODF only through the GSH coefficients up to harmonic degree $l=4$~\cite{bunge2013texture,marki2024nonlinear}. Orientation sets with very different higher-order texture content can therefore produce nearly identical elastic responses. Elastic training data cannot distinguish among such textures, and spurious higher-order components persist even when the stiffness error becomes negligible.

\begin{figure*}[tp]
\centering
\includegraphics[width=0.92\textwidth]{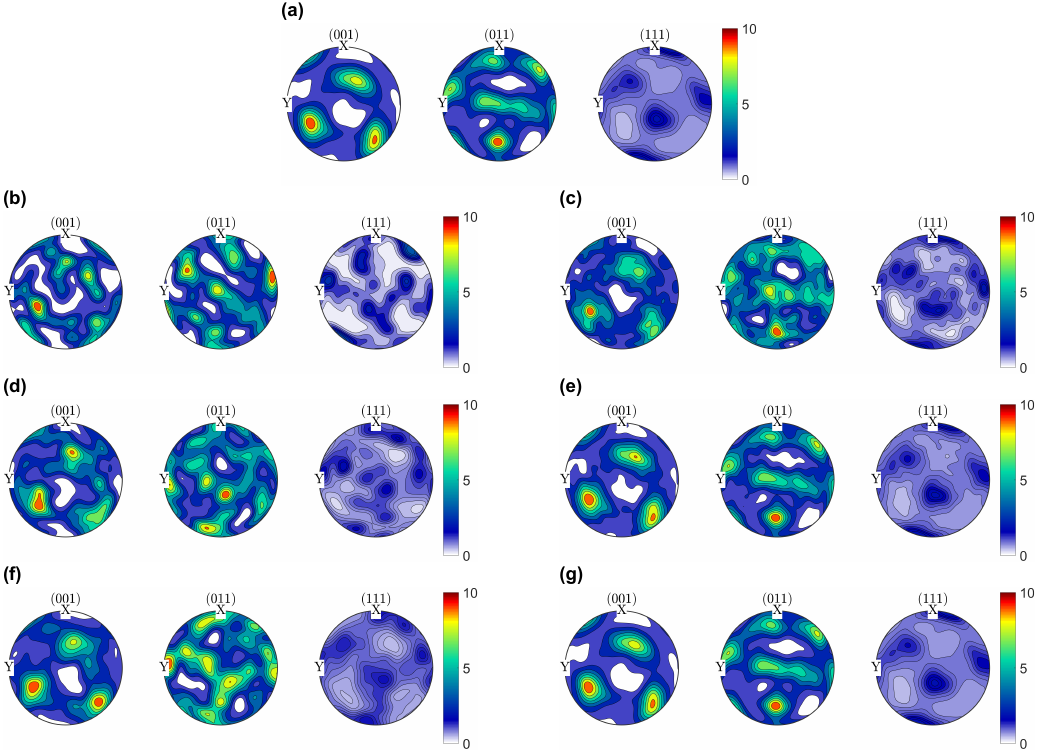}
\caption{Initial textures of the rolled RVE represented by the (001), (011), and (111) pole figures using an equal-area projection and a fixed intensity scale of 0--10 multiples of a random distribution (m.r.d.). Panel (a) shows the volume-weighted 1000-grain reference, and the remaining panels show the trained networks with $N=4$ (b, c), $N=6$ (d, e), and $N=8$ (f, g). The left column (b, d, f) corresponds to the conventional ODMN, and the right column (c, e, g) to the GSH-ODMN.}
\label{fig:PF_initial_rolled}
\end{figure*}

\begin{figure*}[tp]
\centering
\includegraphics[width=0.92\textwidth]{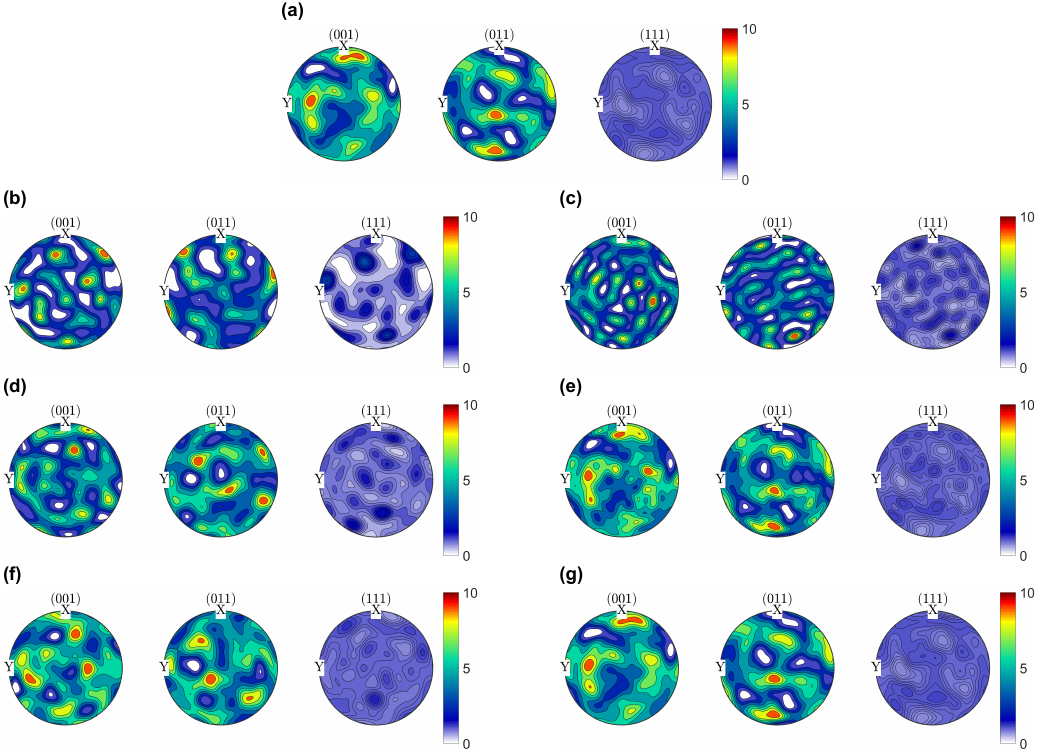}
\caption{Initial textures of the equiaxed RVE represented by the (001), (011), and (111) pole figures using an equal-area projection and a fixed intensity scale of 0--10 multiples of a random distribution (m.r.d.). Panel (a) shows the volume-weighted 1000-grain reference, and the remaining panels show the trained networks with $N=4$ (b, c), $N=6$ (d, e), and $N=8$ (f, g). The left column (b, d, f) corresponds to the conventional ODMN, and the right column (c, e, g) to the GSH-ODMN.}
\label{fig:PF_initial_random}
\end{figure*}

Figs.~\ref{fig:PF_initial_rolled} and \ref{fig:PF_initial_random} show the pole figures of the initial textures. For the rolled RVE, the GSH-ODMN captures the locations of the dominant peaks at $N=4$ and their relative intensities at $N=6$, and it closely reproduces the pole-figure patterns at $N=8$. The conventional ODMN shows spurious maxima at all depths. For the nearly random equiaxed RVE, the conventional ODMN concentrates the orientations into a few sharp components at all depths, whereas the GSH-ODMN gradually recovers the diffuse character of the reference texture and closely reproduces it at $N=8$.

\subsection{Online prediction: homogenized stress response}
\label{sec34}

\begin{figure*}[tp]
\centering
\includegraphics[width=\textwidth]{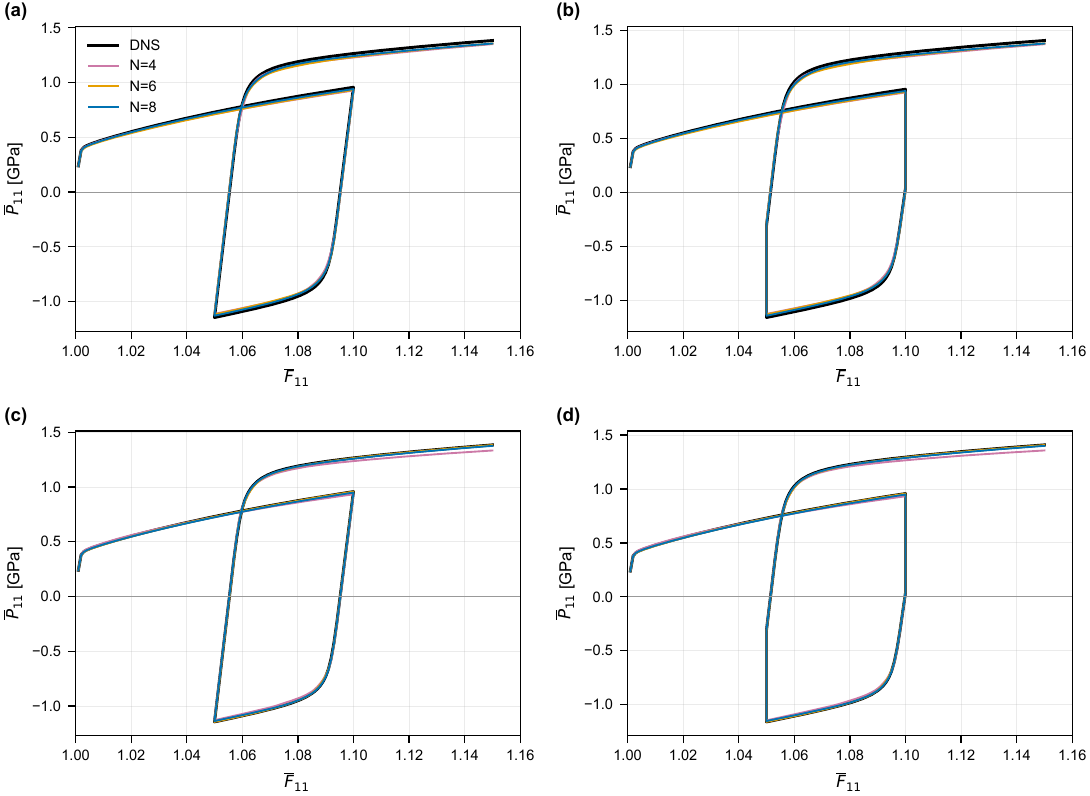}
\caption{Axial stress $\overline P_{11}$ versus $\overline F_{11}$ along the LUR loading path for the rolled RVE, comparing DNS with ODMN predictions at $N=4$, 6, and 8. Panels (a, b) show the conventional ODMN and (c, d) the GSH-ODMN. Panels (a, c) correspond to the isothermal protocol at 293.15\,K and (b, d) to the temperature-cycle protocol.}
\label{fig:lur_rolled}
\end{figure*}

\begin{figure*}[tp]
\centering
\includegraphics[width=\textwidth]{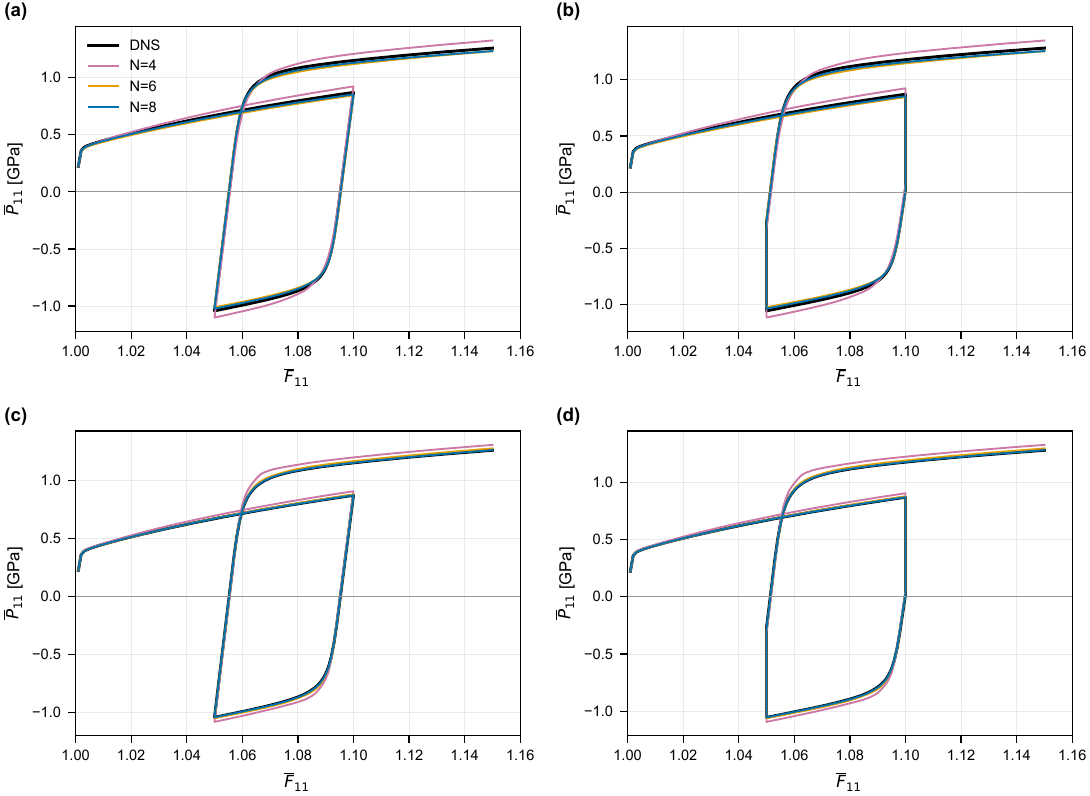}
\caption{Axial stress $\overline P_{11}$ versus $\overline F_{11}$ along the LUR loading path for the equiaxed RVE, comparing DNS with ODMN predictions at $N=4$, 6, and 8. The panel arrangement and loading protocols are the same as in Fig.~\ref{fig:lur_rolled}.}
\label{fig:lur_random}
\end{figure*}

\begin{figure*}[tp]
\centering
\includegraphics[width=\textwidth]{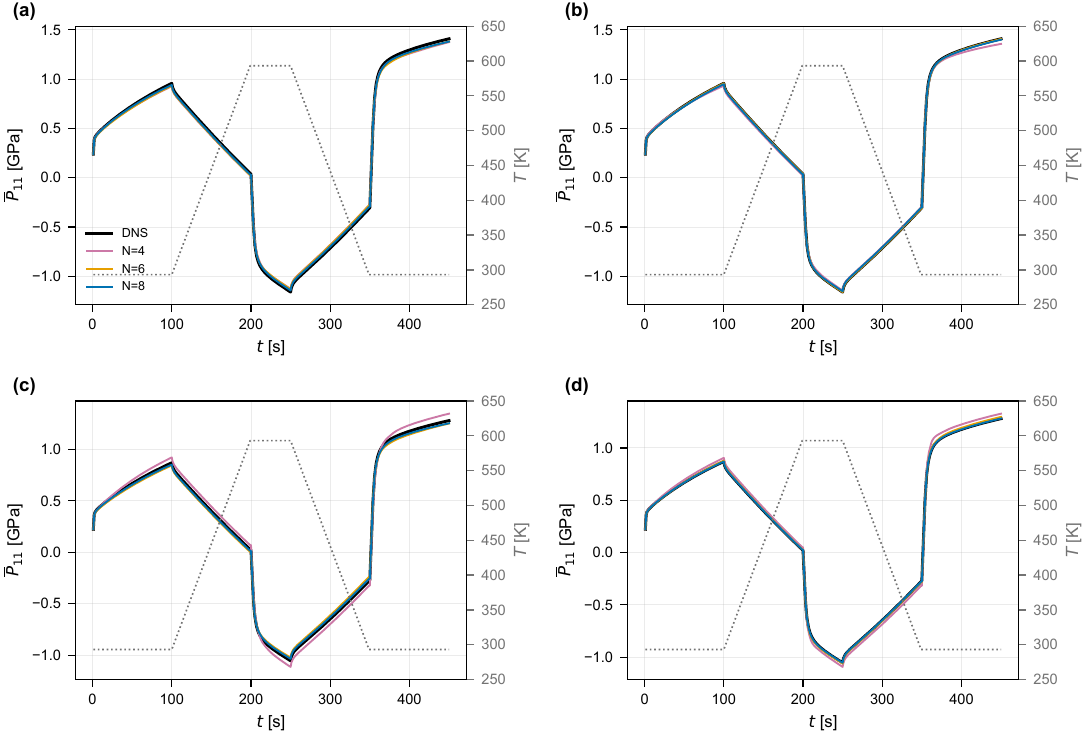}
\caption{Axial stress $\overline P_{11}(t)$ predicted by DNS and by the ODMN with $N=4$, 6, and 8 during the prescribed temperature cycle (dotted line, right axis): (a, c) conventional ODMN and (b, d) GSH-ODMN, with (a, b) corresponding to the rolled RVE and (c, d) to the equiaxed RVE.}
\label{fig:lur_tcycle}
\end{figure*}

Figs.~\ref{fig:lur_rolled} and \ref{fig:lur_random} compare the predicted axial stress $\overline P_{11}$ with the DNS results along the LUR path for both loading protocols. Fig.~\ref{fig:lur_tcycle} shows the response under the temperature-cycle protocol as a function of time, together with the prescribed temperature history. Table~\ref{tab:online_errors} lists the mean and maximum errors of Eq.~\eqref{eq:stress_error} for all 24 combinations of RVE, training strategy, network depth, and loading protocol.

\begin{table*}[tp]
\caption{Mean and maximum of the relative stress error $e(t)$ over the loading history for the isothermal protocol at 293.15\,K and for the temperature-cycle protocol.}
\label{tab:online_errors}
\begin{tabular}{llcrrrr}
\toprule
& & & \multicolumn{2}{c}{293.15\,K} & \multicolumn{2}{c}{temperature cycle} \\
\cmidrule(lr){4-5}\cmidrule(lr){6-7}
RVE & strategy & $N$ & mean [\%] & max [\%] & mean [\%] & max [\%] \\
\midrule
rolled & conventional & 4 & 1.854 & 3.019 & 1.929 & 3.409 \\
rolled & conventional & 6 & 1.826 & 2.684 & 1.809 & 2.707 \\
rolled & conventional & 8 & 1.160 & 1.979 & 1.105 & 1.974 \\
rolled & GSH & 4 & 1.430 & 3.691 & 1.267 & 3.653 \\
rolled & GSH & 6 & 0.185 & 1.192 & 0.115 & 1.236 \\
rolled & GSH & 8 & 0.449 & 0.777 & 0.472 & 0.610 \\
\midrule
equiaxed & conventional & 4 & 3.327 & 5.304 & 3.512 & 5.268 \\
equiaxed & conventional & 6 & 2.012 & 3.216 & 2.135 & 3.118 \\
equiaxed & conventional & 8 & 1.282 & 2.041 & 1.278 & 2.044 \\
equiaxed & GSH & 4 & 2.916 & 5.894 & 2.769 & 5.901 \\
equiaxed & GSH & 6 & 0.683 & 1.419 & 0.565 & 1.280 \\
equiaxed & GSH & 8 & 0.060 & 0.335 & 0.059 & 0.644 \\
\bottomrule
\end{tabular}
\end{table*}

For the isothermal protocol, all networks reproduce the main qualitative features of the response, including the elastic--plastic transition during loading, reverse yielding at about $-1.1$\,GPa during partial unloading, and reloading to 1.3--1.4\,GPa at $\overline F_{11}=1.15$. In the temperature-cycle protocol, heating at fixed deformation reduces the axial stress from about 0.9\,GPa to nearly zero. This relaxation is primarily due to thermal expansion, which nearly offsets the axial elastic strain. The thermal strain is $\alpha_{11}\Delta T=0.38\,\%$, and the elastic strain at 0.9\,GPa is about 0.4\,\% for a polycrystal modulus of 220--240\,GPa. The temperature-induced softening of the elastic constants further reduces the stress. Subsequent cooling at fixed deformation raises the stress again by about 0.8\,GPa. The GSH-ODMN captures these thermomechanical transitions as accurately as the isothermal response. The mean errors of the two protocols differ by less than 0.2 percentage points for every network.

The GSH-ODMN yields a lower mean error than the conventional ODMN in all twelve comparisons in Table~\ref{tab:online_errors}. At $N=4$, the improvement is modest, and the mean errors of both strategies stay above 1\,\%. The 16 prescribed orientations of the GSH-ODMN also lead to slightly larger maximum errors near the elastic--plastic transition. The advantage is much larger at $N=6$ and 8. For the rolled RVE at $N=6$, for example, the mean error decreases from 1.826\,\% to 0.185\,\% under isothermal loading and from 1.809\,\% to 0.115\,\% under the temperature cycle. For the equiaxed RVE at $N=8$, it decreases from 1.282\,\% to 0.060\,\% and from 1.278\,\% to 0.059\,\%, respectively. These GSH-ODMN values are single realizations. Their dependence on the leaf assignment is discussed below. The conventional ODMN improves only gradually with depth, and its mean errors stay above 1\,\% at $N=8$ for both RVEs and both protocols. This is consistent with its persistent error in the initial texture. Even when the elastic response is reproduced accurately, an incorrect texture limits the accuracy of the anisotropic plastic response.

The networks in Table~\ref{tab:online_errors} use the leaf assignment returned by the compaction (Section~\ref{sec22}). Because the orientations and weights of the material nodes are fixed, this assignment is the only arbitrary choice in the construction of the GSH-ODMN. It determines which orientations interact through which building blocks, and the trained interaction directions adapt to it. To quantify its effect, each GSH-ODMN is retrained with five random permutations of the leaf assignment, with the compact texture and all training settings unchanged. Table~\ref{tab:leaf_assignment} summarizes the six networks for each RVE and depth.

The elastic validation error is insensitive to the assignment and decreases steadily with depth, from about 0.9\,\% (rolled) and 0.45\,\% (equiaxed) at $N=4$ to 0.03\,\% at $N=6$ and 0.008\,\% at $N=8$. The stress error under plastic loading is more sensitive. For the rolled RVE, the mean error averaged over the six assignments is 0.46\,\% at $N=6$ and 0.51\,\% at $N=8$. The increase from 0.185\,\% to 0.449\,\% in Table~\ref{tab:online_errors} is therefore within the scatter caused by the assignment: the tabulated $N=6$ network is the most accurate of its six realizations, and the $N=8$ network lies near the median. The mean stress error thus approaches about 0.5\,\% at $N=8$ for both RVEs, and the rolled RVE reaches this level already at $N=6$. The maximum error over the loading history, in contrast, continues to decrease with depth. Averaged over the assignments, it decreases from 2.46\,\% and 4.56\,\% at $N=4$ to 1.34\,\% and 1.81\,\% at $N=6$ and 0.86\,\% at $N=8$ for the rolled and equiaxed RVEs, respectively. On average over the assignments, the GSH-ODMN is more accurate than the conventional ODMN at $N=4$, 6, and 8.

\begin{table*}[tp]
\caption{Sensitivity of the GSH-ODMN to the assignment of the compact orientations to the material nodes: best validation error and mean and maximum stress error $e(t)$ under the isothermal protocol at 293.15\,K over six leaf assignments. Mean $\pm$ standard deviation of the mean error with the range in parentheses; for the validation and maximum errors, the mean with the range.}
\label{tab:leaf_assignment}
\begin{tabular}{lcccc}
\toprule
RVE & $N$ & validation [\%] & mean [\%] & max [\%] \\
\midrule
rolled   & 4 & 0.877 (0.801--0.912) & 0.98 $\pm$ 0.52 (0.30--1.56) & 2.46 (1.20--3.69) \\
rolled   & 6 & 0.029 (0.025--0.034) & 0.46 $\pm$ 0.30 (0.18--1.01) & 1.34 (0.79--1.94) \\
rolled   & 8 & 0.007 (0.007--0.008) & 0.51 $\pm$ 0.19 (0.24--0.81) & 0.86 (0.48--1.41) \\
\midrule
equiaxed & 4 & 0.449 (0.410--0.542) & 2.64 $\pm$ 0.85 (1.62--3.95) & 4.56 (2.64--6.07) \\
equiaxed & 6 & 0.028 (0.023--0.033) & 0.96 $\pm$ 0.36 (0.58--1.51) & 1.81 (1.28--2.83) \\
equiaxed & 8 & 0.008 (0.007--0.008) & 0.44 $\pm$ 0.28 (0.06--0.82) & 0.86 (0.33--1.38) \\
\bottomrule
\end{tabular}
\end{table*}

\subsection{Online prediction: texture evolution}
\label{sec35}

\begin{figure*}[tp]
\centering
\includegraphics[width=0.92\textwidth]{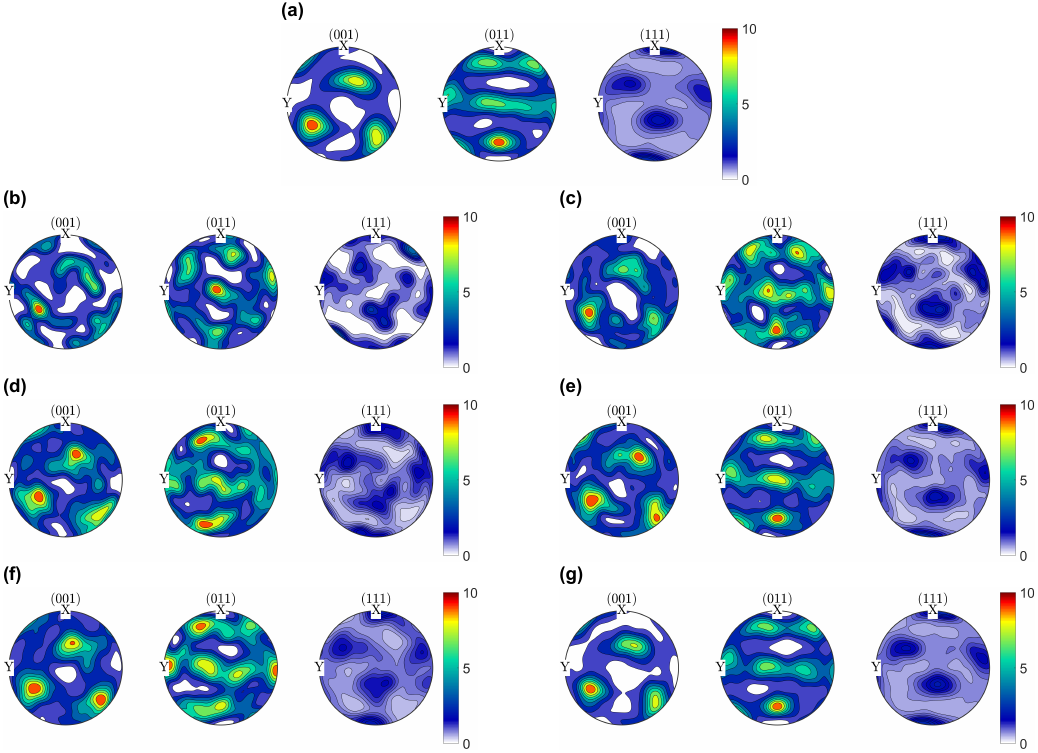}
\caption{Textures of the rolled RVE at the end of the isothermal LUR path ($t=250$\,s, $\overline F_{11}=1.15$). Panel (a) shows the DNS reference, and the remaining panels show the ODMN predictions with $N=4$ (b, c), $N=6$ (d, e), and $N=8$ (f, g). The left column (b, d, f) corresponds to the conventional ODMN, and the right column (c, e, g) to the GSH-ODMN.}
\label{fig:PF_final_rolled}
\end{figure*}

\begin{figure*}[tp]
\centering
\includegraphics[width=0.92\textwidth]{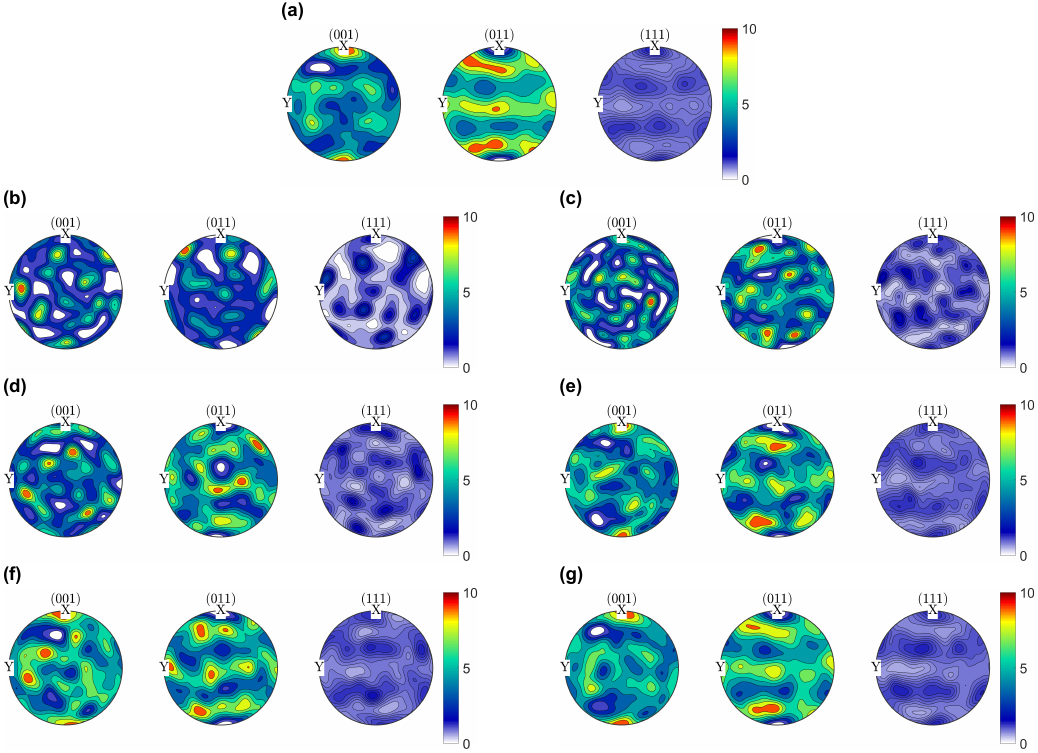}
\caption{Textures of the equiaxed RVE at the end of the isothermal LUR path ($t=250$\,s, $\overline F_{11}=1.15$). Panel (a) shows the DNS reference, and the remaining panels show the ODMN predictions with $N=4$ (b, c), $N=6$ (d, e), and $N=8$ (f, g). The left column (b, d, f) corresponds to the conventional ODMN, and the right column (c, e, g) to the GSH-ODMN.}
\label{fig:PF_final_random}
\end{figure*}

Table~\ref{tab:Td_evolution} lists the texture-difference index between the ODMN and DNS textures at four states along the isothermal loading path ($t=0$, 100, 150, and 250\,s, corresponding to $\overline F_{11}=1.00$, 1.10, 1.05, and 1.15) and at the end of the temperature-cycle protocol. Figs.~\ref{fig:PF_final_rolled} and \ref{fig:PF_final_random} compare the corresponding pole figures at the end of the isothermal path.

\begin{table*}[tp]
\caption{Texture-difference index $\hat{T}^{d}$ between the ODMN and DNS textures along the loading path: isothermal protocol at $t=0$, 100, 150, and 250\,s ($\overline F_{11}=1.00$, 1.10, 1.05, and 1.15) and end of the temperature-cycle protocol at $t=450$\,s.}
\label{tab:Td_evolution}
\begin{tabular}{llcrrrrr}
\toprule
& & & \multicolumn{4}{c}{293.15\,K} & temperature cycle \\
\cmidrule(lr){4-7}\cmidrule(lr){8-8}
RVE & strategy & $N$ & $t=0$ & $t=100$\,s & $t=150$\,s & $t=250$\,s & $t=450$\,s \\
\midrule
rolled & conventional & 4 & 1.0564 & 1.0176 & 1.0566 & 0.9785 & 0.9793 \\
rolled & conventional & 6 & 0.2551 & 0.2263 & 0.2387 & 0.2170 & 0.2174 \\
rolled & conventional & 8 & 0.0964 & 0.0932 & 0.0954 & 0.0892 & 0.0893 \\
rolled & GSH & 4 & 0.4623 & 0.4653 & 0.4648 & 0.4858 & 0.4856 \\
rolled & GSH & 6 & 0.0496 & 0.0574 & 0.0497 & 0.0701 & 0.0697 \\
rolled & GSH & 8 & 0.0075 & 0.0135 & 0.0103 & 0.0185 & 0.0184 \\
\midrule
equiaxed & conventional & 4 & 1.3525 & 1.3247 & 1.3587 & 1.2774 & 1.2797 \\
equiaxed & conventional & 6 & 0.2536 & 0.2661 & 0.2639 & 0.2677 & 0.2676 \\
equiaxed & conventional & 8 & 0.0785 & 0.0766 & 0.0754 & 0.0793 & 0.0793 \\
equiaxed & GSH & 4 & 0.6549 & 0.6556 & 0.6542 & 0.6674 & 0.6680 \\
equiaxed & GSH & 6 & 0.0761 & 0.0902 & 0.0819 & 0.1039 & 0.1037 \\
equiaxed & GSH & 8 & 0.0099 & 0.0194 & 0.0138 & 0.0266 & 0.0264 \\
\bottomrule
\end{tabular}
\end{table*}

The DNS textures evolve moderately along the loading path. The rolling texture sharpens and keeps its principal components, whereas the equiaxed RVE develops a concentration of (001) poles along the tensile axis, that is, a $\langle100\rangle$ fiber component (Fig.~\ref{fig:PF_final_random}(a)). The GSH-ODMN closely reproduces this evolution. At $N=8$, the texture-difference index $\hat{T}^{d}$ between the GSH-ODMN and the DNS remains below 0.019 for the rolled RVE and below 0.027 for the equiaxed RVE throughout the isothermal loading path. At the end of the temperature-cycle protocol, the values are nearly identical to those of the isothermal protocol (0.0184 versus 0.0185 for the rolled RVE and 0.0264 versus 0.0266 for the equiaxed RVE). The GSH-ODMN thus predicts the texture equally well for both loading protocols. The conventional ODMN shows a much larger texture difference, which is dominated by the mismatch in the initial texture.

\subsection{Computational cost}
\label{sec36}

\begin{figure*}[tp]
\centering
\includegraphics[width=\textwidth]{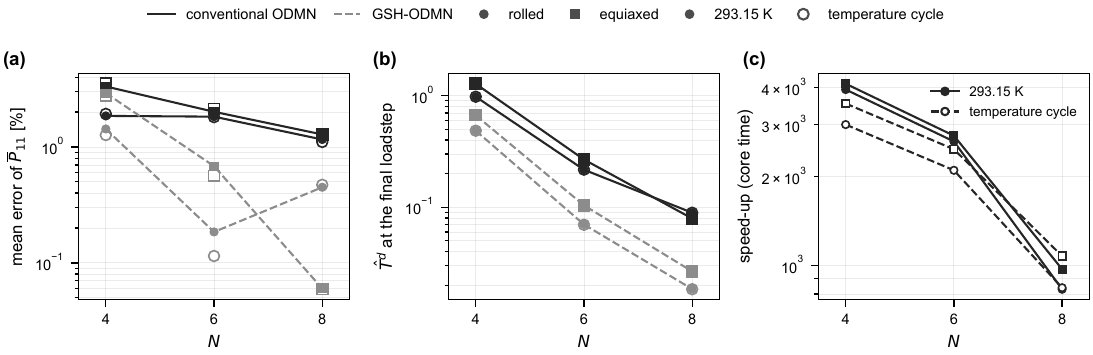}
\caption{Online prediction performance as a function of network depth $N$: (a) mean relative error of $\overline P_{11}$, (b) texture-difference index $\hat{T}^{d}$ at the final load step, and (c) core-time speed-up of the GSH-ODMN relative to the DNS. In (a) and (b), line styles distinguish the training strategies and marker shapes distinguish the RVEs. Filled markers connected by lines denote the isothermal protocol at 293.15\,K, whereas open markers without connecting lines denote the temperature-cycle protocol. In (c), solid and dashed lines denote the isothermal and temperature-cycle protocols, respectively.}
\label{fig:online_summary}
\end{figure*}

Table~\ref{tab:cost} compares the computational costs of the ODMN and the DNS, and Fig.~\ref{fig:online_summary} summarizes the accuracy and efficiency of the online predictions as functions of the network depth. The ODMN cost is the solver wall-clock time reported by Abaqus on a single CPU core, excluding job initialization and user-subroutine compilation. The corresponding DNS wall-clock times on eight CPU threads are 32~min 51~s and 34~min 22~s for the rolled and equiaxed RVEs, respectively, under the isothermal protocol, and 43~min 43~s and 51~min 31~s under the temperature-cycle protocol. Because the two solvers use different numbers of cores, the speed-up is evaluated in terms of core time, that is, wall-clock time multiplied by the number of threads.

\begin{table*}[tp]
\caption{Computational cost of the online predictions in terms of the ODMN solver wall-clock time on one CPU core and the core-time speed-up relative to the DNS on eight CPU threads.}

\label{tab:cost}
\begin{tabular}{llcrrrr}
\toprule
& & & \multicolumn{2}{c}{293.15\,K} & \multicolumn{2}{c}{temperature cycle} \\
\cmidrule(lr){4-5}\cmidrule(lr){6-7}
RVE & strategy & $N$ & ODMN & speed-up & ODMN & speed-up \\
\midrule
rolled & conventional & 4 & 4\,s & 3941 & 7\,s & 2998 \\
rolled & conventional & 6 & 6\,s & 2628 & 10\,s & 2099 \\
rolled & conventional & 8 & 19\,s & 830 & 26\,s & 807 \\
rolled & GSH & 4 & 4\,s & 3941 & 7\,s & 2998 \\
rolled & GSH & 6 & 6\,s & 2628 & 10\,s & 2099 \\
rolled & GSH & 8 & 19\,s & 830 & 25\,s & 839 \\
\midrule
equiaxed & conventional & 4 & 5\,s & 3299 & 7\,s & 3533 \\
equiaxed & conventional & 6 & 6\,s & 2749 & 10\,s & 2473 \\
equiaxed & conventional & 8 & 18\,s & 916 & 26\,s & 951 \\
equiaxed & GSH & 4 & 4\,s & 4123 & 7\,s & 3533 \\
equiaxed & GSH & 6 & 6\,s & 2749 & 10\,s & 2473 \\
equiaxed & GSH & 8 & 17\,s & 970 & 23\,s & 1075 \\
\bottomrule
\end{tabular}
\end{table*}

The computational cost of the ODMN increases with the network depth. Over both loading protocols and both RVEs, the core-time speed-up relative to the DNS decreases from about 3000--4100 at $N=4$ to 2100--2750 at $N=6$ and 810--1080 at $N=8$ (Fig.~\ref{fig:online_summary}(c)). The ODMN solver time increases from 4--7\,s at $N=4$ to 6--10\,s at $N=6$ and 17--26\,s at $N=8$, which is less than a factor of two per additional level. This is broadly consistent with the near-linear scaling with the number of material nodes estimated in Section~\ref{sec252}, bearing in mind that the $N=4$ timings are dominated by the fixed per-increment overhead of the finite-element driver. The two strategies have comparable online costs at a given depth because they use the same network architecture. The small timing differences are mainly due to different numbers of Newton iterations. The temperature-cycle protocol is more expensive than the isothermal protocol because of its additional loading and hold stages, but the dependence of the cost on the network depth is the same.

The network depth is thus a trade-off between accuracy and cost. Increasing $N$ generally improves the predictions of the homogenized stress and of the texture evolution but reduces the speed-up. At $N=4$, the speed-up is several thousand, but the accuracy is lowest, especially for the texture. At $N=6$, the stress and texture predictions are much better and the speed-up is still 2100--2750, which makes this depth a good compromise. At $N=8$, the texture is predicted most accurately, and the speed-up drops to 810--1080. The depth $N=8$ is therefore appropriate when accurate texture predictions are required, and $N=6$ is the more economical choice for large multiscale analyses.

\section{Two-scale thermomechanical simulation}
\label{sec4}

\subsection{Model}
\label{sec41}

\begin{figure*}[tp]
\centering
\includegraphics[width=\textwidth]{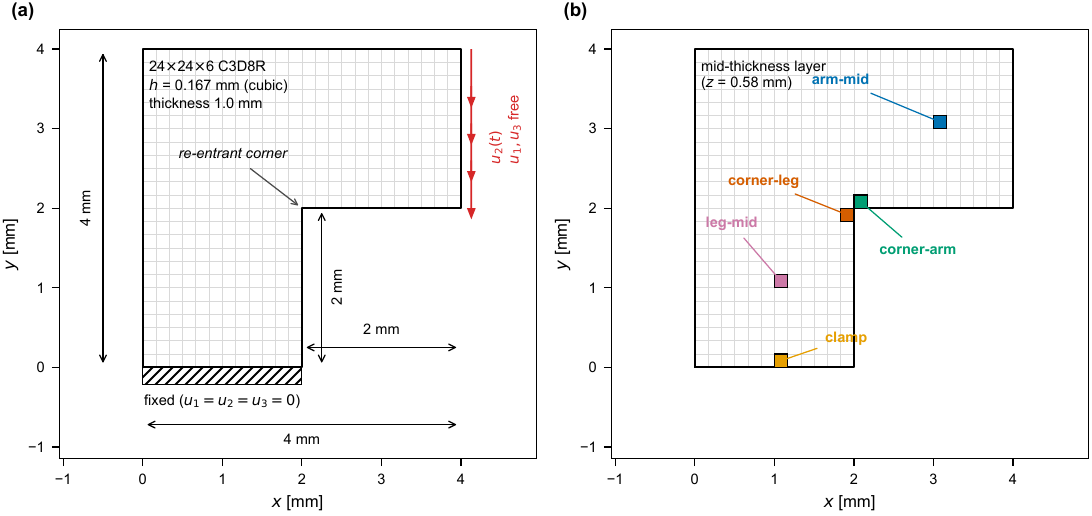}
\caption{Two-scale L-bracket: (a) in-plane geometry, mesh, and boundary conditions and (b) monitored elements in the mid-thickness layer.}
\label{fig:lbracket_setup}
\end{figure*}

\begin{figure}[tp]
\centering
\includegraphics[width=0.7\textwidth]{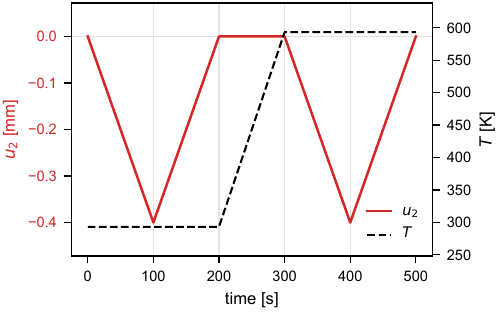}
\caption{Prescribed displacement $u_2$ and temperature $T$ of the L-bracket during cold loading and unloading, heating at fixed displacement, and hot loading and unloading.}
\label{fig:lbracket_loading}
\end{figure}

The structural example is an L-shaped IN625 bracket subjected to displacement-controlled bending at two temperatures (Fig.~\ref{fig:lbracket_setup}). The bracket has a $4\,\mathrm{mm}\times4\,\mathrm{mm}\times1\,\mathrm{mm}$ envelope with a $2\,\mathrm{mm}\times2\,\mathrm{mm}\times1\,\mathrm{mm}$ block removed. It is discretized with cubic elements of 0.167\,mm edge length on a $24\times24\times6$ grid, which gives 2592 reduced-integration hexahedral elements with enhanced hourglass control, 3367 nodes, and 10101 degrees of freedom. The bottom face is fully clamped. On the end face of the horizontal arm, only the vertical displacement $u_2$ is prescribed, and the displacements $u_1$ and $u_3$ are free. The loading history comprises five 100\,s stages (Fig.~\ref{fig:lbracket_loading}): loading to $u_2=-0.4$\,mm at 293.15\,K, unloading to $u_2=0$, heating to 593.15\,K at fixed displacement, reloading to $u_2=-0.4$\,mm, and unloading at 593.15\,K. The temperature is spatially uniform, and geometric nonlinearity is included.

At each integration point, the material response is supplied through the \texttt{UMAT} by the GSH-ODMN trained for the rolled RVE with $N=4$, 6, or 8, using the constitutive parameters listed in Table~\ref{tab:material}. The time increment starts at 0.5\,s and is capped at 2\,s. Each analysis takes 410 increments and 460--467 global equilibrium iterations, with one ODMN localization solve at every integration point in each iteration. The simulations are run with four Message Passing Interface (MPI) processes on a desktop workstation (Intel Core i9-13900K). Five mid-thickness elements are monitored, as shown in Fig.~\ref{fig:lbracket_setup}(b): corner-leg, corner-arm, leg-mid, arm-mid, and clamp.

\subsection{Structural response}
\label{sec42}

\begin{figure*}[tp]
\centering
\includegraphics[width=\textwidth]{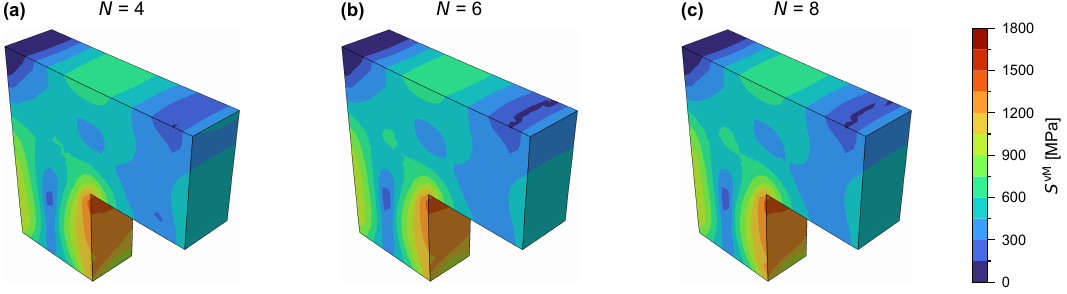}
\caption{Distribution of the equivalent stress $S^{\mathrm{vM}}$, defined as the von Mises equivalent of the macroscopic second Piola--Kirchhoff stress, in the L-bracket at the end of the analysis ($t=500$\,s): (a) $N=4$, (b) $N=6$, and (c) $N=8$.}
\label{fig:lbracket_field}
\end{figure*}

\begin{figure*}[tp]
\centering
\includegraphics[width=\textwidth]{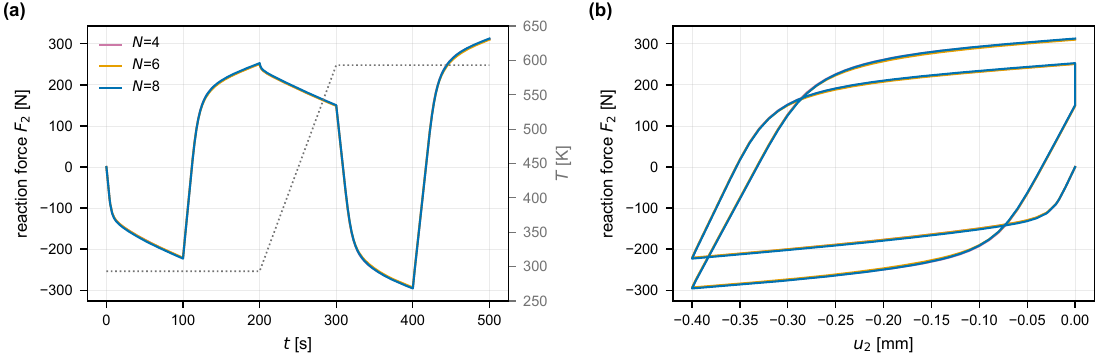}
\caption{Reaction force $F_2$ on the loaded face of the L-bracket for $N=4$, 6, and 8: (a) time history together with the prescribed temperature (dotted line, right axis), and (b) force--displacement response during the cold and hot loading cycles.}
\label{fig:lbracket_rf}
\end{figure*}

Fig.~\ref{fig:lbracket_field} shows the von Mises equivalent of the macroscopic second Piola--Kirchhoff stress $\overline{\bm S}$, $S^{\mathrm{vM}}=\sqrt{\tfrac{3}{2}\operatorname{dev}\overline{\bm S}:\operatorname{dev}\overline{\bm S}}$, where $\operatorname{dev}$ denotes the deviatoric part, at the end of the analysis for the three network depths. Fig.~\ref{fig:lbracket_rf} shows the reaction force $F_2$ on the loaded face as a function of time and of the prescribed displacement, and Table~\ref{tab:lbracket_response} lists characteristic values of the structural response. The stress concentrates at the re-entrant corner, where $S^{\mathrm{vM}}$ reaches about 1.5--1.6\,GPa at the end of the analysis and 1.6--1.7\,GPa at the hot-loading peak (Table~\ref{tab:lbracket_monitored}), compared with about 0.3--0.5\,GPa over most of the leg and arm.

\begin{table}[tp]
\caption{Characteristic L-bracket responses for $N=4$, 6, and 8: reaction forces at the ends of the loading stages and net work over the cold and hot displacement cycles.}
\label{tab:lbracket_response}
\setlength{\tabcolsep}{3pt}
\begin{tabular}{@{}lrrr@{}}
\toprule
quantity & $N=4$ & $N=6$ & $N=8$ \\
\midrule
cold peak $F_2$ [N]              & $-221.8$ & $-221.1$ & $-223.1$ \\
cold residual $F_2$ [N]          & 251.5    & 251.0    & 253.0    \\
$F_2$ after heating [N]           & 149.6    & 148.9    & 150.5    \\
hot peak $F_2$ [N]               & $-294.7$ & $-293.5$ & $-295.7$ \\
hot residual $F_2$ [N]           & 311.8    & 310.3    & 312.9    \\
cold-cycle work [mJ]             & 133.1    & 132.6    & 133.9    \\
hot-cycle work [mJ]              & 154.7    & 153.3    & 154.5    \\
\midrule
range of diff.\ from $N=8$ [\%]  & 0.1--0.6 & 0.7--1.1 & -- \\
\bottomrule
\end{tabular}
\end{table}

The cold and hot cycles form open hysteresis loops (Fig.~\ref{fig:lbracket_rf}, Table~\ref{tab:lbracket_response}). For $N=8$, the reaction force at $u_2=-0.4$\,mm increases in magnitude from $-223$\,N in the cold cycle to $-296$\,N in the hot cycle because the second cycle begins from the hardened, residually stressed state left by the first cycle. Heating at fixed displacement reduces the residual force from 253\,N to 151\,N through thermal expansion and the temperature dependence of the elastic response. The corresponding net work is 134\,mJ for the cold cycle and 155\,mJ for the hot cycle. The shallower networks give nearly the same structural response. Relative to $N=8$, the characteristic forces and the cycle work differ by 0.1--0.6\,\% for $N=4$ and by 0.7--1.1\,\% for $N=6$. The structural response is thus insensitive to the network depth, in line with the RVE-level results.

\subsection{Local response}
\label{sec43}

\begin{figure*}[tp]
\centering
\includegraphics[width=\textwidth]{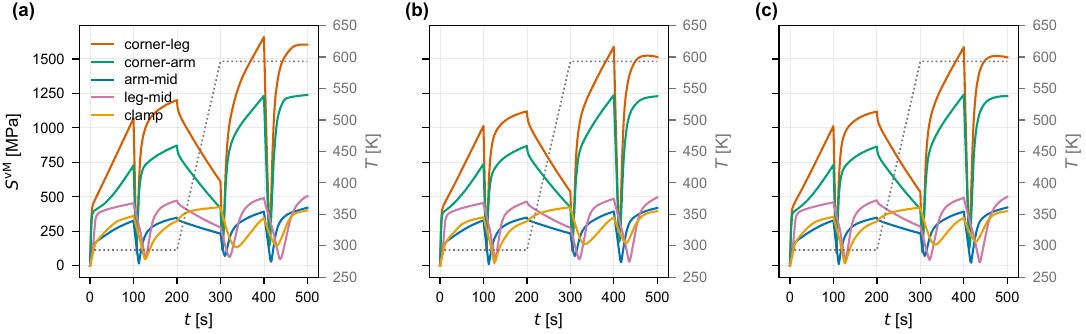}
\caption{Equivalent stress $S^{\mathrm{vM}}$ histories at the five monitored elements of the L-bracket for (a) $N=4$, (b) $N=6$, and (c) $N=8$. The prescribed temperature history is shown by the dotted line on the right axis.}
\label{fig:lbracket_svm}
\end{figure*}

Fig.~\ref{fig:lbracket_svm} compares the equivalent stress histories at the five monitored elements for the three network depths, and Table~\ref{tab:lbracket_monitored} lists the values at the end of each stage. Away from the re-entrant corner, the shallower networks agree with $N=8$ to within 1\,\% at the arm-mid element, 2\,\% at the clamp element, and 3.5\,\% at the corner-arm and leg-mid elements. Larger differences occur for $N=4$ at the corner-leg element, where the stress concentration is strongest. There, $N=4$ deviates from $N=8$ by 4.6--7.4\,\% at the ends of the loading and unloading stages and by 14\,\% after heating (611 versus 535\,MPa). The $N=6$ network agrees with $N=8$ to within 0.4\,\% at this element, so that the stress at the most highly loaded element is converged with respect to the network depth at $N=6$.

The corner-leg history also shows a local feature that the global reaction force does not reveal. After cold unloading, the equivalent stress is 1201, 1120, and 1118\,MPa for $N=4$, 6, and 8, respectively, which exceeds the values of 1068, 1013, and 1013\,MPa at the preceding loading peak. The increase is attributed to stress redistribution and reverse plastic flow near the re-entrant corner during springback, in material that has already hardened. During the hot cycle, the corner-leg stress increases further. The peak equivalent stress over all elements and output times is 1664\,MPa for $N=4$ and 1588\,MPa for both $N=6$ and $N=8$, so that $N=4$ exceeds $N=8$ by 4.8\,\%. The global response is thus nearly insensitive to the network depth, whereas the concentrated stresses near the re-entrant corner are more sensitive. At the corner-leg element, $N=4$ predicts systematically higher stresses than $N=8$, and $N=6$ and $N=8$ nearly coincide.

\begin{table*}[tp]
\centering
\caption{Equivalent stress $S^{\mathrm{vM}}$ [MPa] at the five monitored elements of the L-bracket at the end of each loading stage for $N=4$, 6, and 8.}
\label{tab:lbracket_monitored}
\begin{tabular}{lcrrrrr}
\toprule
element & $N$ & $t=100$\,s & $t=200$\,s & $t=300$\,s & $t=400$\,s & $t=500$\,s \\
\midrule
corner-leg & 4 & 1068 & 1201 & 611 & 1658 & 1603 \\
 & 6 & 1013 & 1120 & 537 & 1587 & 1514 \\
 & 8 & 1013 & 1118 & 535 & 1585 & 1512 \\
\addlinespace
corner-arm & 4 & 727 & 872 & 419 & 1233 & 1238 \\
 & 6 & 736 & 869 & 424 & 1238 & 1229 \\
 & 8 & 731 & 864 & 410 & 1241 & 1230 \\
\addlinespace
arm-mid & 4 & 326 & 349 & 234 & 392 & 421 \\
 & 6 & 329 & 350 & 234 & 393 & 420 \\
 & 8 & 328 & 349 & 235 & 393 & 423 \\
\addlinespace
leg-mid & 4 & 453 & 473 & 278 & 491 & 504 \\
 & 6 & 455 & 465 & 275 & 485 & 495 \\
 & 8 & 460 & 475 & 284 & 492 & 500 \\
\addlinespace
clamp & 4 & 361 & 324 & 425 & 345 & 401 \\
 & 6 & 351 & 320 & 424 & 351 & 396 \\
 & 8 & 354 & 320 & 422 & 347 & 396 \\
\bottomrule
\end{tabular}
\end{table*}

\subsection{Texture evolution at the re-entrant corner}
\label{sec44}

\begin{figure*}[tp]
\centering
\includegraphics[width=\textwidth]{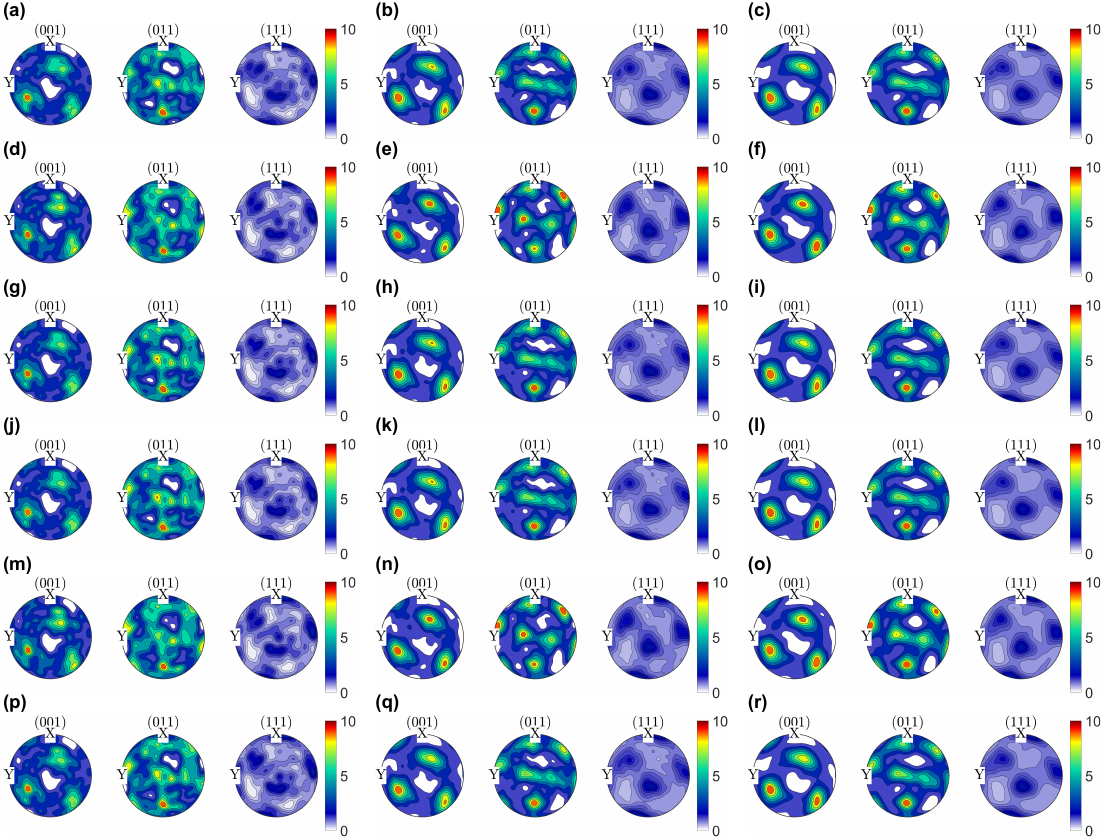}
\caption{Pole figures of the local texture at the corner-leg element of the L-bracket in the co-rotated frame of the element, that is, with the rigid rotation $\overline{\bm R}$ removed, for $N=4$ (left column), $N=6$ (middle column), and $N=8$ (right column) in the initial state at $t=0$ (a--c) and at the end of each stage of the loading history: after cold loading at $t=100$\,s (d--f), after cold unloading at $t=200$\,s (g--i), after heating to 593.15\,K at $t=300$\,s (j--l), after hot loading at $t=400$\,s (m--o), and after hot unloading at $t=500$\,s (p--r).}
\label{fig:lbracket_PF_corner}
\end{figure*}

\begin{figure*}[tp]
\centering
\includegraphics[width=0.75\textwidth]{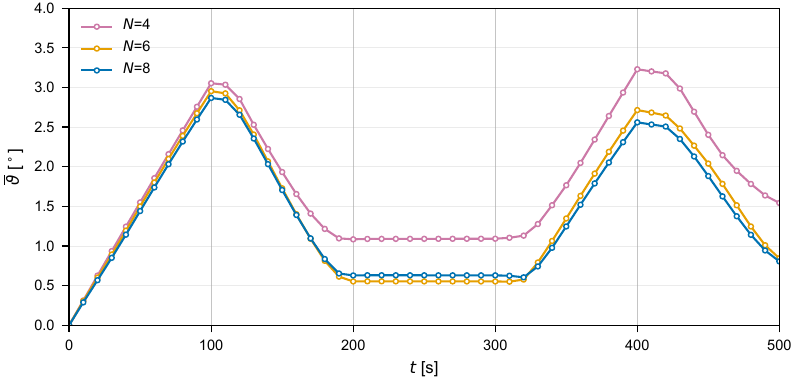}
\caption{Volume-averaged lattice rotation $\overline{\vartheta}$ relative to the initial orientations in the co-rotated frame of the corner-leg element for $N=4$, 6, and 8. Vertical lines indicate loading-stage boundaries.}
\label{fig:lbracket_Td_corner}
\end{figure*}

Because each integration point stores the complete ODMN state, the local texture can be evaluated throughout the loading history. The material-node rotations $\bm R_i$ of Eq.~\eqref{eq:elastic_polar_decomposition} are expressed in the global frame and therefore also contain the rigid rotation of the element. To separate it, the macroscopic deformation gradient at the integration point is decomposed as $\overline{\bm F}=\overline{\bm R}\,\overline{\bm U}$, and the material-node rotations are expressed in the co-rotated frame of the element as $\overline{\bm R}^{\mathsf T}\bm R_i$. Fig.~\ref{fig:lbracket_PF_corner} shows the co-rotated pole figures of the corner-leg element in the initial state and at the end of each loading stage for the three network depths. Fig.~\ref{fig:lbracket_Td_corner} shows the volume-averaged lattice rotation $\overline{\vartheta}(t)=\sum_i w_i\,\vartheta_i(t)$, where $\vartheta_i(t)$ is the angle by which the co-rotated orientation of material node $i$ has rotated from its initial state.

During cold loading, the lattice rotates away from its initial orientation, by about $3^\circ$ on average at the end of loading ($t=100$\,s). Most of this rotation is recovered when the prescribed displacement is returned to zero, because the local deformation at the re-entrant corner is then also nearly reversed, and the texture at $t=200$\,s again closely resembles the initial texture. The texture remains unchanged during heating, and the hot cycle repeats the evolution of the cold cycle. The small residual rotation after each cycle is the irreversible texture evolution caused by plastic slip.

All three networks predict the same timing and the same reversible character of the evolution. For $N=4$, 6, and 8, $\overline{\vartheta}$ is $3.05^\circ$, $2.95^\circ$, and $2.87^\circ$ at the end of cold loading ($t=100$\,s), $1.08^\circ$, $0.55^\circ$, and $0.63^\circ$ after cold unloading ($t=200$\,s), and $3.23^\circ$, $2.71^\circ$, and $2.56^\circ$ at the end of hot loading ($t=400$\,s). The networks with $N=6$ and 8 thus nearly coincide, whereas $N=4$ predicts larger rotations, particularly in the hot cycle, and a larger residual rotation. This mirrors the stress at this element, where $N=4$ deviates from $N=8$, whereas $N=6$ and 8 nearly coincide (Section~\ref{sec43}).

\subsection{Computational cost}
\label{sec45}

Table~\ref{tab:lbracket_resources} lists the computational resources of the three analyses, each run with four processes. All three analyses take 410 increments and between 460 and 467 global equilibrium iterations, so that the differences in cost are due to the network depth alone. The $N=4$ analysis completes in 14\,min of wall-clock time with a peak memory of 1.07\,GB. The $N=6$ analysis takes 44\,min with 1.23\,GB, and the $N=8$ analysis takes 2\,h 54\,min with 1.93\,GB. From $N=4$ to 6 and from $N=6$ to 8, the wall-clock time increases by factors of 3.1 and 4.0, respectively, for each fourfold increase in the number of material nodes. This is consistent with the near-linear scaling estimated in Section~\ref{sec252}.

For comparison, consider a concurrent two-scale analysis that solves the 1000-grain rolled RVE with the spectral solver at each of the 2592 integration points. The temperature-cycle DNS of this RVE in Section~\ref{sec36} takes 43\,min 43\,s on eight threads, or about 5.8 core hours, for a 450\,s loading history of comparable length and complexity. Such an analysis would therefore require roughly $2592\times5.8\approx1.5\times10^{4}$ core hours. The three ODMN analyses require 0.9, 2.9, and 11.6 core hours, respectively, when the wall-clock time is multiplied by the number of processes. The GSH-ODMN thus reduces the cost of the two-scale analysis by three to four orders of magnitude. For the same reason, a full-field reference solution of the L-bracket is not computed, and the comparison among the three network depths in Sections~\ref{sec42}--\ref{sec44} is a convergence study with respect to $N$.

\begin{table*}[tp]
\caption{Computational resources of the two-scale L-bracket analyses.}
\label{tab:lbracket_resources}
\begin{tabular}{crrrrr}
\toprule
$N$ & material nodes & increments & iterations & wall-clock time & peak memory \\
\midrule
4 & 16 & 410 & 467 & 14\,min 6\,s & 1.07\,GB \\
6 & 64 & 410 & 464 & 43\,min 34\,s & 1.23\,GB \\
8 & 256 & 410 & 460 & 2\,h 54\,min 28\,s & 1.93\,GB \\
\bottomrule
\end{tabular}
\end{table*}

\section{Conclusions}\label{sec5}

This work introduced the GSH-ODMN, in which the orientations and weights of the material nodes are prescribed by a volume-weighted GSH compaction of the RVE texture and are not learned from elastic data. A Bunge-angle formulation of the orientation-aware mechanism and a thermomechanical extension of the online prediction were also developed. The framework was verified against full-field CPFFT simulations of two IN625 RVEs under isothermal and thermomechanical LUR paths and then applied to a two-scale simulation of an L-bracket subjected to cold and hot bending cycles. The main findings are as follows.

\begin{enumerate}
\item The GSH truncation order must be matched to the number of material nodes. The texture-difference index of the compact texture varies non-monotonically with $L$, with the optimum shifting from $L=10$ or 12 at $N=4$ to $L=16$ or 18 at $N=8$. With the adopted orders, the compact textures reproduce the volume-weighted RVE textures with $\hat{T}^{d}=0.46$ and 0.65 at $N=4$ and with $\hat{T}^{d}<0.01$ at $N=8$.

\item Elastic training data alone do not uniquely determine the texture. The conventional ODMN reaches homogenized-stiffness validation errors of approximately 0.01\,\% at $N=8$, while its texture error remains close to 0.1 because the elastic response is governed predominantly by the low-order harmonic content of the ODF. The GSH-ODMN removes this ambiguity by prescribing the material-node orientations. It reduces the number of trainable parameters by about two-thirds and the training time by roughly one-half to two-thirds. Its offline accuracy approaches that of the conventional ODMN at $N=6$ and slightly surpasses it at $N=8$.

\item In the online predictions, the GSH-ODMN yields a lower mean stress error than the conventional ODMN across all combinations of RVE, network depth, and loading protocol, averaged over six orientation assignments. At $N=8$, the assignment-averaged mean errors are 0.44\,\% and 0.51\,\% for the equiaxed and rolled RVEs, respectively, compared with 1.3\,\% and 1.2\,\% for the conventional ODMN.

\item The GSH-ODMN tracks the texture evolution of the DNS. At $N=8$, the final texture-difference index remains below 0.03 for both RVEs and both loading protocols. This is three to five times smaller than for the conventional ODMN, whose error is dominated by the mismatch in the initial texture.

\item The Newton Jacobian of the online solve inherits the binary-tree sparsity of the network and is factorized by a tree-structured LU decomposition without fill-in. The cost per Newton iteration then scales as $O(2^N)$ for the constitutive updates plus $O(2^N N^2)$ for the network algebra, compared with $O(8^N)$ for a dense formulation. For the depths considered, the online cost grows nearly linearly with the number of material nodes, and the core-time speed-up relative to the DNS is still of the order of $10^3$ at $N=8$.

\item When embedded at every integration point of a 2592-element L-bracket, the $N=4$ GSH-ODMN completes the five-stage thermomechanical loading history in 14\,min using four processes, whereas the $N=8$ network completes it in about 3\,h. Relative to $N=8$, the global structural responses of $N=4$ and $N=6$ agree to within about 1\,\%. The stress at the corner-leg element, where the stress concentration is strongest, converges with the network depth. There, $N=4$ deviates from $N=8$ by up to 14\,\%, whereas $N=6$ agrees with $N=8$ to within 0.4\,\%.
\end{enumerate}

Two limitations remain. First, the framework assumes an isothermal RVE with the temperature prescribed uniformly at all material nodes. Extending the localization to the thermal field would resolve the local temperature and its gradient, so that the heat flux could be predicted and problems with strong thermal gradients or with heating by plastic dissipation could be treated. Second, the temperature dependence of the constitutive model is limited to the elastic constants and the thermal expansion. Temperature-dependent slip resistance and thermally activated flow can be added in the material-node constitutive update without changing the localization framework. Both extensions are needed for thermomechanical manufacturing processes with large plastic deformation and texture evolution, such as deep drawing, machining, and cold rolling.

\backmatter

\bmhead{Acknowledgements}

The authors gratefully acknowledge the computational resources and technical support provided by the National Center for Research on Earthquake Engineering (NCREE), National Institutes of Applied Research (NIAR), Taiwan, as well as by the NTUCE-NCREE Joint Artificial Intelligence Research Center and the National Center for High-performance Computing (NCHC).

\section*{Declarations}

\bmhead{Funding}

This work was supported by the National Science and Technology Council, Taiwan, under Grant 114-2221-E-002-010-MY3.

\bmhead{Conflict of interest}

The authors declare that they have no conflict of interest.

\bmhead{Data availability}

The data supporting the findings of this study are available from the corresponding author upon reasonable request.

\bmhead{Code availability}

The code supporting the findings of this study is available from the corresponding author upon reasonable request.

\bmhead{Author contributions}

\textbf{Ting-Ju Wei}: Conceptualization, Methodology, Software, Validation, Formal analysis, Investigation, Visualization, Writing -- original draft. \textbf{Chuin-Shan Chen}: Conceptualization, Supervision, Resources, Funding acquisition, Writing -- review and editing.

\begin{appendices}
\providecommand{\theHequation}{}\renewcommand{\theHequation}{app\arabic{equation}}

\section{Constitutive derivations}
\label{app:constitutive_derivations}

The constitutive formulation used at the material nodes is that of DAMASK~\cite{roters2019damask}, and the expressions in this appendix follow its implementation. They are not new results of this work. The appendix writes out, in the notation of this paper, the forms that the DAMASK implementation evaluates, so that the material-node update of Section~\ref{sec251} is documented in closed form.

\subsection{Mandel stress driving plastic flow}
\label{app:mandel_stress}

Under the multiplicative decomposition
\begin{equation}
\bm F
=
\bm F^{\mathrm e}
\bm F^{\mathrm i}
\bm F^{\mathrm p},
\label{eq:app_F_decomposition}
\end{equation}
it is convenient to introduce the combined elastic--eigen deformation gradient
\begin{equation}
\bm F^{\star}
=
\bm F^{\mathrm e}\bm F^{\mathrm i}
=
\bm F
\left(\bm F^{\mathrm p}\right)^{-1},
\label{eq:app_Fstar}
\end{equation}
such that
$\bm F=\bm F^{\star}\bm F^{\mathrm p}$.
The elastic Green--Lagrange strain introduced in
Eq.~\eqref{eq:cp_hooke} can then be expressed as
\begin{equation}
\bm E^{\mathrm e}
=
\frac{1}{2}
\left[
\left(\bm F^{\star}\right)^{\mathsf T}
\bm F^{\star}
-
\left(\bm F^{\mathrm i}\right)^{\mathsf T}
\bm F^{\mathrm i}
\right].
\label{eq:app_Ee_Fstar}
\end{equation}

The plastic velocity gradient is defined as
\begin{equation}
\bm L^{\mathrm p}
=
\dot{\bm F}^{\mathrm p}
\left(\bm F^{\mathrm p}\right)^{-1}.
\label{eq:app_Lp}
\end{equation}
Taking the time derivative of
Eq.~\eqref{eq:app_Fstar} gives
\begin{equation}
\begin{aligned}
\dot{\bm F}^{\star}
&=
\dot{\bm F}
\left(\bm F^{\mathrm p}\right)^{-1}
+
\bm F
\frac{\mathrm d}{\mathrm dt}
\left(\bm F^{\mathrm p}\right)^{-1}
\\
&=
\dot{\bm F}
\left(\bm F^{\mathrm p}\right)^{-1}
-
\bm F
\left(\bm F^{\mathrm p}\right)^{-1}
\dot{\bm F}^{\mathrm p}
\left(\bm F^{\mathrm p}\right)^{-1}
\\
&=
\dot{\bm F}
\left(\bm F^{\mathrm p}\right)^{-1}
-
\bm F^{\star}\bm L^{\mathrm p}.
\end{aligned}
\label{eq:app_Fstar_rate_full}
\end{equation}
To identify the thermodynamic force associated with plastic
flow, the contribution proportional to
$\bm L^{\mathrm p}$ is isolated while the total deformation
$\bm F$ and the inelastic deformation
$\bm F^{\mathrm i}$ are held fixed with respect to this variation.
Equation~\eqref{eq:app_Fstar_rate_full} therefore gives
\begin{equation}
\left.\dot{\bm F}^{\star}\right|_{\mathrm p}
=
-\bm F^{\star}\bm L^{\mathrm p}.
\label{eq:app_Fstar_rate}
\end{equation}
Accordingly, the plastic contribution to the rate of the elastic strain is
\begin{equation}
\begin{aligned}
\left.\dot{\bm E}^{\mathrm e}\right|_{\mathrm p}
&=
\frac{1}{2}
\left[
\left(
\left.\dot{\bm F}^{\star}\right|_{\mathrm p}
\right)^{\mathsf T}
\bm F^{\star}
+
\left(\bm F^{\star}\right)^{\mathsf T}
\left.\dot{\bm F}^{\star}\right|_{\mathrm p}
\right]
\\
&=
-\frac{1}{2}
\left[
\left(\bm L^{\mathrm p}\right)^{\mathsf T}
\left(\bm F^{\star}\right)^{\mathsf T}
\bm F^{\star}
+
\left(\bm F^{\star}\right)^{\mathsf T}
\bm F^{\star}
\bm L^{\mathrm p}
\right].
\end{aligned}
\label{eq:app_Ee_rate_plastic}
\end{equation}
For convenience, define
\begin{equation}
\bm C^{\star}
=
\left(\bm F^{\star}\right)^{\mathsf T}
\bm F^{\star},
\label{eq:app_Cstar}
\end{equation}
which is symmetric by construction. Let $\psi$ denote the elastic
strain-energy density. Since
$\bm S=\partial\psi/\partial\bm E^{\mathrm e}$ is work-conjugate to
$\bm E^{\mathrm e}$ and is also symmetric, the plastic contribution to
the elastic energy rate becomes
\begin{equation}
\begin{aligned}
\left.\dot{\psi}\right|_{\mathrm p}
&=
\bm S:
\left.\dot{\bm E}^{\mathrm e}\right|_{\mathrm p}
\\
&=
-\frac{1}{2}
\bm S:
\left[
\left(\bm L^{\mathrm p}\right)^{\mathsf T}\bm C^{\star}
+
\bm C^{\star}\bm L^{\mathrm p}
\right].
\end{aligned}
\label{eq:app_plastic_power_intermediate}
\end{equation}
Using the symmetry of $\bm S$ and $\bm C^{\star}$,
\begin{equation}
\bm S:
\left(\bm L^{\mathrm p}\right)^{\mathsf T}\bm C^{\star}
=
\bm S:
\bm C^{\star}\bm L^{\mathrm p}
=
\left(\bm C^{\star}\bm S\right):\bm L^{\mathrm p},
\label{eq:app_mandel_contraction}
\end{equation}
and hence
\begin{equation}
\left.\dot{\psi}\right|_{\mathrm p}
=
-
\bm M^{\mathrm p}:\bm L^{\mathrm p},
\label{eq:app_plastic_power}
\end{equation}
where the Mandel stress thermodynamically conjugate to the plastic
velocity gradient is
\begin{equation}
\begin{aligned}
\bm M^{\mathrm p}
&=
\bm C^{\star}\bm S
\\
&=
\left(\bm F^{\star}\right)^{\mathsf T}
\bm F^{\star}\bm S
\\
&=
\left(\bm F^{\mathrm i}\right)^{\mathsf T}
\left(\bm F^{\mathrm e}\right)^{\mathsf T}
\bm F^{\mathrm e}
\bm F^{\mathrm i}\bm S.
\end{aligned}
\label{eq:app_mandel_exact}
\end{equation}
Equation~\eqref{eq:app_mandel_exact} gives the exact driving stress
associated with the evolution of $\bm F^{\mathrm p}$ for the adopted
kinematic decomposition.

From the definition of the elastic strain in
Eq.~\eqref{eq:app_Ee_Fstar},
\begin{equation}
2\bm E^{\mathrm e}
=
\left(\bm F^{\mathrm i}\right)^{\mathsf T}
\left[
\left(\bm F^{\mathrm e}\right)^{\mathsf T}
\bm F^{\mathrm e}
-
\bm I
\right]
\bm F^{\mathrm i},
\label{eq:app_Ee_expanded}
\end{equation}
or, equivalently,
\begin{equation}
\left(\bm F^{\mathrm i}\right)^{\mathsf T}
\left(\bm F^{\mathrm e}\right)^{\mathsf T}
\bm F^{\mathrm e}
\bm F^{\mathrm i}
=
\left(\bm F^{\mathrm i}\right)^{\mathsf T}
\bm F^{\mathrm i}
+
2\bm E^{\mathrm e}.
\label{eq:app_mandel_identity}
\end{equation}
The exact Mandel stress may therefore also be written as
\begin{equation}
\bm M^{\mathrm p}
=
\left[
\left(\bm F^{\mathrm i}\right)^{\mathsf T}
\bm F^{\mathrm i}
+
2\bm E^{\mathrm e}
\right]\bm S.
\label{eq:app_mandel_Ee}
\end{equation}

In metals, the elastic stretch is small,
although the lattice rotation may be finite. The elastic right
Cauchy--Green tensor $\bm C^{\mathrm e}$ therefore satisfies
\begin{equation}
\bm C^{\mathrm e}
=
\left(\bm F^{\mathrm e}\right)^{\mathsf T}
\bm F^{\mathrm e}
\simeq
\bm I,
\qquad
\left\|
\bm C^{\mathrm e}-\bm I
\right\|
\ll 1.
\label{eq:app_small_elastic_strain}
\end{equation}
Accordingly,
\begin{equation}
\left(\bm F^{\mathrm i}\right)^{\mathsf T}
\left(\bm F^{\mathrm e}\right)^{\mathsf T}
\bm F^{\mathrm e}
\bm F^{\mathrm i}
\simeq
\left(\bm F^{\mathrm i}\right)^{\mathsf T}
\bm F^{\mathrm i},
\end{equation}
and Eq.~\eqref{eq:app_mandel_exact} reduces to
\begin{equation}
\bm M^{\mathrm p}
\simeq
\left(\bm F^{\mathrm i}\right)^{\mathsf T}
\bm F^{\mathrm i}\bm S,
\label{eq:app_mandel_approx}
\end{equation}
which is the Mandel stress used in Eq.~\eqref{eq:cp_mandel} and the form evaluated in DAMASK~\cite{roters2019damask}.

\subsection{Mapping of the eigen-deformation velocity gradient}
\label{app:eigen_rate_mapping}

The multiplicative decomposition adopted in the present CP
formulation,
\begin{equation}
\bm F
=
\bm F^{\mathrm e}
\bm F^{\mathrm i}
\bm F^{\mathrm p},
\end{equation}
introduces the sequence of configurations
\begin{equation}
\mathcal B_{0}
\xrightarrow{\;\bm F^{\mathrm p}\;}
\mathcal B_{\mathrm p}
\xrightarrow{\;\bm F^{\mathrm i}\;}
\mathcal B_{\mathrm i}
\xrightarrow{\;\bm F^{\mathrm e}\;}
\mathcal B_{t},
\label{eq:app_configuration_sequence}
\end{equation}
where $\mathcal B_{0}$, $\mathcal B_{\mathrm p}$,
$\mathcal B_{\mathrm i}$, and $\mathcal B_{t}$ denote the reference,
plastic, intermediate, and current configurations, respectively.
In particular, $\bm F^{\mathrm p}$ maps the reference configuration to
the isoclinic plastic configuration, while $\bm F^{\mathrm i}$ maps the
plastic configuration to the intermediate configuration.

The constitutive laws are evaluated in the plastic configuration. Let
$\widehat{\bm L}^{\mathrm i}$ denote the eigen-deformation rate supplied
by the constitutive model in $\mathcal B_{\mathrm p}$. The evolution of
the eigen-deformation gradient is written as
\begin{equation}
\dot{\bm F}^{\mathrm i}
=
\bm L^{\mathrm i}\bm F^{\mathrm i},
\label{eq:app_Fi_evolution}
\end{equation}
where $\bm L^{\mathrm i}$ acts in the intermediate configuration
$\mathcal B_{\mathrm i}$. The constitutive eigen rate must therefore be
mapped from $\mathcal B_{\mathrm p}$ to $\mathcal B_{\mathrm i}$.
DAMASK~\cite{roters2019damask} performs this push-forward as
\begin{equation}
\bm L^{\mathrm i}
=
J^{\mathrm i}
\bm F^{\mathrm i}
\widehat{\bm L}^{\mathrm i}
\left(\bm F^{\mathrm i}\right)^{-1},
\qquad
J^{\mathrm i}
=
\det\!\left(\bm F^{\mathrm i}\right),
\label{eq:app_Li_pushforward}
\end{equation}
with the inverse relation
\begin{equation}
\widehat{\bm L}^{\mathrm i}
=
\frac{1}{J^{\mathrm i}}
\left(\bm F^{\mathrm i}\right)^{-1}
\bm L^{\mathrm i}
\bm F^{\mathrm i}
.
\label{eq:app_Li_pullback}
\end{equation}
Equation~\eqref{eq:app_Li_pushforward} is adopted here as it is
implemented in DAMASK.

The mapping can be read as the dual of a Piola-type pull-back of the
conjugate stress. Let $\bm M^{\mathrm i}$ denote the stress measure
paired with $\bm L^{\mathrm i}$ in the intermediate configuration, and
define its counterpart in the plastic configuration as
\begin{equation}
\widehat{\bm M}^{\mathrm i}
=
J^{\mathrm i}
\left(\bm F^{\mathrm i}\right)^{\mathsf T}
\bm M^{\mathrm i}
\left(\bm F^{\mathrm i}\right)^{-\mathsf T}.
\label{eq:app_Mi_pullback}
\end{equation}
For any second-order tensors $\bm A$ and $\bm B$ and any invertible
$\bm F$,
\begin{equation}
\left(
\bm F^{\mathsf T}
\bm A
\bm F^{-\mathsf T}
\right)
:
\bm B
=
\bm A
:
\left(
\bm F
\bm B
\bm F^{-1}
\right).
\label{eq:app_double_contraction_identity}
\end{equation}
Applying this identity to Eqs.~\eqref{eq:app_Li_pushforward} and
\eqref{eq:app_Mi_pullback} shows that the two mappings preserve the
pairing between stress and rate,
\begin{equation}
\widehat{\bm M}^{\mathrm i}
:
\widehat{\bm L}^{\mathrm i}
=
\bm M^{\mathrm i}
:
\left[
J^{\mathrm i}
\bm F^{\mathrm i}
\widehat{\bm L}^{\mathrm i}
\left(\bm F^{\mathrm i}\right)^{-1}
\right]
=
\bm M^{\mathrm i}
:
\bm L^{\mathrm i}.
\label{eq:app_power_conjugacy}
\end{equation}

For thermal expansion, the constitutive model evaluates the eigen rate
in the plastic configuration as
\begin{equation}
\widehat{\bm L}^{\mathrm i}
=
\dot T\,\bm\alpha(T),
\label{eq:app_Li_thermal_constitutive}
\end{equation}
where $\bm\alpha(T)$ is expressed in the crystal basis associated with
the plastic configuration. Substitution into
Eq.~\eqref{eq:app_Li_pushforward} yields the eigen-deformation velocity
gradient entering the evolution of $\bm F^{\mathrm i}$,
\begin{equation}
\bm L^{\mathrm i}
=
\det\!\left(\bm F^{\mathrm i}\right)
\bm F^{\mathrm i}
\left[
\dot T\,\bm\alpha(T)
\right]
\left(\bm F^{\mathrm i}\right)^{-1}
.
\label{eq:app_Li_thermal}
\end{equation}

\subsection{Consistent constitutive tangent}
\label{app:consistent_tangent}

This appendix summarizes the time integration of the constitutive equations of Section~\ref{sec251} and the consistent tangent $\partial\bm P/\partial\bm F$ entering Eq.~\eqref{eq:local_tangent_matrix}. Both the integration scheme and the tangent are those of the crystallite update in DAMASK~\cite{roters2019damask}. They are restated here in index notation, and the node index is omitted.

\paragraph{Implicit update}

Consider an increment $[t_n,t_{n+1}]$ of length $\Delta t$ with the deformation gradient $\bm F$ prescribed at $t_{n+1}$. The flow rules of Eq.~\eqref{eq:cp_flow_rules} are integrated with a backward Euler scheme, which, written for the inverse deformation gradients, reads
\begin{equation}
\begin{aligned}
\left(\bm F^{\mathrm p}\right)^{-1}
&=
\left(\bm F^{\mathrm p}_{n}\right)^{-1}
\left(\bm I-\Delta t\,\bm L^{\mathrm p}\right),
\\
\left(\bm F^{\mathrm i}\right)^{-1}
&=
\left(\bm F^{\mathrm i}_{n}\right)^{-1}
\left(\bm I-\Delta t\,\bm L^{\mathrm i}\right),
\end{aligned}
\label{eq:app_implicit_update}
\end{equation}
where quantities without subscript refer to $t_{n+1}$. The elastic deformation gradient follows from Eq.~\eqref{eq:multiplicative_decomposition},
\begin{equation}
\bm F^{\mathrm e}
=
\bm F
\left(\bm F^{\mathrm p}\right)^{-1}
\left(\bm F^{\mathrm i}\right)^{-1},
\label{eq:app_Fe}
\end{equation}
the stress $\bm S$ from Eq.~\eqref{eq:cp_hooke}, the Mandel stress from Eq.~\eqref{eq:cp_mandel}, and the velocity gradients from Eqs.~\eqref{eq:cp_Lp} and \eqref{eq:cp_Li_thermal}. The two velocity gradients are the unknowns of the update. They are determined from the residuals
\begin{equation}
\begin{aligned}
\bm R^{\mathrm p}
&=
\bm L^{\mathrm p}
-
\bm L^{\mathrm p}\!\left(\bm M^{\mathrm p},\bm\xi\right)
=
\bm 0,
\\
\bm R^{\mathrm i}
&=
\bm L^{\mathrm i}
-
\bm L^{\mathrm i}\!\left(\bm F^{\mathrm i},T,\dot T\right)
=
\bm 0,
\end{aligned}
\label{eq:app_residuals}
\end{equation}
by nested Newton iterations, the inner one for $\bm L^{\mathrm p}$ at fixed $\bm L^{\mathrm i}$ and the outer one for $\bm L^{\mathrm i}$. The slip resistances $\bm\xi$ are integrated in a staggered fixed-point iteration with the stress update. Within the stress update they are held at their current iterate. The converged state gives the first Piola--Kirchhoff stress
\begin{equation}
\bm P
=
\bm F
\left(\bm F^{\mathrm p}\right)^{-1}
\bm S
\left(\bm F^{\mathrm p}\right)^{-\mathsf T}.
\label{eq:app_P}
\end{equation}

\paragraph{Linearization}

The tangent is obtained by differentiating the converged update with respect to $\bm F$, holding the state at $t_n$ and the slip resistances $\bm\xi$ at their converged values fixed. For the thermal eigen deformation, $\partial\bm L^{\mathrm i}/\partial\bm M^{\mathrm i}=\bm 0$ (Section~\ref{sec251}), so that $\bm L^{\mathrm i}$ and $\bm F^{\mathrm i}$ do not depend on $\bm F$ within the increment, that is, $\mathrm d\bm F^{\mathrm i}/\mathrm d\bm F=\bm 0$. A perturbation of $\bm F$ therefore enters $\bm S$ only through $\bm F^{\mathrm e}$. From Eqs.~\eqref{eq:app_implicit_update} and \eqref{eq:app_Fe},
\begin{equation}
\begin{aligned}
\mathrm d\bm F^{\mathrm e}
={}&
\mathrm d\bm F
\left(\bm F^{\mathrm p}\right)^{-1}
\left(\bm F^{\mathrm i}\right)^{-1}
\\
&-
\Delta t\,
\bm F
\left(\bm F^{\mathrm p}_{n}\right)^{-1}
\mathrm d\bm L^{\mathrm p}
\left(\bm F^{\mathrm i}\right)^{-1}.
\end{aligned}
\label{eq:app_dFe}
\end{equation}
The thermoelastic law of Eq.~\eqref{eq:cp_hooke} gives $\mathrm d\bm S=\mathbb{Y}:\mathrm d\bm F^{\mathrm e}$ with
\begin{equation}
Y_{ijkl}
=
\frac{\partial S_{ij}}{\partial F^{\mathrm e}_{kl}}
=
C_{ijpq}\,
F^{\mathrm i}_{lp}
\left(\bm F^{\mathrm e}\bm F^{\mathrm i}\right)_{kq},
\label{eq:app_dSdFe}
\end{equation}
where $C_{ijpq}$ are the components of $\mathbb C^{\mathrm c}(T)$ and the minor symmetry $C_{ijpq}=C_{ijqp}$ has been used. The flow rule of Eqs.~\eqref{eq:cp_Lp} to \eqref{eq:cp_flow_rule} gives $\mathrm d\bm L^{\mathrm p}=\bm\Lambda:\mathrm d\bm S$ with
\begin{equation}
\Lambda_{mnij}
=
\frac{\partial L^{\mathrm p}_{mn}}{\partial S_{ij}}
=
\sum_{\alpha=1}^{N_{\mathrm s}}
\frac{\partial\dot\gamma^{\alpha}}{\partial\tau^{\alpha}}\,
P^{\alpha}_{mn}
\left[
\left(\bm F^{\mathrm i}\right)^{\mathsf T}
\bm F^{\mathrm i}
\bm P^{\alpha}
\right]_{ij},
\label{eq:app_dLpdS}
\end{equation}
with
\begin{equation}
\frac{\partial\dot\gamma^{\alpha}}{\partial\tau^{\alpha}}
=
\frac{n\,\dot\gamma_0}{\xi^{\alpha}}
\left|
\frac{\tau^{\alpha}}{\xi^{\alpha}}
\right|^{n-1},
\label{eq:app_dgamma_dtau}
\end{equation}
where the factor in square brackets in Eq.~\eqref{eq:app_dLpdS} is the derivative of the resolved shear stress with respect to $\bm S$ through the Mandel stress of Eq.~\eqref{eq:cp_mandel}. Combining Eqs.~\eqref{eq:app_dFe} to \eqref{eq:app_dgamma_dtau} yields a linear system for the sensitivity of the stress,
\begin{equation}
\begin{aligned}
&\left(
\delta_{ic}\delta_{jd}
+
\Delta t\,
Y_{ijkl}
B_{klcd}
\right)
\frac{\mathrm d S_{cd}}{\mathrm d F_{ab}}
\\
&\qquad=
Y_{ijal}
\big(
(\bm F^{\mathrm p})^{-1}
(\bm F^{\mathrm i})^{-1}
\big)_{bl},
\end{aligned}
\label{eq:app_dSdF}
\end{equation}
with
\begin{equation}
B_{klcd}
=
\big(\bm F(\bm F^{\mathrm p}_{n})^{-1}\big)_{km}
\big((\bm F^{\mathrm i})^{-1}\big)_{nl}
\Lambda_{mncd},
\label{eq:app_B}
\end{equation}
which is a $9\times9$ system solved once per material node at every evaluation of the constitutive update, that is, at every Newton iteration of Section~\ref{sec252}. For a stress-dependent eigen velocity gradient, the system would also contain the sensitivities $\mathrm d\bm F^{\mathrm i}/\mathrm d\bm F$ and $\partial\bm L^{\mathrm p}/\partial\bm F^{\mathrm i}$. For the thermal eigen deformation, $\mathrm d\bm F^{\mathrm i}/\mathrm d\bm F=\bm 0$ and these contributions vanish. The sensitivity of the inverse plastic deformation gradient follows from Eq.~\eqref{eq:app_implicit_update},
\begin{equation}
\frac{\mathrm d(\bm F^{\mathrm p})^{-1}_{kl}}{\mathrm d F_{ab}}
=
-\Delta t
\big((\bm F^{\mathrm p}_{n})^{-1}\big)_{km}
\Lambda_{mlcd}
\frac{\mathrm d S_{cd}}{\mathrm d F_{ab}}.
\label{eq:app_dFpinv}
\end{equation}
Finally, differentiating Eq.~\eqref{eq:app_P} gives the consistent tangent
\begin{equation}
\begin{aligned}
\frac{\mathrm d P_{ij}}{\mathrm d F_{ab}}
={}&
\delta_{ia}
\big(
(\bm F^{\mathrm p})^{-1}
\bm S
(\bm F^{\mathrm p})^{-\mathsf T}
\big)_{bj}
\\
&+
F_{ik}\,
\frac{\mathrm d(\bm F^{\mathrm p})^{-1}_{kl}}{\mathrm d F_{ab}}
\big(
\bm S
(\bm F^{\mathrm p})^{-\mathsf T}
\big)_{lj}
\\
&+
\big(
\bm F
(\bm F^{\mathrm p})^{-1}
\big)_{ik}
\frac{\mathrm d S_{kl}}{\mathrm d F_{ab}}
\big((\bm F^{\mathrm p})^{-1}\big)_{jl}
\\
&+
\big(
\bm F
(\bm F^{\mathrm p})^{-1}
\bm S
\big)_{ik}
\frac{\mathrm d(\bm F^{\mathrm p})^{-1}_{jk}}{\mathrm d F_{ab}}.
\end{aligned}
\label{eq:app_dPdF}
\end{equation}
The fourth-order tensor of Eq.~\eqref{eq:app_dPdF}, arranged according to the row-wise vectorization of Section~\ref{sec252}, is the matrix $\bm{\mathsf L}_i$ of Eq.~\eqref{eq:local_tangent_matrix}. 

\end{appendices}

\bibliographystyle{unsrt}

\bibliography{sn-bibliography}

@article{PEIRCE19831951,
title = {Material rate dependence and localized deformation in crystalline solids},
journal = {Acta Metallurgica},
volume = {31},
number = {12},
pages = {1951-1976},
year = {1983},
issn = {0001-6160},
doi = {https://doi.org/10.1016/0001-6160(83)90014-7},
url = {https://www.sciencedirect.com/science/article/pii/0001616083900147},
author = {D. Peirce and R.J. Asaro and A. Needleman},
}

@article{BECKER19952701,
title = {Effects of grain interactions on deformation and local texture in polycrystals},
journal = {Acta Metallurgica et Materialia},
volume = {43},
number = {7},
pages = {2701-2719},
year = {1995},
issn = {0956-7151},
doi = {https://doi.org/10.1016/0956-7151(94)00460-Y},
url = {https://www.sciencedirect.com/science/article/pii/095671519400460Y},
author = {R. Becker and S. Panchanadeeswaran}
}

@article{LEBENSOHN19932611,
title = {A self-consistent anisotropic approach for the simulation of plastic deformation and texture development of polycrystals: Application to zirconium alloys},
journal = {Acta Metallurgica et Materialia},
volume = {41},
number = {9},
pages = {2611-2624},
year = {1993},
issn = {0956-7151},
doi = {https://doi.org/10.1016/0956-7151(93)90130-K},
author = {R.A. Lebensohn and C.N. Tomé},
}

@book{tome2023material,
  title={Material modeling with the visco-plastic self-consistent ({VPSC}) approach},
  author={Tom{\'e}, Carlos N and Lebensohn, Ricardo A},
  year={2023},
  publisher={Elsevier},
  address={Amsterdam},
  doi={10.1016/C2019-0-01217-4}
}

@article{berveiller1978extension,
  title={An extension of the self-consistent scheme to plastically-flowing polycrystals},
  author={Berveiller, M and Zaoui, A},
  journal={Journal of the Mechanics and Physics of Solids},
  volume={26},
  number={5--6},
  pages={325--344},
  year={1978},
  doi={10.1016/0022-5096(78)90003-0},
  publisher={Elsevier}
}

@article{zhao2004texture,
  title={A texture optimization study for minimum earing in aluminium by use of a texture component crystal plasticity finite element method},
  author={Zhao, Zi-Su and Mao, Wei-Min and Roters, F and Raabe, D},
  journal={Acta Materialia},
  volume={52},
  number={4},
  pages={1003--1012},
  year={2004},
  doi={10.1016/j.actamat.2003.03.001},
  publisher={Elsevier}
}

@article{ROTERS20101152,
title = {Overview of constitutive laws, kinematics, homogenization and multiscale methods in crystal plasticity finite-element modeling: Theory, experiments, applications},
journal = {Acta Materialia},
volume = {58},
number = {4},
pages = {1152--1211},
year = {2010},
issn = {1359-6454},
doi = {https://doi.org/10.1016/j.actamat.2009.10.058},
author = {F. Roters and P. Eisenlohr and L. Hantcherli and D.D. Tjahjanto and T.R. Bieler and D. Raabe},
}

@article{tjahjanto2015multiscale,
doi = {https://doi.org/10.1088/0965-0393/23/4/045005},
year = {2015},
month = {apr},
publisher = {IOP Publishing},
volume = {23},
number = {4},
pages = {045005},
author = {Tjahjanto, D D and Eisenlohr, P and Roters, F},
title = {Multiscale deep drawing analysis of dual-phase steels using grain cluster-based {RGC} scheme},
journal = {Modelling and Simulation in Materials Science and Engineering},
}

@article{eisenlohr2013spectral,
title = {A spectral method solution to crystal elasto-viscoplasticity at finite strains},
journal = {International Journal of Plasticity},
volume = {46},
pages = {37-53},
year = {2013},
issn = {0749-6419},
doi = {https://doi.org/10.1016/j.ijplas.2012.09.012},
author = {P. Eisenlohr and M. Diehl and R.A. Lebensohn and F. Roters}
}

@article{ shanthraj2015numerically,
Author = {Shanthraj, P. and Eisenlohr, P. and Diehl, M. and Roters, F.},
Title = {Numerically robust spectral methods for crystal plasticity simulations of heterogeneous materials},
Journal = {International Journal of Plasticity},
Year = {2015},
Volume = {66},
Pages = {31-45},
Month = {MAR},
DOI = {https://doi.org/10.1016/j.ijplas.2014.02.006},
ISSN = {0749-6419},
EISSN = {1879-2154},
ResearcherID-Numbers = {Diehl, Martin/A-2831-2016
   Eisenlohr, Philip/E-6866-2010
   },
ORCID-Numbers = {Shanthraj, Pratheek/0000-0002-6324-0306
   Diehl, Martin/0000-0002-3738-7363
   Eisenlohr, Philip/0000-0002-8220-5995
   Roters, Franz/0000-0002-9098-9566},
Unique-ID = {WOS:000349879500003},
}

@article{prabhu2026incorporation,
  title={Incorporation of Physics-Based Strengthening Coefficients into Phenomenological Crystal Plasticity Models},
  author={Prabhu, Nikhil and Diehl, Martin},
  journal={Advanced Engineering Materials},
  volume={28},
  number={6},
  pages={e202500390},
  year={2026},
  doi={10.1002/adem.202500390},
  publisher={Wiley Online Library}
}

@article{madec2017dislocation,
  title={Dislocation strengthening in {FCC} metals and in {BCC} metals at high temperatures},
  author={Madec, Ronan and Kubin, Ladislas P},
  journal={Acta Materialia},
  volume={126},
  pages={166--173},
  year={2017},
  doi={10.1016/j.actamat.2016.12.040},
  publisher={Elsevier}
}

@article{gierden2022review,
  title={A review of {FE}-{FFT}-based two-scale methods for computational modeling of microstructure evolution and macroscopic material behavior},
  author={Gierden, Christian and Kochmann, Julian and Waimann, Johanna and Svendsen, Bob and Reese, Stefanie},
  journal={Archives of Computational Methods in Engineering},
  volume={29},
  number={6},
  pages={4115--4135},
  year={2022},
  doi={10.1007/s11831-022-09735-6},
  publisher={Springer}
}

@article{ lebensohn2020spectral,
Author = {Lebensohn, Ricardo A. and Rollett, Anthony D.},
Title = {Spectral methods for full-field micromechanical modelling of
   polycrystalline materials},
Journal = {Computational Materials Science},
Year = {2020},
Volume = {173},
Pages = {109336},
Month = {FEB 15},
DOI = {https://doi.org/10.1016/j.commatsci.2019.109336},
ISSN = {0927-0256},
EISSN = {1879-0801},
ResearcherID-Numbers = {Rollett, Anthony/A-4096-2012
   Lebensohn, Ricardo/A-2494-2008},
ORCID-Numbers = {Lebensohn, Ricardo/0000-0002-3152-9105},
Unique-ID = {WOS:000506172700006},
}

@book{roters2011crystal,
  title={Crystal plasticity finite element methods: in materials science and engineering},
  author={Roters, Franz and Eisenlohr, Philip and Bieler, Thomas R and Raabe, Dierk},
  year={2010},
  publisher={Wiley-VCH},
  address={Weinheim},
  doi={10.1002/9783527631483}
}

@article{sachs1928ableitung,
  title={Zur {A}bleitung einer {F}liessbedingung},
  author={Sachs, G},
  journal={Zeitschrift des Vereines Deutscher Ingenieure},
  volume={72},
  pages={734--736},
  year={1928}
}

@article{taylor1938plastic,
  title={Plastic strain in metals},
  author={Taylor, Geoffrey Ingram},
  journal={Journal of the Institute of Metals},
  volume={62},
  pages={307--324},
  year={1938}
}

@article{ibragimova2021new,
title = {A new {ANN} based crystal plasticity model for {FCC} materials and its application to non-monotonic strain paths},
journal = {International Journal of Plasticity},
volume = {144},
pages = {103059},
year = {2021},
issn = {0749-6419},
doi = {https://doi.org/10.1016/j.ijplas.2021.103059},
url = {https://www.sciencedirect.com/science/article/pii/S0749641921001327},
author = {Olga Ibragimova and Abhijit Brahme and Waqas Muhammad and Julie Lévesque and Kaan Inal}
}

@article{IBRAGIMOVA2022103374,
title = {A convolutional neural network based crystal plasticity finite element framework to predict localised deformation in metals},
journal = {International Journal of Plasticity},
volume = {157},
pages = {103374},
year = {2022},
issn = {0749-6419},
doi = {https://doi.org/10.1016/j.ijplas.2022.103374},
url = {https://www.sciencedirect.com/science/article/pii/S0749641922001541},
author = {Olga Ibragimova and Abhijit Brahme and Waqas Muhammad and Daniel Connolly and Julie Lévesque and Kaan Inal}
}

@article{hu2024temporal,
title = {A temporal graph neural network for cross-scale modelling of polycrystals considering microstructure interaction},
journal = {International Journal of Plasticity},
volume = {179},
pages = {104017},
year = {2024},
issn = {0749-6419},
doi = {https://doi.org/10.1016/j.ijplas.2024.104017},
author = {Yuanzhe Hu and Guowei Zhou and Myoung-Gyu Lee and Peidong Wu and Dayong Li}
}

@article{roters2019damask,
title = {{DAMASK} – The {Düsseldorf} Advanced Material Simulation Kit for modeling multi-physics crystal plasticity, thermal, and damage phenomena from the single crystal up to the component scale},
journal = {Computational Materials Science},
volume = {158},
pages = {420-478},
year = {2019},
issn = {0927-0256},
doi = {https://doi.org/10.1016/j.commatsci.2018.04.030},
author = {F. Roters and M. Diehl and P. Shanthraj and P. Eisenlohr and C. Reuber and S.L. Wong and T. Maiti and A. Ebrahimi and T. Hochrainer and H.-O. Fabritius and S. Nikolov and M. Friák and N. Fujita and N. Grilli and K.G.F. Janssens and N. Jia and P.J.J. Kok and D. Ma and F. Meier and E. Werner and M. Stricker and D. Weygand and D. Raabe},
}

@article{BONATTI2022104697,
title = {On the importance of self-consistency in recurrent neural network models representing elasto-plastic solids},
journal = {Journal of the Mechanics and Physics of Solids},
volume = {158},
pages = {104697},
year = {2022},
issn = {0022-5096},
doi = {https://doi.org/10.1016/j.jmps.2021.104697},
author = {Colin Bonatti and Dirk Mohr}
}

@article{BONATTI2022103430,
title = {From {CP-FFT} to {CP-RNN}: Recurrent neural network surrogate model of crystal plasticity},
journal = {International Journal of Plasticity},
volume = {158},
pages = {103430},
year = {2022},
issn = {0749-6419},
doi = {https://doi.org/10.1016/j.ijplas.2022.103430},
author = {Colin Bonatti and Bekim Berisha and Dirk Mohr}
}

@article{wei2025orientation,
  title={Orientation-aware interaction-based deep material network in polycrystalline materials modeling},
  author={Wei, Ting-Ju and Su, Tung-Huan and Chen, Chuin-Shan},
  journal={Computer Methods in Applied Mechanics and Engineering},
  volume={441},
  pages={117977},
  year={2025},
  doi={10.1016/j.cma.2025.117977},
  publisher={Elsevier}
}

@article{wei2026parametric,
  title={A parametric multiscale surrogate framework based on texture-generalizable Deep Material Networks for polycrystal modeling},
  author={Wei, Ting-Ju and Su, Tung-Huan and Chen, Chuin-Shan},
  journal={Computer Methods in Applied Mechanics and Engineering},
  volume={461},
  pages={119250},
  year={2026},
  doi={10.1016/j.cma.2026.119250},
  publisher={Elsevier}
}

@misc{wei2026efficientnonlinearmultiscaleprediction,
      title={Efficient nonlinear multiscale prediction for unseen polycrystalline textures via self-supervised microstructure pretraining},
      author={Ting-Ju Wei and Chuin-Shan Chen},
      year={2026},
      eprint={2512.06770},
      archivePrefix={arXiv},
      primaryClass={cs.CE},
      url={https://arxiv.org/abs/2512.06770}, 
}

@article{wei2027deep,
  title={Deep material network for homogenization of piezoelectric composites},
  author={Wei, Ting-Ju and Lu, Yen-Ming and Chen, Chuin-Shan},
  journal={Computer Methods in Applied Mechanics and Engineering},
  volume={464},
  pages={119536},
  year={2027},
  doi={10.1016/j.cma.2026.119536},
  publisher={Elsevier}
}

@article{marki2024nonlinear,
  title={Nonlinear optimization for compact representation of orientation distributions based on generalized spherical harmonics},
  author={Marki, Russell E and Knezevic, Marko},
  journal={Journal of the Mechanics and Physics of Solids},
  volume={187},
  pages={105609},
  year={2024},
  doi={10.1016/j.jmps.2024.105609},
  publisher={Elsevier}
}

@article{WeiOverview,
author = {Ting-Ju Wei and Wen-Ning Wan and Chuin-Shan Chen},
title = {Deep Material Network: Overview, Applications and Current Directions},
journal = {Multiscale Science and Engineering},
volume = {8},
pages = {1--19},
doi = {https://doi.org/10.1007/s42493-026-00146-4},
year = {2026}
}

@article{wan2024decoding,
  title={Decoding material networks: exploring performance of deep material network and interaction-based material networks},
  author={Wan, Wen-Ning and Wei, Ting-Ju and Su, Tung-Huan and Chen, Chuin-Shan},
  journal={Journal of Mechanics},
  volume={40},
  pages={796--807},
  year={2024},
  publisher={Oxford University Press},
  doi = {https://doi.org/10.1093/jom/ufae053}
}

@article{noels2022micromechanics,
title = {Micromechanics-based material networks revisited from the interaction viewpoint; robust and efficient implementation for multi-phase composites},
journal = {European Journal of Mechanics - A/Solids},
volume = {91},
pages = {104384},
year = {2022},
issn = {0997-7538},
doi = {https://doi.org/10.1016/j.euromechsol.2021.104384},
author = {Van Dung Nguyen and Ludovic Noels},
}

@article{haddad2026deep,
  title={A deep material network for multiscale modeling of Viscoelastic-Viscoplastic porous structures with stochastic porosity level distribution},
  author={Haddad, Mohamed and Nguyen, Van Dung and Noels, Ludovic and Doghri, Issam},
  journal={European Journal of Mechanics - A/Solids},
  volume={120},
  pages={106243},
  year={2026},
  doi={10.1016/j.euromechsol.2026.106243},
  publisher={Elsevier}
}

@article{noels2022interaction,
title = {Interaction-based material network: A general framework for (porous) microstructured materials},
journal = {Computer Methods in Applied Mechanics and Engineering},
volume = {389},
pages = {114300},
year = {2022},
issn = {0045-7825},
doi = {https://doi.org/10.1016/j.cma.2021.114300},
author = {Van Dung Nguyen and Ludovic Noels},
}

@article{shin2024deep2,
title = {Deep material network for thermal conductivity problems: Application to woven composites},
journal = {Computer Methods in Applied Mechanics and Engineering},
volume = {431},
pages = {117279},
year = {2024},
issn = {0045-7825},
doi = {https://doi.org/10.1016/j.cma.2024.117279},
author = {Dongil Shin and Peter Jefferson Creveling and Scott Alan Roberts and Rémi Dingreville},
}

@article{srinivas2026rapid,
  title={Rapid Offline Training for Deep Material Networks via a displacement-based laminate formulation and a novel sampling technique for a compliance-based fatigue model},
  author={Srinivas, Pavan Bhat Keelanje and Kabel, Matthias and Schneider, Matti},
  journal={Computer Methods in Applied Mechanics and Engineering},
  volume={449},
  pages={118517},
  year={2026},
  doi={10.1016/j.cma.2025.118517},
  publisher={Elsevier}
}

@article{shin2024deep,
  title={A deep material network approach for predicting the thermomechanical response of composites},
  author={Shin, Dongil and Alberdi, Ryan and Lebensohn, Ricardo A and Dingreville, R{\'e}mi},
  journal={Composites Part B: Engineering},
  volume={272},
  pages={111177},
  year={2024},
  doi={10.1016/j.compositesb.2023.111177},
  publisher={Elsevier}
}

@book{bunge2013texture,
  title={Texture analysis in materials science: mathematical methods},
  author={Bunge, H-J},
  year={2013},
  publisher={Elsevier},
  address={Amsterdam}
}

@article{ma2006dislocation,
  title={A dislocation density based constitutive model for crystal plasticity {FEM} including geometrically necessary dislocations},
  author={Ma, A and Roters, F and Raabe, D},
  journal={Acta Materialia},
  volume={54},
  number={8},
  pages={2169--2179},
  year={2006},
  doi={10.1016/j.actamat.2006.01.005},
  publisher={Elsevier}
}

@article{ liu2019deep,
Author = {Liu, Zeliang and Wu, C. T. and Koishi, M.},
Title = {A deep material network for multiscale topology learning and accelerated
   nonlinear modeling of heterogeneous materials},
Journal = {Computer Methods in Applied Mechanics and Engineering},
Year = {2019},
Volume = {345},
Pages = {1138-1168},
Month = {MAR 1},
DOI = {10.1016/j.cma.2018.09.020},
ISSN = {0045-7825},
EISSN = {1879-2138},
ResearcherID-Numbers = {Liu, Zeliang/AAB-7221-2019
   Liu, Zeliang/U-8269-2018},
ORCID-Numbers = {Liu, Zeliang/0000-0002-3264-8907},
Unique-ID = {WOS:000456953900047},
}

@article{li2024micromechanics,
title = {Micromechanics-informed parametric deep material network for physics behavior prediction of heterogeneous materials with a varying morphology},
journal = {Computer Methods in Applied Mechanics and Engineering},
volume = {419},
pages = {116687},
year = {2024},
issn = {0045-7825},
doi = {https://doi.org/10.1016/j.cma.2023.116687},
author = {Tianyi Li},
}

@article{ liu2019exploring,
Author = {Liu, Zeliang and Wu, C. T.},
Title = {Exploring the {3D} architectures of deep material network in data-driven
   multiscale mechanics},
Journal = {Journal of the Mechanics and Physics of Solids},
Year = {2019},
Volume = {127},
Pages = {20-46},
Month = {JUN},
DOI = {10.1016/j.jmps.2019.03.004},
ISSN = {0022-5096},
EISSN = {1873-4782},
ResearcherID-Numbers = {Liu, Zeliang/AAB-7221-2019
   Liu, Zeliang/U-8269-2018},
ORCID-Numbers = {Liu, Zeliang/0000-0002-3264-8907},
Unique-ID = {WOS:000466833900002},
}

@article{jean2024graph,
  title   = {Graph-enhanced deep material network: multiscale materials modeling with microstructural informatics},
  journal = {Computational Mechanics},
  volume  = {75},
  number  = {1},
  pages   = {113--136},
  year    = {2025},
  doi     = {https://doi.org/10.1007/s00466-024-02493-1},
  issn    = {1432-0924},
  author  = {Jean, Jimmy Gaspard and Su, Tung-Huan and Huang, Szu-Jui and Wu, Cheng-Tang and Chen, Chuin-Shan},
}

@article{gajek2020micromechanics,
title = {On the micromechanics of deep material networks},
journal = {Journal of the Mechanics and Physics of Solids},
volume = {142},
pages = {103984},
year = {2020},
issn = {0022-5096},
doi = {https://doi.org/10.1016/j.jmps.2020.103984},
author = {Sebastian Gajek and Matti Schneider and Thomas Böhlke},
}

@article{wei2025foundation,
title = {Foundation model for composite microstructures: Reconstruction, stiffness, and nonlinear behavior prediction},
journal = {Materials \& Design},
volume = {257},
pages = {114397},
year = {2025},
issn = {0264-1275},
doi = {https://doi.org/10.1016/j.matdes.2025.114397},
author = {Ting-Ju Wei and Chuin-Shan Chen},
}

@article{bachmann2010texture,
  title={Texture analysis with MTEX--free and open source software toolbox},
  author={Bachmann, F and Hielscher, Ralf and Schaeben, Helmut},
  journal={Solid state phenomena},
  volume={160},
  pages={63--68},
  year={2010},
  publisher={Trans Tech Publ}
}

@article{li2005texture,
title = {Texture evolution during multi-pass equal channel angular extrusion of copper: Neutron diffraction characterization and polycrystal modeling},
journal = {Acta Materialia},
volume = {53},
number = {7},
pages = {2111-2125},
year = {2005},
issn = {1359-6454},
doi = {https://doi.org/10.1016/j.actamat.2005.01.023},
author = {Saiyi Li and Irene J. Beyerlein and David J. Alexander and Sven C. Vogel},
}

@article{han2020microstructure,
  title={Microstructure-based multiscale modeling of large strain plastic deformation by coupling a full-field crystal plasticity-spectral solver with an implicit finite element solver},
  author={Han, Fengbo and Roters, Franz and Raabe, Dierk},
  journal={International Journal of Plasticity},
  volume={125},
  pages={97--117},
  year={2020},
  doi={10.1016/j.ijplas.2019.09.004},
  publisher={Elsevier}
}

@article{terada2000simulation,
  title={Simulation of the multi-scale convergence in computational homogenization approaches},
  author={Terada, Kenjiro and Hori, Muneo and Kyoya, Takashi and Kikuchi, Noboru},
  journal={International Journal of Solids and Structures},
  volume={37},
  number={16},
  pages={2285--2311},
  year={2000},
  doi={10.1016/S0020-7683(98)00341-2},
  publisher={Elsevier}
}

@article{terada2001class,
  title={A class of general algorithms for multi-scale analyses of heterogeneous media},
  author={Terada, Kenjiro and Kikuchi, Noboru},
  journal={Computer Methods in Applied Mechanics and Engineering},
  volume={190},
  number={40--41},
  pages={5427--5464},
  year={2001},
  doi={10.1016/S0045-7825(01)00179-7},
  publisher={Elsevier}
}

@article{watanabe2008characterization,
  title={Characterization of macroscopic tensile strength of polycrystalline metals with two-scale finite element analysis},
  author={Watanabe, Ikumu and Terada, Kenjiro and {de Souza Neto}, Eduardo Alberto and Peri{\'c}, Djordje},
  journal={Journal of the Mechanics and Physics of Solids},
  volume={56},
  number={3},
  pages={1105--1125},
  year={2008},
  doi={10.1016/j.jmps.2007.06.001},
  publisher={Elsevier}
}

@article{terada2013method,
  title={A method of two-scale analysis with micro-macro decoupling scheme: application to hyperelastic composite materials},
  author={Terada, K. and Kato, J. and Hirayama, N. and Inugai, T. and Yamamoto, K.},
  journal={Computational Mechanics},
  volume={52},
  number={5},
  pages={1199--1219},
  year={2013},
  doi={10.1007/s00466-013-0872-5},
  publisher={Springer}
}

\end{document}